\documentclass[onecolumn]{aastex631}
\usepackage[figuresright]{rotating}

\usepackage{amsmath}

\def\ee #1 {\times 10^{#1}}          
\def\ut #1 #2 { \, \textrm{#1}^{#2}} 
\def\un #1 { \, \textrm{#1}}          
\newcommand{\msol}{\,\textrm{M}_\odot}                
\newcommand{\kms}{km\,s$^{-1}$}

\def\kms {\,\textrm{km\,s}^{-1}}

\def\la{\lower.4ex\hbox{$\;\buildrel <\over{\scriptstyle\sim}\;$}}
\def\ga{\lower.4ex\hbox{$\;\buildrel >\over{\scriptstyle\sim}\;$}}

\newcommand{\J}{\bm{J}}
\newcommand{\A}{\bm{A}}
\newcommand{\B}{\bm{B}}
\newcommand{\E}{\bm{E}}            

\begin{document}

\def\msol{\hbox{$\hbox{M}_\odot$}}
\def\lsol{\hbox{$\hbox{L}_\odot$}}
\def\kms{km s$^{-1}$}

\title{Statistical Properties of the Population of the Galactic Center\\
 Filaments II: The Spacing Between Filaments
}

\author[0000-0001-8551-9220]{F. Yusef-Zadeh} 
\affiliation{Dept Physics and Astronomy, CIERA, Northwestern University, 2145 Sheridan Road, Evanston , IL 60207, USA
(zadeh@northwestern.edu)}

\author[0000-0001-8403-8548]{R. G. Arendt} 
\affiliation{Code 665, NASA/GSFC, 8800 Greenbelt Road, Greenbelt, MD 20771, USA}
\affiliation{UMBC/CRESST 2 (Richard.G.Arendt@nasa.gov)}

\author[0000-0002-1737-0871]{M. Wardle} 
\affiliation{School of Mathematical and Physical Sciences,  Research Centre for Astronomy, Astrophysics
and Astrophotonics, Macquarie University, Sydney NSW 2109, Australia, (mark.wardle@mq.edu.au)}

\author[0000-0001-6252-5169]{S. Boldyrev} 
\affiliation{Dept of Physics,  University of Wisconsin, WI, (boldyrev@wisc.edu)}

\author[0000-0001-6864-5057]{I. Heywood} 
\affiliation{Astrophysics, Department  of Physics, University of Oxford, Keble Road, Oxford, OX1 3RH, UK}
\affiliation{Department of Physics and Electronics, Rhodes University, PO Box 94, Makhanda, 6140, South Africa (ian.heywood@physics.ox.ac.uk)}

\author[0000-0001-7363-6489]{W. Cotton} 
\affiliation{National Radio Astronomy Observatory, Charlottesville, VA, USA, (bcotton@nrao.edu)}

\author[0000-0002-1873-3718]{F. Camilo} 
\affiliation{South African Radio Astronomical Observatory, 2 Fir Street, Black River Park, Observatory,
Cape Town, 7925, South Africa, (fernando@ska.ac.za)}

\begin{abstract} 
We carry out  a population study of magnetized radio filaments in the Galactic center using MeerKAT data by focusing on the spacing between the filaments that are grouped. The  
morphology of a sample of 43 groupings containing 174 magnetized radio filaments are presented. Many grouped filaments show harp-like, fragmented cometary tail-like, or 
loop-like structures in contrast to many straight filaments running mainly perpendicular to the Galactic plane. There are many striking examples of a single filament splitting 
into two prongs at a junction, suggestive of a flow of plasma along the filaments. Spatial variations in spectral index, brightness, bending and sharpening along the filaments 
indicate that they are evolving on a $10^{5-6}$-year time scale. The mean spacings between parallel filaments in a given grouping peaks at $\sim16''$. Modeling of the spacing between the 
filaments indicates that the filaments in a  group  all lie on a same plane and are isotropically oriented in 3D space. One candidate for the origin of filamentation is interaction with an obstacle, which could be a 
compact radio source, before a filament splits and bends into multiple filaments. In this picture, the obstacle or sets the length scale of the separation between the 
filaments. Another possibility is synchrotron cooling instability occurring in cometary tails formed as a result of the interaction of cosmic-ray driven Galactic center 
outflow with obstacles such as stellar winds. 
In this picture, the mean spacing and the mean width of the filaments are expected to be a fraction of a parsec, consistent with  observed spacing.  
\end{abstract}


\section{Introduction} 
\label{sec:intro}


Apart from the Sgr A complex where the  supermassive black hole lies at the center of the Galaxy,  
the first hint that the nucleus of our Galaxy harbored energetic activity was the discovery of the prototype 
magnetized radio filaments in the Galactic center Radio Arc near $l\sim0.2^\circ$ more than three decades ago 
\citep{zadeh84}. Since then, VLA observations have shown linearly polarized synchrotron emission tracing 
nucleus-wide cosmic ray activity throughout the inner few hundred pc of the Galaxy 
\citep{liszt85,zadeh86,bally89,gray91,haynes92,staguhn98,lang99a,lang99b,larosa01,larosa04,zadeh04,nord04,law08,pound18,staguhn19,arendt19}. 
MeerKAT observations have recently provided a remarkable mosaic of the inner few degrees of the Galactic center 
revealing a radio bubble of $\sim400$ pc in exquisite detail with  $4''-6''$ spatial resolution 
\citep{heywood19,heywood22,zadeh22}. Each of the filaments tracks a source of 
cosmic rays suggesting high cosmic-ray flux in the Galactic center. 
H$_3^+$ absorption line observations \citep{oka19} 
also indicate high cosmic ray ionization rates permeating the 
Central Molecular Zone (CMZ) at levels a thousand times that in the solar neighborhood. One of the consequences of the  interaction of 
cosmic ray particles with the gas in the CMZ is to heat the gas to higher temperatures \citep{zadeh13}. Multiple observational 
studies show  that warm gas temperatures characterize molecular clouds in the CMZ \citep{henshaw22}. 


{\it Chandra}, {\it XMM} 
and {\it NuSTAR} have  detected X-ray emission from a handful of nonthermal radio filaments. Five prominent 
filaments have been studied in detail, G359.89--0.08 (Sgr A-E), G359.54+0.18 (ripple), G359.90-0.06 
(Sgr A-F), G0.13-0.11, and  G0.173-0.420 \citep{sakano03,lu03,lu08,zadeh05,zhang14,zhang20,zadeh21}. Unlike the 
long and distinct radio filaments, there are also several short, linear X-ray features  identified 
within  6$'$ of Sgr A*, the supermassive black hole at the Galactic center \citep{muno08,johnson09,lu08}.


Although numerous  models have been proposed to explain the origin of the filaments, there is 
no consensus on how these filaments are produced 
\citep{nicholls95,bicknell01,dahlburg02,zadeh03, ferriere09,barragan16}.
Broadly speaking nonthermal radio filament (NRF) models invoke  
an energetic event or  compact source 
accelerating cosmic rays to high energies \citep{rosner96,shore99,zadeh19}  
or else a process that involves the  
environment of the Galactic center hosting 
high cosmic-ray flux,  highly turbulent medium and highly organized large-scale magnetic fields 
\citep{boldyrev06,thomas20,sofue20,coughlin21}.

The abundance of NRFs within  the radio bubble indicated that some  filaments are 
isolated but others are grouped together  \citep{heywood22,zadeh22}.  A 
statistical study characterizing the mean properties of the spectral index and the equipartition magnetic field of 
NRFs based on MeerKAT data were recently discussed  \citep{zadeh22}. 
The mean magnetic field strengths along the filaments  ranged
 between $\sim100$ to 400 $\mu$G depending on the assumed ratio of cosmic-ray protons to 
electrons. The large equipartition magnetic field along the filaments is different than 
large-scale  magnetic flux averaged over the  entire  Galactic center region. This suggests that the  magnetic field 
along the filaments is significantly amplified locally when compared to pervasive magnetic field in this region. 
The mean spectral indices ($\alpha$) of the 
filaments, where flux density $S_\nu\, \propto\, \nu^{\alpha}$, were also shown to be steeper than supernova remnants 
(SNRs). This characteristic of the filaments and its wide range in the spectral index variation may
 indicate different mechanism accelerating particles in the filaments than typical SNRs. 

Here we continue our  total intensity and spectral index studies but focus on  groups of filaments observed throughout  the 
Galactic center.  We present images of about 43  groupings 
 of filaments consisting of 174  individual filaments
using unfiltered and filtered images. 
We have not studied  single filaments that are not members of any  group of filaments. 
We show a new characteristic of the filaments in that they are spaced equally 
from each other with the mean value of $\sim16''$. We also determine the mean width span of all  filament groupings  is $\sim27''$. 
We interpret that filament separation is due to synchrotron cooling instability. 
Another possibility that we  discuss  is 
an interaction of  cosmic-ray driven outflow with an obstacle with a size scale of the width of a filament or a group of 
filaments triggering  filamentation.

\section{Observations}

\subsection{MeerKAT}

Details of MeerKAT observations including spectral index determination of the entire region and the procedure in  
making  high-pass filtering of mosaic 
images can be found in \cite{heywood22} and \cite{zadeh22}.  Briefly, the mosaic image is filtered using a difference of
 Gaussians to smooth out noise and remove large scale backgrounds in order to enhance the visibility of narrow filaments. The 
narrower, smoothing Gaussian function has $\sigma = 2.75''$ 
and the wider, background-subtracting Gaussian function has $\sigma = 4.95''$.  
Unfiltered images have FWHM $\sim 4''$ resolutions. Point sources in the final filtered image have FWHM $\sim 6.4''$, 
and negative side lobes that are $<11\%$ of the filtered peak value. 
The process reduces the numerical value of the brightness in filament pixels by a factor of $\sim$0.13, with large variations due to the effect 
of the background removal. This factor is not applied to the filtered image intensities, presented in Figure 2a to 29a, 
which need to be increased  by $\sim7.7$. However, this factor does not effect the calculation of 
spectral indices. The spectral index is determined using cubes of 16 filtered frequency channels. 
Pixels lacking sufficient signal are assigned ``not a number''  (NaN) values, and appear black in later figures.
To improve the signal-to-noise ratio (S/N), the median spectral index of each filament is determined from the median of 
the pixel spectral indices along its length. In cases where an insufficient number of pixels have valid spectral indices, the median 
spectra index may be reported as NaN. The statistical uncertainty of the mean spectral index is typically $\sim 0.1$, but 
can drop as low as $0.01$ for long bright filaments. Part of the uncertainty may be caused by intrinsic variation of the 
spectral index, but in most cases it is dominated by measurement uncertainties.

\subsection{VLA}

We  used  the Karl G. Jansky Very Large Array (VLA)
to obtain a higher resolution image of the region toward  Sgr C. 
 The observation was 
carried out at  L-band (1–2 GHz) with the array in the
most extended A-configuration and centered at 
Sgr C (J2000 17$^{\rm h}$44$^{\rm m}$35$\fs$0, $-29^\circ 29'\,00\farcs0$) for 2.5 hours. 
The initial flagging and reference calibration was performed using the VLA casa pipeline8 and processed 
with  the wsclean multiscale clean algorithm with a resolution of $\sim1''$. 
Further details on the use of the VLA data can be found in \cite{heywood22}.


\section{Results}


The new MeerKAT data reveal new  morphological and spectral details of NRFs.  
We identified a large sample of prominent groupings of  filaments with memberships 
ranging from pairs to dozens.  We define  a grouping 
as a set of filaments with   similar orientations, similar curvature or bending,  and spatially close to each other.
In some cases, the filaments in a grouping converge to a point,  
 shift sideways together,  
change direction coherently implying 
that they are parts of the same system of filaments with similar origin. 
These are distinct from  filaments along the line of sight
at large physical distances from each other but which appear close in projection. 
Here, we only focus on some of the most spectacular groupings  of filaments and  their spectral indices. 
The results are summarized below. 

\vskip 0.05in
{$\indent\bullet$}
Many  filaments in groupings  are  approximately equally separated from each other and run parallel to each other, giving 
the appearance of a harp. In some groupings, the brightest filament lies at one edge, suggesting  harmonics of 
the brightest filament followed by fainter filaments running parallel to the bright filament. 
In other   cases, 
the filament lengths in a grouping
progressively diminish toward zero. 

\vskip 0.05in
{$\indent\bullet$} 
Most filaments in a grouping  are  parallel to each other, change direction coherently,   
and  appear to converge  to an external point, similar to the shape of a cometary tail or meteor trail. 
The  filaments in many groupings  are bent gently, and  
some show loop-like structures bending  by $\sim90^\circ$.  
These morphological details suggest that grouping  filaments are threaded by the magnetic field and have similar origin. 

\vskip 0.05in
{$\indent\bullet$}
The spectral index  between individual grouping  filaments trends between  flat (hard) and steep (soft)
for short and long filaments. The change in the spectral index, filamentation  and morphological changes such as 
the width,  and brightness of the filament give us clues to   the evolution of the  relativistic electron population 
along the filaments.


\vskip 0.05in
{$\indent\bullet$} 
In some  instances, a single filament encounters a compact source or a region of enhanced emission  and 
then splits   into two parallel fainter filamentary components. Spectral variations are noted across 
 the length of filaments before and after splitting. 
This suggests  filaments carry a flow of plasma along their lengths and that 
 an interaction  with an obstacle such as a stellar bubble or a planetary nebula 
 triggers filamentation. 
In a  	grouping  with a number of parallel filaments,  
we   explore the possibility that 
filamentation occurs because of synchrotron cooling instability. 


\vskip 0.05in






\subsection{Total intensity  and spectral index images of filament groups} 

\begin{sidewaysfigure}
\epsscale{1.2}
\plotone{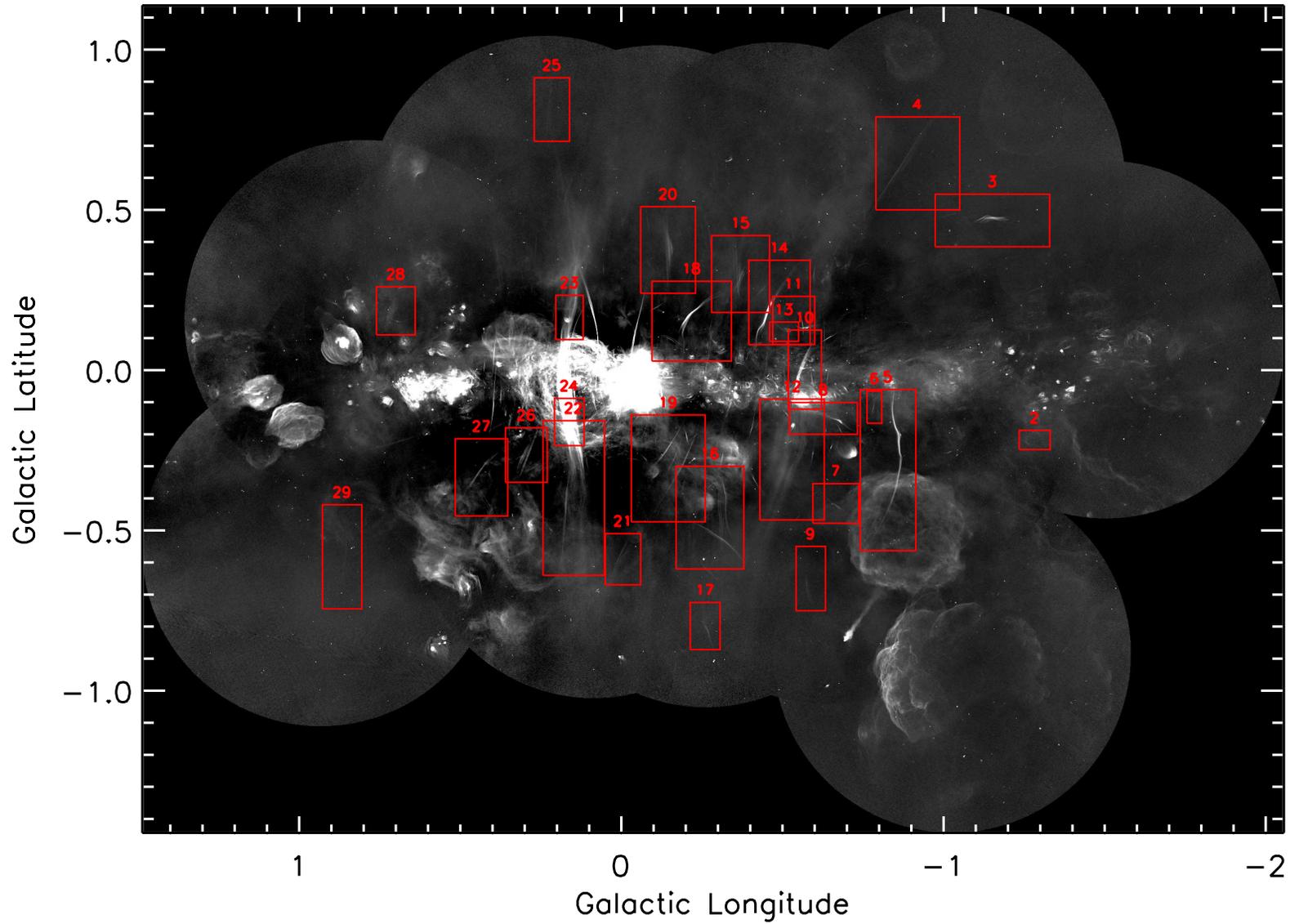}
\caption{A 1.28 GHz mosaic image of the  Galactic center region \citep{heywood22}. Red boxes show groupings that 
are described here. There are 27 boxes corresponding to the number of figures  between 2 and 28. 
\label{fig:regions}}
\end{sidewaysfigure}

\begin{sidewaysfigure}
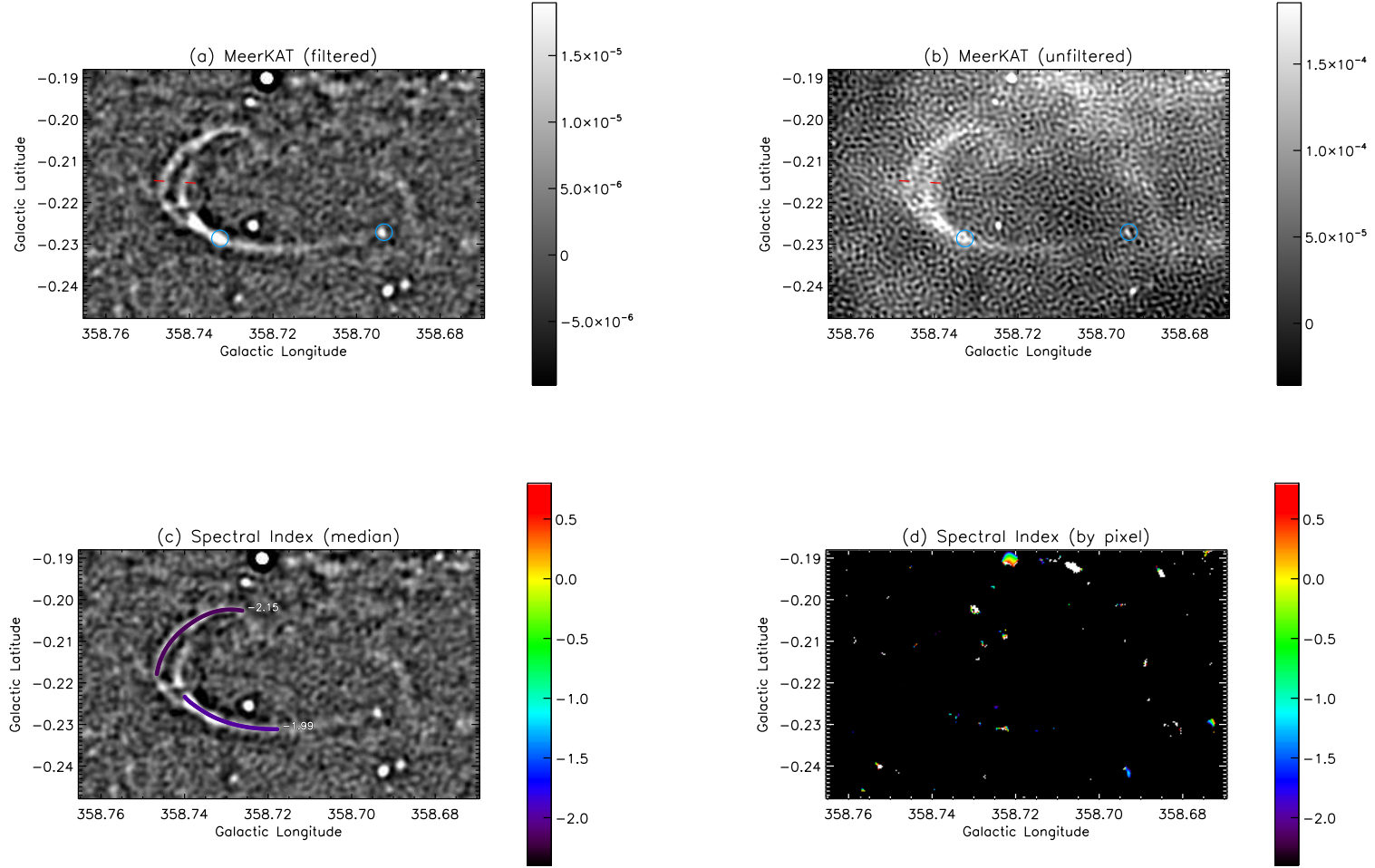

\plottwo{{cross_section_filtered_G358.72-0.22}.eps}{{cross_section_G358.72-0.22}.eps}
\plottwo{{detected_G358.72-0.22}.eps}{{alpha_G358.72-0.22}.eps}
\caption{(
(The Horseshoe) Filtered (a) and unfiltered (b) 1.28 GHz continuum images of a
horseshoe-shaped grouping G358.743-0.215 are shown in the top two panels. Mean
(c) and pixel-by-pixel (d) spectral indices of the filaments are displayed in 
the bottom two panels. The red markings seen as two short line segments in unfiltered images
indicate where the filament spacing was measured, as listed in Table 1. The
tabulated coordinates are where the center of the marking lies. Blue and
red circles represent point (or compact) sources and general locations of
interest, respectively. The scale bars with the filtered and unfiltered images
show the intensity grayscale Jy per $4''\times 4''$ beam. The intensity of the filtered
image does not account for changes in beam which is roughly a filtering factor of
7.1. The color scale bars show spectral index values in panels (c) and (d). 
As a more precise guide, the median spectral index value is printed at one end
of each detected filament in panel (c). ``NaN'' values indicate filaments that are
detected in intensity but are too faint to have reliable spectral indices.
\label{fig:horseshoe}}
\end{sidewaysfigure}

Figure \ref{fig:regions} shows the MeerKAT mosaic image of Galactic center \citep{heywood22} 
and the red boxes highlight identified groupings of filaments. 
With the exception of the Radio Arc near $l\sim0.2^\circ$,  
most prominent groupings   lie at  negative longitudes.  
Figures \ref{fig:horseshoe} to \ref{fig:meteor} display structural details of each grouping  of filaments in each
 box as  well as their spectral indices. 
Each of these figures consists of four sub-images. The top panels
show  the surface  brightness  using   filtered and  unfiltered  data (a, b). 
The bottom panels  show the median and  pixel-by-pixel spectral index images  of individual filaments (c, d).   Positions  of individual compact sources and other locations of interest (discussed below) 
are marked  as  blue circles. Because of limited spatial resolution, it is not clear if any of 
the identified compact sources are related to filaments, nevertheless, the apparent association is suggestive of interaction of compact radio sources  with filaments.  
Opposing pairs of red tick marks overlaid on the  filtered and unfiltered images designate the separation 
between the inner and outer filaments The mean spacing was evaluated by counting the number of individual 
filaments.  Table \ref{tab:table1} tabulates the separations measured for various filament groupings. Nicknames  are given to individual 
groupings  based on their appearance, 
Galactic coordinates $(l,b)$ at the center of the 
tick marks drawn on filtered and unfiltered images, the angular separation between the inner and outer identified 
filaments, the number of filaments, and the mean separation of individual filaments in columns 2 to 7, 
respectively. The last column lists the relevant figure number of enlarged figures as indicated in Figure \ref{fig:regions}.
We identified a number of compact sources and sources of interest,  their  positions  are drawn on figures as blue and 
red circles, respectively. Table \ref{tab:table2} lists the coordinates of sources,  where their names and  Galactic coordinates $(l,b)$  
 are tabulated in columns 1 to 3. 
Column 4 identifies   whether  the source is compact
or a position of interest, as drawn as red and blue circles on Figures 2 to 29, respectively.  The last column 
gives the figure number that compact sources are identified.

\subsubsection{Figure \ref{fig:horseshoe}: G358.743-0.215 (The Horseshoe)} 

358.743-0.215with its horseshoe-shaped structure is one of the few filament groupings   that is bent substantially and lies  along 
the Galactic plane, unlike typical  filaments 
that run perpendicular to the Galactic plane. 
The 
western side of the grouping  consists of a single filament, which  splits  into two components 
where there is a compact source  358.733-0.229
and  enhanced emission along the filament. 
The split filaments are separated by $\sim18''$ and  converge to the north suggesting that the magnetic 
field along the filaments is continuous and is threaded through both separated filamentary components. 
The spectral index of the brightest region where the filament splits into two loop-like filaments 
 is flatter by $\delta\alpha\sim0.16$
 than the  loops, suggesting that re-acceleration of particles took place in brighter region. If the 
magnetic field is initially straight, it is possible that 
large-scale  motion of surrounding plasma is responsible for significant bending of the magnetic field along the filament.   
There is a compact source 
at the western edge at G358.693-0.227, although it  is not clear if it is associated with the filament.

\begin{sidewaysfigure}
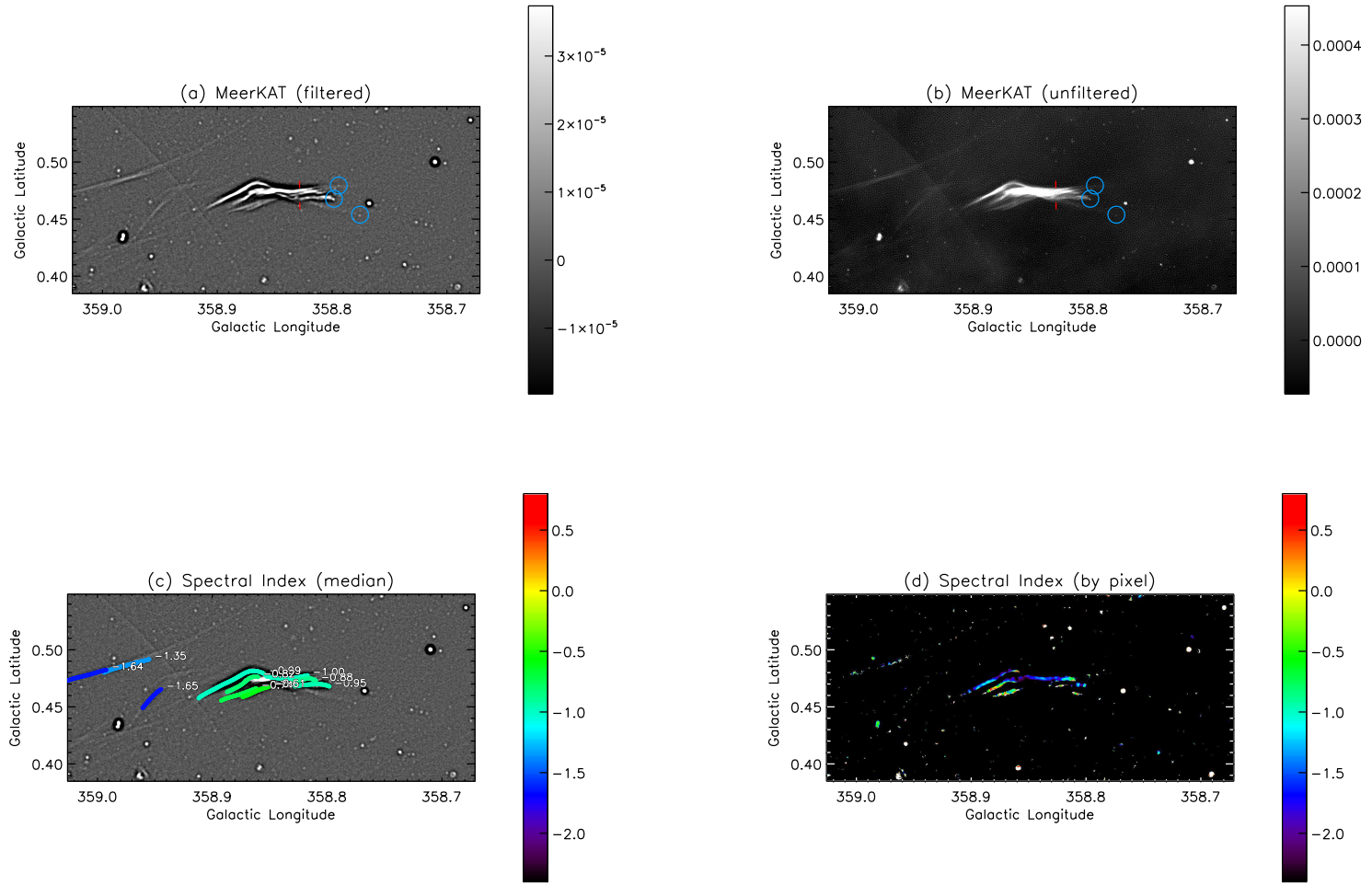

\plottwo{{cross_section_filtered_G358.85+0.47}.eps}{{cross_section_G358.85+0.47}.eps}
\plottwo{{detected_G358.85+0.47}.eps}{{alpha_G358.85+0.47}.eps}
\caption{(The Pelican) Same as Figure \ref{fig:horseshoe} except G358.828+0.471 filaments are displayed.  
\label{fig:pelican}}
\end{sidewaysfigure}

\subsubsection{Figure \ref{fig:pelican}: G358.828+0.471  (The Pelican)} 

G358.828+0.471, aka the Pelican, was the first example of a group  of filaments running 
parallel to the Galactic plane 
\citep{anantha99,lang99a}. Detailed VLA observations of this grouping   detected  linearly polarized 
emission, consistent with synchrotron emission  \citep{lang99a}. Figure \ref{fig:pelican} shows four separate filaments that 
run parallel to each other along the $\sim7'$ length and 45$''$ width of the Pelican. 

The mean spectral index  of the individual filaments in the grouping   is close to $\sim-1$. 
The spectrum  of the western filaments is steeper 
than the eastern half, suggesting that the western structure is  older. We also notice that 
the filaments become fainter, more diffuse and spread wider towards  the west 
whereas the eastern filaments 
are bright, have narrow widths and are sharp in their appearance.  This is consistent with 
the spectral index gradient in suggesting  that the western half is more evolved. If  the magnetic field was initially vertical and 
perpendicular to the Galactic plane, the  large-scale  motion of the surrounding plasma 
in this  region has a component along the Galactic plane, bends the magnetic field and runs from east to west. 
We also note three  compact sources at the 
ends of the filaments in the western half: G358.776+0.454, G358.794+0.479 and G358.798+0.468. 
 
The rotation measure (RM) distribution also shows 
a gradient with the eastern half having RM$\sim500$ rad m$^{-2}$ whereas the western half RMs range  between 
-1000 and -500 rad m$^{-2}$ \citep{lang99a}. If the RM gradient is intrinsic, the orientation of  the magnetic field 
along the line of sight must be  switching  by $\sim180^\circ$.

\begin{sidewaysfigure}
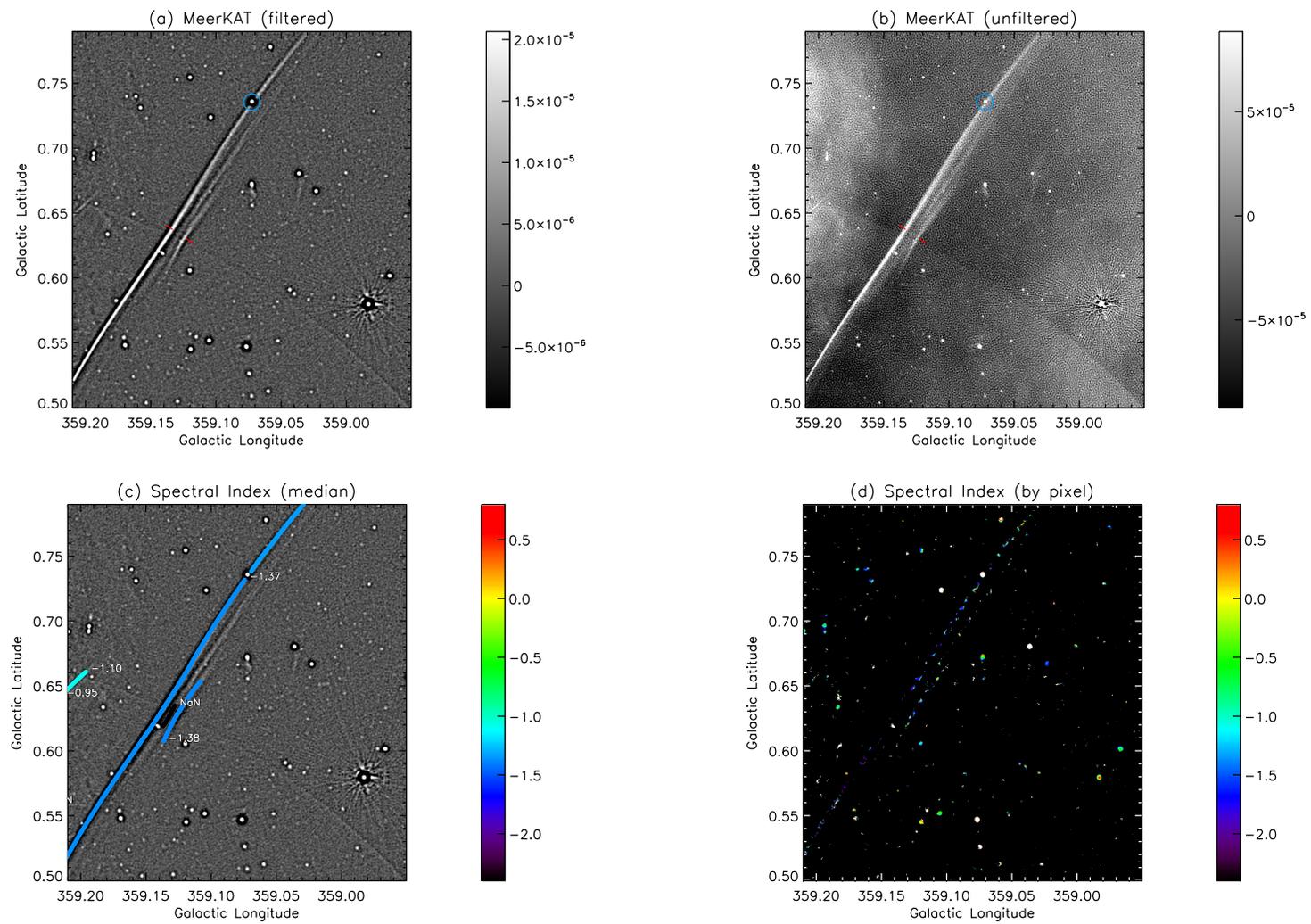

\plottwo{{cross_section_filtered_G359.08+0.64}.eps}{{cross_section_G359.08+0.64}.eps}
\plottwo{{detected_G359.08+0.64}.eps}{{alpha_G359.08+0.64}.eps}
\caption{(The Arrow) Same as Figure \ref{fig:horseshoe} except that G359.128+0.634 filaments are displayed.  
\label{fig:arrow}}
\end{sidewaysfigure}

\subsubsection{Figure \ref{fig:arrow}: G359.128+0.634 (The Arrow)} 

This grouping  consists of three parallel filaments in which the  longest and brightest filament extends for $\sim30'$. 
The brightest filament shows a gentle curvature bending by few degrees toward northwest, becoming brighter, more 
diffuse and wider in its width.  Figure \ref{fig:arrow}b shows diffuse emission between filaments in the NW direction. 
These morphological details suggest that the bright filament is evolving from SE to NW. 

G359.128+0.634 lies in projection at the western boundary of the extended Galactic Center Lobe (GCL) \citep{sofue84} or 
and the western edge of the radio 
bubble \citep{heywood19}. 
We note a clear change in the brightness of extended and diffuse emission across the long filament.
This clear demarcation across the  Arrow 
suggests that it coexists with the western edge  of the large-scale  Galactic center bubble. 

The two 
brightest filaments in G359.128+0.634 show similar spectral index values,  $\alpha\sim-1.3$ to   $-1.4$. 
The compact source 359.072+0.735
 lies 
along the filament and shows a flat spectrum consistent with  a thermal source.  The deviation from a straight filament toward NW 
near the compact source is noted.  It is not clear if the compact source is a 
foreground object or is interacting with the long filament.

\begin{sidewaysfigure}
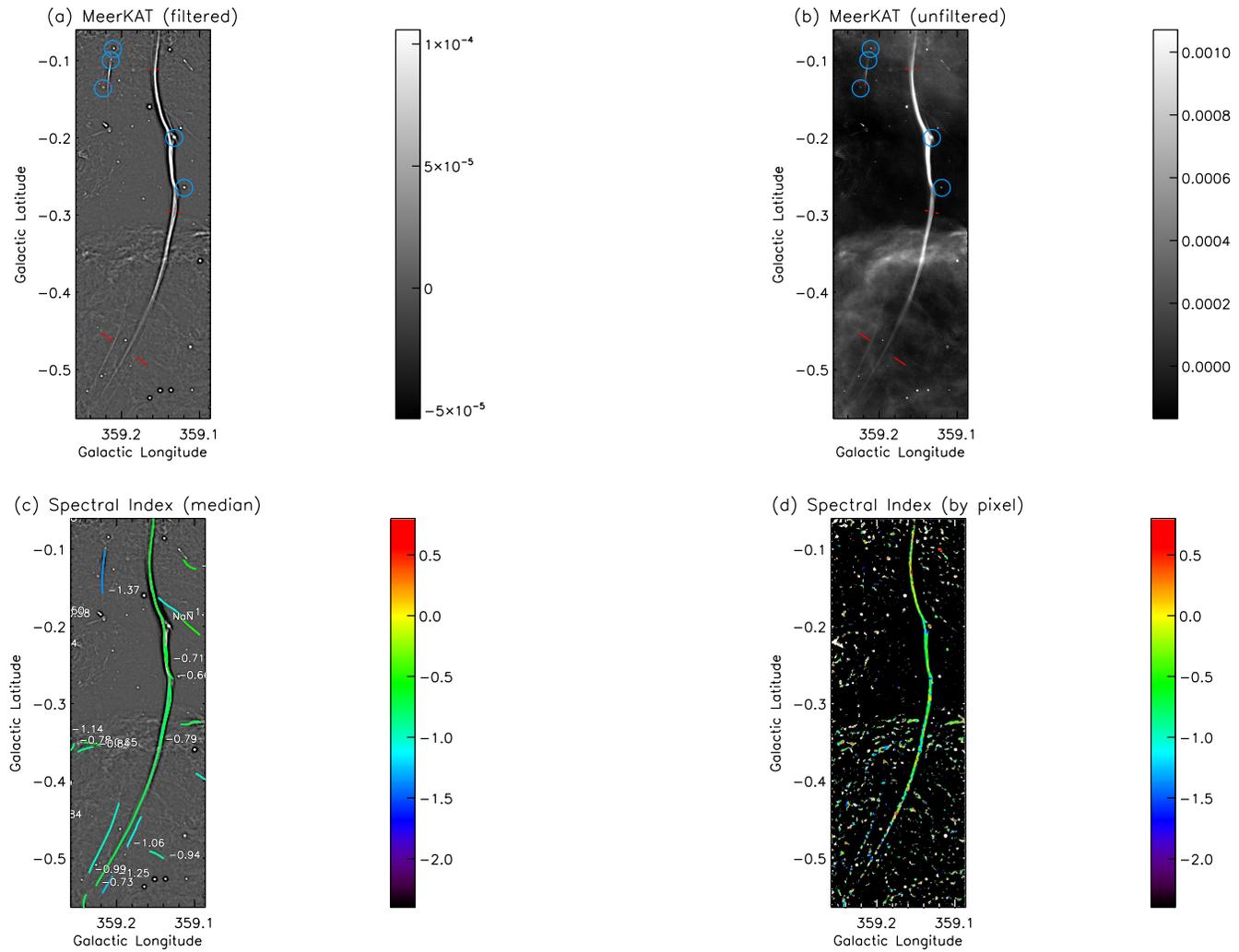

\plottwo{{cross_section_filtered_G359.17-0.31}.eps}{{cross_section_G359.17-0.31}.eps}
\plottwo{{detected_G359.17-0.31}.eps}{{alpha_G359.17-0.31}.eps}
\caption{(The Snake) Same as Figure \ref{fig:horseshoe} except that G359.132-0.296, G359.159-0.111 and G359.196-0.474 filaments 
are displayed. 
\label{fig:snake}}
\end{sidewaysfigure}

\subsubsection{Figure \ref{fig:snake}: G359.132-0.296, G359.159-0.111, G359.196-0.474 (The Snake)} 

The well-known group of filaments in the Snake with a length of $\sim30'$ ($\sim70$ pc), is one of the longest 
in the Galactic center. 
Figure \ref{fig:snake} overlays  three  locations where the spacing  between the 
filaments have been measured.  Unlike most other Galactic 
center filaments that exhibit smooth curvature like the Arrow, 
the Snake  is unique in that it is distinguished by its three different curvatures and two kinks, northern and 
southern, along its length.  It is intriguing that there are two compact radio sources G359.132-0.200 and G359.120-0.265 located to 
the west of the northern and southern kinks, respectively. 
 Similarly  the equipartion  magnetic field, which scales with the 
total intensity, has a maximum of $\sim0.15$ mG as it decreases from north to south \citep{zadeh22}. Faraday 
rotation measures of  5500 and 1400 rad m$^{-2}$ have been reported to be due to an external and internal 
medium, respectively \citep{gray95}. 

The Snake 
has a nonthermal radio spectrum with typical spectral index $\alpha\sim-0.7$. 
The region to the north 
of the southern  kink is  brighter, 
and has a  spectrum of close  to $\sim-0.6$ and becoming stepper to a value of $\sim-0.9$ at the southern end.
This suggests that cosmic ray particles 
are  more recently accelerated to the north of  the northern kink, than to the  south of the southern  kink.



\begin{sidewaysfigure}
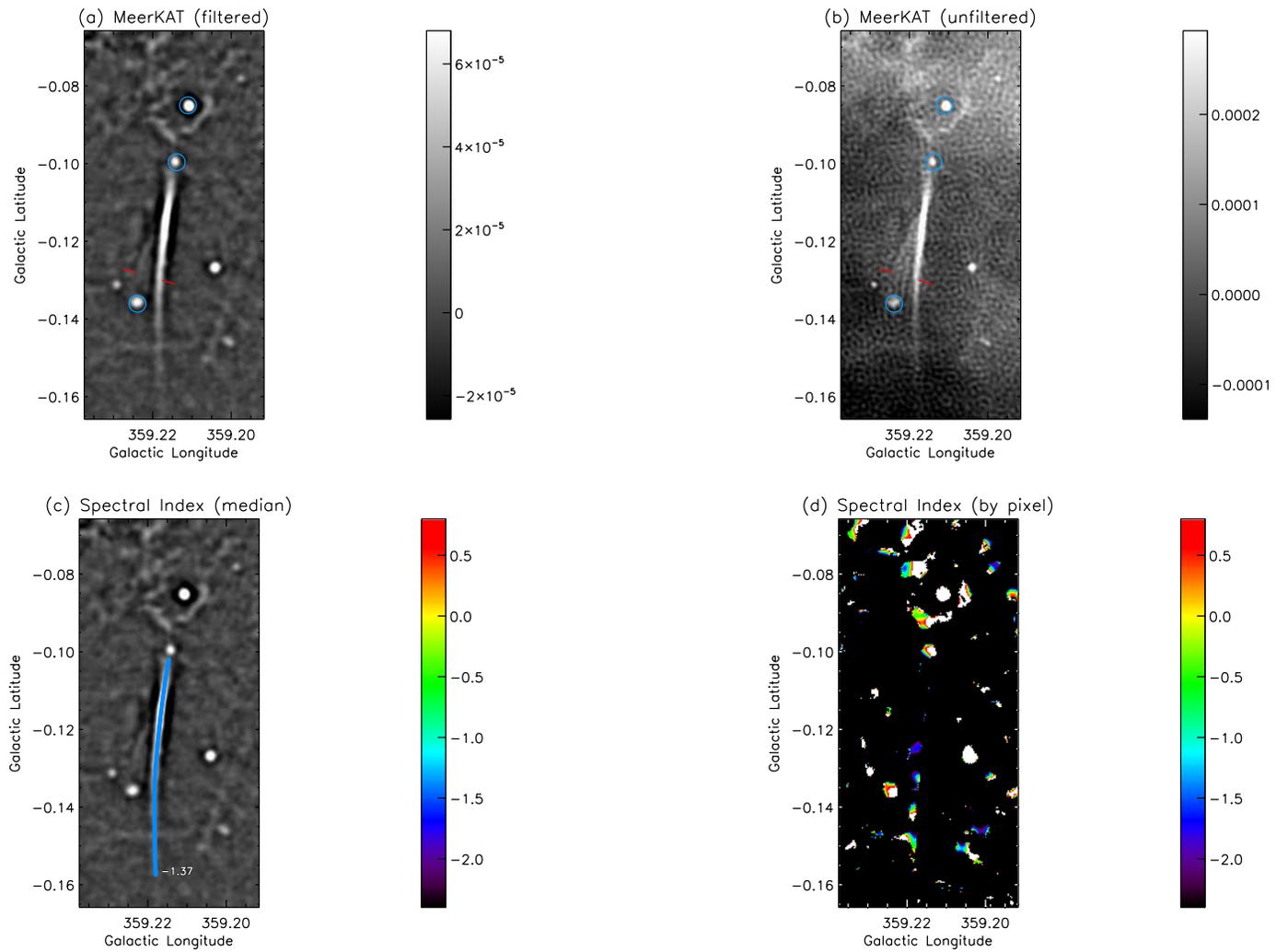

\plottwo{{cross_section_filtered_G359.21-0.12}.eps}{{cross_section_G359.21-0.12}.eps}
\plottwo{{detected_G359.21-0.12}.eps}{{alpha_G359.21-0.12}.eps}
\caption{(The Candle) Same as Figure \ref{fig:horseshoe} except G359.221-0.129 filaments 
are displayed. 
\label{fig:candle}}
\end{sidewaysfigure}

\subsubsection{Figure \ref{fig:candle}: G359.221-0.129 (The Candle)} 
 
359.214-0.100 consists of a number of faint filaments 
and a single bright filament all running vertically. 
A compact source G359.212-0.087 is detected at the northern end of the bright filament. 
 The northern extension of the bright filament splits
into multiple faint components  followed by a 
 a shell-like structure within which a bright compact source 
359.211-0.085  lies (see Figure \ref{fig:candle}b). 
The bright vertical filament and the shell-like structure give the appearance of a flaming candle. 
We also note a number of faint filaments on the eastern side of the vertical filament giving the appearance of a fan. 
The faint  filaments are  terminated
at a resolved compact source 359.224-0.136  with its bow-shock appearance. 
The spectral index of G359.221-0.129 is steep when compared to the  Snake. 
 The shell-like source  and the compact sources  have a thermal spectrum.

\subsubsection{Figure \ref{fig:sausage}: G359.317-0.430 (The Sausage)} 

G359.317-0.430 consists of three components, the brightest and narrowest segment is in the middle and two wider and fainter 
filamentary structures to the NE and SW.  The bright middle section of this filamentary structure with its sausage-like appearance 
is unresolved but the northeastern and southwestern extensions are split into multiple components.  The SW extension splits into 
two components, becoming fainter, giving the appearance of a  two-pronged fork. Two discrete sources, G359.324-0.430 and 359.321-0.431, 
are detected at the junction where the SW extension of the filament splits.  Similarly, the NE extension is wider, fainter, is more 
diffuse with smoother curvature, and splits at the position of the bright radio source G359.344-0.416 (peak intensity of 
$\sim1.6\times10^{-4}$ Jy beam$^{-1}$).

The compact sources in the SW, where the filament splits into two, are possibly acting as an obstacle responsible for structural 
change in the filament; this is suggestive of a flow running along the filament. The change in the brightness and deviations from a 
straight line also suggests that an interaction is taking place. We note that the spectral index of the bright sausage-like 
filament is steeper than the filament to the southwest by $\delta\alpha\sim 0.45$. This mean spectral index value includes the 
faint and strong emission, so it is not an accurate estimate of the spectral index of the bright region.  We note in Figure 
\ref{fig:sausage}d, a hint that the spectral index of the bright region is flatter than the faint feature to the SW. This suggests 
that cosmic ray particles are re-accelerated at the location of the bright sausage-like structure. The characteristics noted in the 
NE direction can also be explained by the bulk motion of large-scale plasma flow, that is diffuse in NE, compressing the magnetic 
field along the sausage-like structure and then breaking into two parallel components in SW.
 

\subsubsection{Figure \ref{fig:hummingbird}: G359.300-0.175 and  G359.366-0.140  (The Hummingbird)}  

The region to the south of Sgr C (G359.45-0.08) is rich with filamentary structures, lying at the western 
boundary of  the large-scale Galactic center lobe  and the radio bubble.   Figure \ref{fig:hummingbird} 
exhibits  parallel filaments  connect to a prominent  HII region in the Sgr C complex.
The filaments appear to be dragged from Sgr C by large-scale plasma motion with a strong component running SW, away from the Galactic plane. The 
grouping 
 shows wobbly and diffuse  structure closer  to
the HII region  and then  straightens while bending by a few degrees to SW at G359.321-0.158, giving the appearance
of   hummingbird beak. 
The spectral index of the filaments is somewhat flatter than most nonthermal radio filaments. Chandra observations of this source shows X-ray emission from 
the region where the filament bends most \citep{zadeh07}.  
Spitzer's  MIPS image shows a ridge of 24$\mu$m dust emission 
running  parallel to the eastern half of the filaments. 
It is possible that the flatter spectral index of the filament 
arises from the surface of  dust cloud, though  is not  clear if the two, the filament and the ridge of 24$\mu$m emission,  
are associated with each other.

\subsubsection{Figure \ref{fig:feather}: G359.411-0.709 (The Feather)} 

Figure \ref{fig:feather} shows a beautiful grouping 
 of two long eastern and western filaments running with similar curvature. This grouping  
gives the appearance of a feather in which the spacing between the filaments becomes narrower to the north and diverges to the south. 
The curvature of parallel filaments suggests that they both have 
similar origin where they converge to the NE.  
The extension of these filaments to the north  appears to converge at a 
location where a  compact source G359.419-0.583  lies. The spectral index of the western, fainter  filament 
is steeper than the eastern filament.  

There are two substructures noted along the filaments. One is to the south where  the eastern filament splits into two components. This 
is another striking example of two-pronged forked filament with the junction 
at the location of a compact source G359.416-0.706. 
 Similar to the Sausage  G359.317-0.430, as discussed above (Fig. \ref{fig:sausage}), 
the compact source  may  be an obstacle that  
splits the bright eastern filament into two parallel, fainter filaments with a steeper 
spectral index 
 The spectral index variation  is consistent with re-acceleration of 
particles at the junction where the filaments separate. This suggests that large-scale bulk  flow is  compressing 
the magnetic field  from NE to SW before the flow breaks up into two components after encountering the obstacle. 

The second substructure 
is a distortion along the western filament at 359.420-0.660.  A compact 
source lies $\sim14''$ SE of the depression at 359.423-0.664
with a peak intensity of 36 $\mu$Jy 
beam$^{-1}$.  
The shape of the 
distorted filament suggests a wind-blown outflow from the compact source is responsible for distorting  
 the adjacent filament into a cavity. Another possibility is the structure is due to a background contaminating radio source 
superimposed at the location where the distortion is noted.


\subsubsection{Figures \ref{fig:sgrc}: G359.425+0.043 and G359.446-0.005 (SgrC)}  

Sgr C is one of the most well-known massive star-forming regions in the CMZ showing thermal and 
 nonthermal radio continuum emission and hosting dense molecular gas 
\citep{zadeh07,pound18,chuard18,martinez20}. Figure \ref{fig:sgrc} shows the circular-shaped Sgr~C HII region and prominent 
nonthermal vertical filaments on the eastern side where there is a hole  in the HII region. The bright
vertical  filament extends not only to the N but to the S,  as discussed below (see Fig. \ref{fig:cataract}). 
VLA image of the vertical filament at  8 GHz showed $\sim 4$  knot-like structure along the filament (see Figure 10b of  \citep{zadeh07)} before the filament bends to 
NW. The spectral indices  of the knots along the filament in \ref{fig:sgrc}(d) show flatter spectrum than the filament.   
The vertical filament N of $b=-0.04^\circ$ , where the brightest knots is noted, consists of E and W filaments running  parallel to each other.  

The western component is bright and  straight,  unlike the eastern filament which   appears wobbly and appears to be 
 dragged out of the 
diffuse Sgr C HII region, deviating from the bright filament and curving  to NW near $b\sim0.0^\circ$. 


The brightest segment of the Sgr C vertical grouping   lies between 
G359.455-0.055 and G359.453-0.033 with an extent of $\sim2'$. The northern extension of the filament 
 from the compact source G359.455-0.055 (saturated in Fig. \ref{fig:sgrc}) 
bends NE by few degrees and becomes brighter by a factor of $\sim2$.
This strongly suggests  that there is an interaction between the compact source,  with a peak intensity  of 
$\sim0.3$ mJy beam$^{-1}$,  and the filament. 
 We note that the NW extension of the vertical component of the Sgr C filament 
is bent again to NW at G359.455-0.033, 
becomes wider and breaks up into a  large number of parallel  components with steeper spectrum. 
We note a bright  thermal feature with a hook-shape appearance is oriented  parallel to the Galactic plane  and 
crosses the Sgr C filament at G359.439+0.006.   
A high-resolution $\sim1''$  VLA image of the same region is shown in the bottom panel of Figure  \ref{fig:sgrc}(e)   where we 
note clearly 
that the  direction of the vertical  filaments
deviate  at the intersection where the horizontal  structure G359.439+0.006 is noted. 
In addition, multiple fainter filaments are created to the north of G359.439+0.006.  
It is most likely that the  filamentation is triggered by the 
interaction of two  vertical bright filaments with the horizontal  hook-shaped structure, 
as we have observed in a number of filament groupings.   
Another  example is  the northern extension of the Radio Arc where there is 
ring-shaped structure, as  noted in Figure \ref{fig:ring}, causing filamentation. 
The mean spectral index of the filaments in Sgr C ranges between 
$\alpha\sim-0.5$ to $\sim-0.6$.



\subsubsection{Figure \ref{fig:fknife}: G359.429+0.132 and G359.495+0.188 (The French Knife and the Concorde)}

Figure \ref{fig:fknife} displays three groupings 
 of filaments, one of which is G359.484+0.122, as  discussed below (Fig. 
\ref{fig:bentharp}). The red markings show where three groupings 
 are located. G359.429+0.132  lies to the right of the image 
and resembles a French knife. This structure  consists of a loop-like structure that bends by $\sim90^\circ$ and a bright vertical 
filament. Low-resolution observations 
 had detected this grouping  
at 6 and 20 cm, called C6 and C7 \citep{zadeh04,law08}. The 
horizontal component  runs along the Galactic plane. Adjacent to the bright vertical filament, two 
additional fainter, but longer filaments run parallel to the bright vertical filament. The vertical component of G359.429+0.132 shows 
a  pattern noted in a number of filament groupings 
 where multiple parallel filaments with different brightness, length and 
spectral index run adjacent to each other, as shown Figure \ref{fig:fknife}. The spectral index of the vertical filaments is  
steeper for the fainter  filaments. We also notice that the loop-like filament has a flatter spectrum than the vertical 
filaments by $\delta\alpha\sim0.3$. Bending of the loop-like filament 
is perhaps  caused by a large-scale plasma motion in the NW direction.

Lastly, G359.495+0.188, 
 which is located to the NE of Figure \ref{fig:fknife},  shows three filaments with spacing that becomes narrower toward the 
north giving the appearance of the Concorde aircraft. Again, the structure of G359.495+0.188 is similar to converging 
fragmented cometary tail-like pattern noted in a  number of filament groupings. 



\subsubsection{Figure \ref{fig:cataract}: G359.504-0.321, G359.399-0.213, and G359.397-0.187 (The Cataract, the Forceps and 
the River)} 

The  grouping 
 of filaments in  G359.504-0.321 
appears like a cataract, as shown in Figure \ref{fig:cataract}, and is characterized as a broad  grouping 
 of 
closely-spaced filaments that are difficult to distinguish because of limited resolution, confusion, and/or intrinsic diffuse 
emission. This system consists of both diffuse and filamentary structure. 
The morphology of 
the cataract system  is unusual in that the northern part of the system is wider  and more diffuse 
 than the  south. 
The  faint filaments run parallel to each other, shift together to  the east and to the west 
suggesting  that the magnetic field of the filaments change direction coherently implying 
that they are parts of a larger magnetic structure.  
The large-scale structure consists of  three parallel components 
separated by  gaps  with a width of $1-2'$ 
with weaker emission between them. 
One gap is prominently shown to the south of 
the extended cometary source 
G359.467-0.171. 
 The spectral index distribution of the Cataract grouping 
appears to be flatter than typical 
steep spectrum of filaments, as shown in Figure \ref{fig:cataract}d.  This perhaps suggests that 
they are relatively younger 
or that the relativistic electron population has recently been re-accelerated.

The exact relationship between the Cataract and the Sgr C HII region  is not clear but there appears to be an 
association with  the
prominent  nonthermal Sgr C filaments.  
We note that  a narrow and long filament that is clearly the southern extension of the vertical  bright  Sgr C filaments, 
bends  $\sim b=-0.107^\circ$ and continues to the south. Another bend is seen on the western edge 
of the extended cometary HII region at 
G359.467-0.171. The wavy  pattern of the filament 
has a wavelength of $\sim15'$.

The northern extension of this  grouping is a single, slightly bent  filament
that merges with the 
Sgr C filament,  as discussed above (Fig. \ref{fig:sgrc}),
and displays a wavy pattern 
which appears to be distorted near the Galactic equator 
and  the cometary HII region. 
The southernmost  extension of this 
single filament becomes brighter, narrower  in its width and bends at the closest approach to 
the cometary feature. This suggests  that there is an interaction because the HII region shows a linear feature at its western edge  
and  appears to be  torn  by the filament. Furthermore, the variation of the spectral index 
from north to south shows flattening of the spectrum, implying re-acceleration of particles along the filaments to the south 
of the cometary feature.  
These support a picture  that these features are embedded in a large-scale 
flow directed to 
 the south of the Galactic plane. 

Figure \ref{fig:cataract} 
also shows another filament grouping 
G359.399-0.213 
with an appearance of  two-pronged forceps running vertically.   
We note a curious structure consisting  of two filaments that are 
separated by 18$''$ and run vertically to within a degree  perpendicular to the Galactic plane. 
This faint and  narrow filaments give the appearance of a  narrow river.   
The River grouping, 
(Fig. \ref{fig:hummingbird}), lies to the north of the Forceps.
This vertical filament splits into two fainter diverging components at G359.401-0.238 along its northern extension. 
The  spacing between these two components is similar to that of filaments  in the River.   
It is not clear if these two  
structures are  associated with each other. 
The southern extension of the filament becomes brightest with  a kink at G359.405-0.259. 
The spectral index of the Forceps is $\alpha\sim-0.59$. 

\subsubsection{Figure \ref{fig:bentharp}: G359.484+0.122 (The Bent Harp)} 


G359.484+0.122 is a remarkable 
grouping  consisting of five regularly spaced filaments with the appearance of a bent harp. A trend is 
noted between the length of the filaments and their brightness. These harp-like systems ordered by length have 
previously been described by \cite{thomas20}. The spectral index distribution shown in Figure \ref{fig:bentharp}c,d suggests another 
trend between the length of the filament and the steepening of the spectral index. Shorter filaments have flatter spectral index 
values. The regular spacing between the filaments becomes narrower to the SW to the point that the grouping 
 appears to converge 
at 359.439+0.103 shown in the wider view of Fig. \ref{fig:fknife}.
 The SW extension of the filaments appears sharper and narrower than 
their NE extension. It is possible that filamentation occurred as a result of an interaction in which multiple components are produced. 
Bending of the filaments can also be explained in this picture by assuming that an 
 obstacle is moving 
through the  cosmic-ray driven nuclear wind. We also note a compact source at G359.485+0.121 with a peak intensity  of $\sim 60\, 
\mu$Jy beam$^{-1}$ that lies between two of the longest filaments  as well as a compact source at G359.475+0.127 
adjacent to the shortest filament.  Lastly, we note a compact source along the longest filament at 
G359.491+0.121  with a peak intensity  of $\sim12\, \mu$Jy beam$^{-1}$.  The NE extension of this filament beyond the compact source 
G359.491+0.121 is split into two components (see Fig. \ref{fig:bentharp}b).  Again, the compact source 
may act as an obstacle embedded in  a large-scale flow 
from SW to NE and 
be responsible for splitting the long filament into two components. 

A bright source  G359.522+0.097 with a peak intensity  $\sim300\, \mu$Jy beam$^{-1}$ is also noted to the SE of the image. 
This bright source is 
resolved into an asymmetric shell  and a short cometary tail running to the NE. The spectral index of this source is
consistent with an HII region.  The orientation of the cometary HII region and the regularly spaced 
  filaments both suggest
large-scale flow in the direction away from the Galactic plane.  


\subsubsection{Figure \ref{fig:ripple}: G359.359.520+0.237 (The Ripple)} 

G359.520+0.237 is another  group
of  filaments with a length of $\sim0.3^\circ$, as shown in Figure \ref{fig:ripple}. 
The figure shows contrasting details between the bright, distorted structures in the SE and straight, faint filaments in the NW.  
The NW end of the grouping 
 shows multiple straight filaments running parallel to each other.  However, the grouping  is distorted 
toward the SE end,  exhibiting  ripple-like appearance on the brightest side of the grouping. The ripple-like distortions are brighter than the straight 
NW  filaments by a factor of $\sim$10. The cause of the ripple is thought to be 
an interaction with clumps of a molecular cloud in the SE 
\citep{bally89,staguhn98}. 
We also note splitting of the bright filament in the SE into two distorted components at a junction 
with a  compact source at
 G359.557+0.140. This appearance is reminiscent of other filament groupings 
 showing filamentation at a junction where there is a compact source. 

Earlier polarization and spectral index studies of the brightest segment of this grouping  indicate that the intrinsic magnetic 
field runs along the filament \citep{zadeh97} and the spectral index at multiple frequencies is between -0.5 and -0.08 
\citep{zadeh04,anantha91,law08}. Figure \ref{fig:ripple}d shows spectral index variations along the filaments. The region near the 
ripple shows a flatter spectrum than to the NW suggesting that re-acceleration has occurred.  This behavior is 
consistent with an interaction hypothesis, perhaps with a molecular clump \citep{bally89,staguhn98}. It is possible that the  NW filaments
were initially extended to the SE along its straight direction  before a molecular clump interacted with the filaments and distorted its
structure.  
The rotation measure between 
-4200 and -3700 rad m$^{-2}$ is reported to vary on scales of several arcseconds, implying that the ionized, magnetized medium is 
less than 0.1 pc thick \citep{zadeh97}. This is another example in which filamentation and distortion of straight filaments
 occur at a junction where there is an obstacle implying that there is flow of material along the Ripple. Lastly, another 
structure to the NE,  G359.563+0.249,  
consists of two filaments 
is better seen, and discussed, as part of the next figure.


\subsubsection{Figure \ref{fig:brokenharp}: G359.640+0.308, G359.625+0.298, and G359.563+0.249 (The Broken Harp and the Edge-on Spiral)} 

Figure \ref{fig:brokenharp} shows a grouping  of filaments that appear like a broken harp or two harps that are crossing each other. 
G359.640+0.308 runs roughly perpendicular to the Galactic plane with mean spacing of 15.5$''$ whereas G359.625+0.298 is tilted by 
10-20$^\circ$ with mean spacing of 10.6$''$.  Spectral indices of vertical and tilted filaments show a relatively flat 
distribution,  which is unusual when compared to typical steep spectrum of filaments found in this region \citep{zadeh22}. 
Low-resolution studies of this region indicate spectral index values ranging between -0.3 and -0.1 \citep{law08}, consistent with 
some of the filaments shown in Figure \ref{fig:brokenharp}c,d.  The 
diffuse emission between the filaments 
 may have contributed to spectral index estimates measured at low resolution.

We also note two parallel filaments G359.563+0.249 to the SW (also present in  Fig. \ref{fig:ripple}) that appear to be twisted about 
each
other, giving the appearance of an edge-on Spiral. Figure \ref{fig:brokenharp}b shows excess diffuse emission above the 
background is      between the filaments suggesting that the two filaments are physically related to each other.  G359.563+0.249 has a steep 
spectrum ($\alpha\sim -0.99$)
compared to the adjacent Broken Harp.

\subsubsection{Figure \ref{fig:waterfall}: G359.687-0.435 and  G359.702-0.517  (The Waterfall and the Hilltop)} 

Figure \ref{fig:waterfall} shows another striking grouping  of filaments, 
G359.687-0.435,  which resembles a waterfall, 
and consists of 
at least six filaments with a mean separation of 18$''$. The brightest filament lies at the western edge. 
The 
fainter filaments run parallel to each other, shift to the east  near $b\sim-0.43^\circ$ and then to the west  near 
$b\sim-0.45^\circ$.  This suggests that the magnetic field of the filaments change direction coherently implying 
that they are parts of a large-scale magnetic structure.  
The spectral index of the grouping  is flatter where thermal and diffuse features lie to the 
north, and becomes steeper to the south. 


The extension of the filaments to the N and NE shows that this grouping  fades out and becomes confused 
with the strong thermal background along  the Galactic plane.  The remarkable morphology of this grouping  
suggests  that  diffuse structure 
is dragged out of the Galactic plane by  a large-scale flow, becoming filamentary 
and forming  a network of equally spaced vertical filaments running to the south. 

There is a diffuse circular-shaped structure 
centered on an extended, thermal HII region  G359.734-0.409, 
 having a  flat thermal spectrum
(not marked in Table 2; Fig. \ref{fig:waterfall}c,d).  The 
grouping  of parallel filaments runs on one side of G359.734-0.409 which itself is edge-brightened 
 on the side  facing the waterfall filaments. 
The asymmetry of the HII region  maybe explained by the relative motion of the 
circular-shaped HII cloud, acting as an obstacle, with respect to the direction of cosmic-ray driven wind arising from 
the Galactic plane. This grouping  shows  similar  morphology to that of Cataract  G359.504-0.321 (Fig. \ref{fig:cataract}). 

Another system of filamentary structure G359.702-0.517 lies to the south of the Waterfall  
consisting   of a long and bright  
structure with the appearance of a hilltop, as shown in Figure \ref{fig:waterfall}. The bright  filament is 
surrounded by three  additional faint  filaments projected on its sides. 
The spectral indices of the   filaments in G359.702-0.517 is $\sim-0.5$ 
which is  flatter than steep spectrum of typical filaments with $\alpha<-1$.

\subsubsection{Figure \ref{fig:knot}: G359.738-0.792  (The Knot)} 

Figure \ref{fig:knot} shows an example of two entangled filaments that appear to twist about each other in an 
elongated knot-shaped structure with an extent of $\sim0.6'$. The bright segment of the filament to the south splits into two 
 at G359.729-0.811 and then rejoins at G359.734-0.801. The spectral index variation, bending and brightening of the 
filaments near the knot  indicate that G359.738-0.792 is where the  intersecting  filaments are interacting 
and where   re-acceleration  of cosmic ray particles is taking place. 
Two compact sources are noted at one end of the 
eastern filament at G359.748-0.781 and 359.746-0.784.

\subsubsection{Figure \ref{fig:flamingo}: G359.808+0.130 and G359.717+0.228 (The Flamingo and the  Eyebrow)}

Figure \ref{fig:flamingo} shows a group of filaments where two bright, gently curving filaments bend and break up into multiple 
components along their NW and SE extensions. This complex group  gives the appearance of a Flamingo that is standing on one leg, 
perpendicular to the Galactic plane.  
The NE extension is bent by about $\sim10^\circ$ and deviate from each other at G359.758+0.198
where there is no obvious compact source. 
The spectral index of the northern component of the deviated filament is flatter than the that of southern 
component by $\delta\alpha\sim0.2$.

The northern extension of the Flamingo leg near a source  G359.808+0.117 breaks up into multiple filaments running to the NE. The 
spectral index distribution indicates that the Flamingo has a relatively flat spectrum compared to typical filaments with 
$\alpha<-1.0$. The characteristics  of this grouping  is similar to 
 the Ripple where large-scale bending, 
brightening and spectral index variation are detected.    We note an isolated  CO  clump at 
G359.822+0.107 (\cite{schuller20}) 
at the southern  end 
of the grouping suggesting that the distortion and brightening of the grouping 
 is due to a physical interaction with this clump.   

At the SW corner of the image, a grouping 
 with the appearance of an eyebrow is noted. This consists of two filaments, one on top of each other,   
running  with the same curvature suggesting that they are related to each other. 
The top filament breaks up into multiple components whereas the bottom filament is a single bright filament with flatter spectral 
index than the top filament.   The tendency that filaments that are  longer and fainter have steep spectral index values is consistent 
with  those of the Harp, as discussed below. We note a compact source G359.886+0.102 at the eastern end of the the Eyebrow where the filaments converge.  
It is not clear if this pp
is associated with the Eyebrow.

\subsubsection{Figure \ref{fig:cleaver}: G359.890-0.289 (The Cleaver)} 

359.890-0.289  is located in a rich region of the Galactic center hosting a large concentration of groupings  and isolated filaments 
criss-crossing each other, shown in  Figure \ref{fig:cleaver}. A  bright single filament to the NW with multiple fainter filaments 
running to the SE parallel to each other.  The width of the single filament thickens  from $\sim4''$ to $\sim7''$ and 
brighter by a factor of 5. The grouping 
 of filaments and a square-shaped diffuse structure to the SE gives the appearance of a 
cleaver. The SE extension of the bright filament bends by a few degrees at a bright unresolved compact source G359.874-0.264.  
A more detailed close-up of this unresolved compact source indicates a faint filament splitting into two components running from NW 
to SE. The spectral index of the bright filament is $\alpha\sim-1.13$ and that of the 
 square-shaped structure is consistent with being 
thermal. There is no obvious variation in  the spectral index along the long filaments in the grouping.
 The decreasing sharpness and  thickening of the width of the filament, deviation of parallel filaments, bending, and splitting of 
the filaments from NW to 
SE all suggest that the grouping  is evolving and  
supporting  a picture  that these features are embedded in a large-scale 
flow directed to  the south of the Galactic plane. 

We also note to the NE, there is 
a striking head-tail cometary source which consists of a bright compact thermal 
source at one end of G359.950-0.259 and an elongated tail. 
The narrow tail becomes brighter by a factor of five and thicker by a factor of two from NE 
to SW. The tail is also split into two faint components  near G359.921-0.283.  
The spectral index of the narrow segment of the tail has a flat spectral index but 
becomes steep $\alpha\sim-0.71$ where the tail is  brightest. 
These   characteristics suggest that 
cosmic-ray electrons are populated close to the source which is 
embedded in a large-scale flow. 


\subsubsection{Figure \ref{fig:harp}: G359.851+0.374  (The Harp)} 

G359.851+0.374 is the best example of harp-like structure in which all the strings  run perpendicular to the Galactic plane 
and show no visible signs of distortion  with the exception of the westernmost filament with its gentle concave bend, as 
displayed in Figure \ref{fig:harp}.
The initially parallel filaments appear to 
converge for some pairs either to the north or to the south.  
 The morphology of this grouping  is similar to the Radio Arc where a number of 
parallel filaments with different lengths run perpendicular to the Galactic plane (see Fig. \ref{fig:radioarc}).  
The harp-like group
 of filaments are 
ordered by their lengths \citep{thomas20} and diffuse emission is detected from the regions between the filaments (see Fig. 
\ref{fig:harp}b).
Low-resolution 
radio observations of G359.851+0.374 (source N10 in \cite{zadeh04}) had previously been detected with the VLA at multiple 
frequencies \citep{larosa01,zadeh04,law08} before MeerKAT data became available \citep{heywood19}. 

At the leading edge of the Harp where the shortest filament is detected, there is a compact source at G359.837+0.375 with a peak 
intensity of 25 $\mu$Jy beam$^{-1}$.  Four of the shortest filaments appear to have a compact source at the southern end of the 
filaments G359.841+0.366, G359.844+0.358, G359.848+0.298,  and G359.851+0.293
  It is not clear if these sources are associated with 
the filaments or they are background radio sources. We also searched for compact sources to the northern end of the filaments but found
only one compact radio source. It is perhaps suggestive that 
the  southern compact sources are  associated with the Harp, thus supporting a picture in which 
 the Harp is embedded within a large-scale flow driven away  from the Galactic plane.  

Another striking result is the spectral index distribution of harp filaments, as shown in Figure \ref{fig:harp}c,d. There is a 
tendency that the filaments with the longest lengths have steeper spectra. If such  a correlation exists between 
length and spectral index and assuming that 
 the spectral index traces synchrotron cooling time scale, then it implies that cosmic ray 
particles in short filaments are younger  than in the longer filaments.

\subsubsection{Figure \ref{fig:comet}: G359.992-0.572 (The Comet Tail)} 

The grouping  G359.992-0.572 seen in Figure \ref{fig:comet} consists of six  linear structures diverging towards the 
NW,  resembling a fragmented comet tail. 
The mean angular spacing of the filaments in G359.992-0.572 is 27.5$''$ (Fig. \ref{fig:comet}) which is 
relatively large compared to the mean angular spacing of $\sim16''$ found towards all groupings (see below).   
The brightest filament in the grouping 
 branches out into two fainter 
components at two locations in the NW and SE at 359.992-0.555  and G0.023-0.626, respectively.

We also note three compact sources at the SE ends of the grouping 
 G0.030-0.660, G0.034-0.661 and G0.045-0.668. 
The spectral index of individual filaments in the grouping  vary and show 
steep spectra. This is consistent with a trend that the filaments with steeper spectral indices lie at high latitudes 
\citep{zadeh22}.

\subsubsection{Figure \ref{fig:radioarcs}: G0.138-0.299 and G0.172-0.423 (The Radio Arc S.)}  

The filament grouping 
 Radio Arc S-1, G0.138-0.299,  is  shown in Figure \ref{fig:radioarcs}
 and is labelled  S5 in \citep{zadeh04}. The grouping  
lies towards the eastern boundary of a diffuse, large-scale linearly polarized Eastern  Galactic Center Lobe  or the  radio bubble 
\citep{tsuboi95,heywood19}. 
The northern extension  of this grouping  is associated with the  Radio Arc, as discussed below 
(see Fig. \ref{fig:radioarc}). The grouping 
 of the Radio Arc has a lateral  span of $\sim249''$. The bright filaments 
in the Radio Arc become faint and diffuse toward  southern latitudes.  At the eastern boundary of 
diffuse structure lies a long, bright filament at  an oblique angle to  the Radio Arc. 
This grouping, G0.172-0.423,  runs vertically for more than 0.3$^\circ$ and is one of the few long 
filaments that has an X-ray counterpart with 
an extent of $\sim2'$  detected along the brightest segment
\citep{ponti15,zadeh21}.  Spectral index distribution of this region indicates that the filament spectra 
 at more negative latitudes are 
steeper than  in the Radio Arc.  This is consistent with single-dish measurements indicating that the 
extensions of the Radio Arc away from the Galactic plane have  steeper spectra 
than close to the plane 
\citep{tsuboi95}.



\subsubsection{Figure \ref{fig:radioarc}:  G0.160-0.115 (The Radio Arc )} 

Figure \ref{fig:radioarc} shows a segment of one of the longest and 
most well-known filament groupings 
 associated with the Radio Arc,  
a prototype magnetized radio filament \citep{zadeh84}. The southern extension of this grouping 
 was discussed above 
(Fig.  \ref{fig:radioarcs}). 
This grouping 
 consists of
the largest number of filaments,  17,  that are parallel to each other 
and run perpendicular to the Galactic plane with average spacing of 17$''$.  Its polarization, rotation 
measure and spectral indices have been measured  \citep{pare19} indicating that intrinsic  magnetic field runs along the filaments. 
 The spectral index of individual filaments in the 
Radio Arc is flat  when compared to other Galactic center filaments,  as noted in Figure \ref{fig:radioarc}c,d. 
The Radio Arc's unusual spectral index values are  consistent with earlier measurements using wider frequency bands  
at multiple frequencies \citep{zadeh86}. 

\subsubsection{Figure \ref{fig:ring}: G0.168+0.142 (The Ring)} 

Figure \ref{fig:ring} shows a grouping 
 of nonthermal filaments surrounding a ring-shaped structure G0.17+0.15 \citep{zadeh88}.  
The southern extensions of the filaments run into the Radio Arc, as discussed in the previous section. G0.17+0.15 is a 
dusty HII region embedded within the highly polarized radio lobe of the Galactic center \citep{sofue84,tsuboi95} where 
a number of radio filaments meander through the HII region, as shown in Figure \ref{fig:ring}. 

A detection of a hydrogen 
recombination line with a radial velocity exceeding 130 \kms has been reported toward the ring \citep{royster11}. The filaments to 
the south of the HII region are generally brighter and sharper compared to those to the north. 
There are 7 harp-like filaments distributed to the south of the ring G0.17+0.15 with mean spacing of 16.2$''$. 
which is larger than that of 
the 12$''$ filament spacing in the Radio Arc. 
The spectral index of 
the filaments in the Ring is  flat but  slightly steeper than those found toward the Radio Arc 
G0.160-0.115.  
 The interaction of nonthermal filaments  with a thermal HII region G0.17+0.15  
is very similar
to two HII regions,  G359.467-0.171 and G359.439+0.006, noted to the south and north of the Sgr C HII region, 
as discussed above (Figs.  \ref{fig:cataract} and \ref{fig:sgrc}), respectively. 
In both circumstances, the filaments appear to change their characteristics as a result of an interaction with 
HII regions. 

\subsubsection{Figure \ref{fig:shuttle}: G0.228+0.812 (The Space Shuttle)} 

Figure \ref{fig:shuttle} shows a group  of three filaments that appear to converge to a single bright filament to the north near a compact source at 
G0.225+0.843. The structure of the filaments 
resembles Shuttle rocket with boosters on the sides. These  filaments  follow similar pattern noted previously in 
fragmented cometary tail-like  groupings  in which the  separation between the strings  decrease, giving the appearance of converging filaments. We note a compact source at 
G0.226+0.792.  
with a peak intensity of 53 $\mu$Jy beam$^{-1}$ at the southern end of one of the side filaments. 
This group of  filament  lies at the easternmost edge of the large-scale 
Galactic center bubble and along the northern extension  of the Radio Arc.
Figure \ref{fig:shuttle}c,d indicate  that fainter filaments to the south have steeper spectral indices than the bright filament to the 
north away from the Galactic plane.  

\subsubsection{Figure \ref{fig:porcupine}: G0.288-0.259, G0.306-0.273, and  G0.297-0.266 (The Porcupine)} 
 
Figure \ref{fig:porcupine} shows two systems of  fragmented cometary tail-like  structures,  G0.228-0.259 and G0.306-0.273,  
with mean  spacings  6.8$''$ and 14.3$''$, respectively. 
We note that the longest filaments lie at the edges of these two systems.  
The mean spacing of the combined  grouping  G0.297-0.266 is estimated to be 10.5$''$. 
This group  of filaments gives the appearance of 
of  porcupine  quills converging to a bright compact source  toward NW 
at G0.264-0.197 with a peak intensity of 1.5$\times10^{-4}$ Jy beam$^{-1}$. 
The spectral indices of individual filaments are   fairly flat with typical values of $\sim-0.2$ similar to the Radio Arc. 

\subsubsection{Figure \ref{fig:contrail}: G0.405-0.277 (The Contrail)} 

Figure \ref{fig:contrail} shows a group 
 of filaments which consists of 
two parallel filaments lying on top of each other. 
The grouping 
appears like a contrail of a jet running diagonally. The bright top filament is split into two components at
a position G0.420-0.297. The change in the brightness and deviations from a straight line is also noted at this junction. 
The long bottom  filament is very faint  in the middle and then becomes bright by a factor of few along the NW and SE directions. 
The positions  at which the filament becomes brighter to NW and SE  are G0.421-0.290 and G0.443-0.324, respectively.  
Spectral index maps  show steeper values for the top filament than the bottom filament.

\subsubsection{Figure \ref{fig:bentfork}: G0.687+0.147 (The Bent Fork)} 

Figure \ref{fig:bentfork} shows a loop-like  filamentary  structure 
that splits into two filaments, giving the appearance of  a bent 2-pronged fork, with filament spacing of 
7.5$''$. At the junction where the eastern filament splits at a location 
G0.697+0.148, the brightness increases  from 
$\sim$0.2 to $\sim$0.8 mJy beam$^{-1}$. At the junction where typical filaments split into multiple components, a compact source or 
enhanced emission is generally detected. 
This source  is  elongated and runs in the same 
direction as the filaments. 
As such the resolved structure shows clear  evidence that the 
filament is resolved into two components at the 
bright and extended juncture and is unlikely to be due to a background source along the line of sight. 
 In addition, the spectral index of the elongated structure appears to be flat compared to  the rest of 
the filaments. These characteristics are very similar to those of the Horseshoe (see Fig. \ref{fig:horseshoe}) 
 show clear signatures that 359.697+0.148  is physically associated with the grouping 
 and is 
responsible for splitting the filament and re-accelerating particles. 

\subsubsection{Figure \ref{fig:meteor}: G0.859-0.597 (The Meteor Trail)} 

Figure \ref{fig:meteor} shows a vertical filament running perpendicular to the Galactic plane and gives the appearance of a meteor trail. 
The filament breaks up into two components and diverges to the south, away from the Galactic plane. 
 We also note that split
 filaments show slight  curvature 
to the east.  The spectral index is steep and is consistent with filaments at high latitudes having  steeper spectral index 
than those closer  to the Galactic plane.


\section{Discussion} 
\label{sec:filter}

We have studied the detailed morphology and spectral index distributions of close to 40 prominent groupings of filaments in the 
Galactic center region. The dominant morphology of the groupings that stands out can be characterized as multiple filaments that 
either run parallel to each other (harp-like), converge to a point (fragmented cometary tail-like), or bend together and 
form a partial loop. In some cases, a group of filaments such as the Waterfall, Cataract and Hummingbird appear to be dragged out from 
an  HII region and then turning into long, narrow and parallel filaments. In almost all filaments in groupings, we note 
deviations from a straight verticality running perpendicular to the Galactic plane. We determine below the characteristic size 
scales associated with the spacing between filaments in a grouping and the width span of groupings, thus providing constraints in 
the geometry of the filaments.  We will then discuss the origin of filamentation as a consequence of either the interaction of 
Galactic center wind with obstacles or as synchrotron cooling instability in the Galactic center ISM.


\subsection{The Geometry of the filaments}

To characterize the structure of groups of filaments, we select 43 cross sections where the mean separation of the filaments is 
calculated simply as the total span of the cross section divided by the number of spacings. 
The cross sections are chosen at one 
or two locations in each grouping where the filaments are well defined. However in some cases,  filaments may be very faint or 
poorly resolved, and the number of filaments may be undercounted (and the separations overestimated). 
The results are listed in Table 1.

The mean angular separation of the filaments in groupings typically falls in the range $10-22''$ with peak value of $\sim16''$ (0.6 
pc).  Outliers extend up  to $\sim50''$. The spacing between any two filaments in a given grouping  varies 
slightly. In groupings with a large number of filaments, the spacing varies by  a factor of two. 


Figure \ref{fig:separation}a shows a histogram of the filament spacings. This histogram which has used a sample 
of 43 grouping cross sections is fitted by a Gaussian with a peak of 16$''$ ($\sim$0.65 pc at the 8 kpc Galactic center distance). 
Large spacings are outliers and could suffer from a lack of  sensitivity and confusion. 
Lack of counts  at small angular sizes, as shown in Figure \ref{fig:separation}a,  is likely  due to limited spatial resolution
$\sim6''$. We also determined the widths of filament grouping or ``spans'' giving a mean value of 
$\sim27'' (\sim1.1$ pc).  This is the size scale on which  filaments are bunched together in a grouping. A histogram of spans of individual
groupings  is shown in Figure \ref{fig:separation}b. The peak of the Gaussian is $\sim27''$ which corresponds to 
$\sim$1.2pc.


These angular scales (and the implied linear scales) are new physical parameters that characterize the population of Galactic 
center filament groupings. The linear separation of the filaments assumes that all filaments are on the plane of the sky. We 
modeled and reproduced the observed spacing distribution of filament spacing by considering that the filaments have randomly distributed 
inclinations with respect to the line of sight assuming that the filaments are strand-like (so are not sheet-like structures viewed 
edge-on). They are arranged in regularly spaced comb-like structures viewed at random inclinations. Then, the observed distribution 
of the spacings between adjacent 1D filaments is  projection of their true spacings onto the plane of the sky.

As described in the appendix, 
the distribution of apparent spacings
for $h$ for an isotropically oriented  population of filament groupings  with a single true spacing $h_0$ is:
\begin{equation}
    P(h|h_0) = \int_0^{2\pi} P(\phi,h)\, d\phi \;= 
    \begin{cases}
    	{\displaystyle \frac{2/\pi}{\sqrt{h_0^2-h^2}}} & \textrm{when $h<h_0$, } \\[18 pt]
	    \qquad0 & \textrm{otherwise.}
	 \end{cases}
    \label{eqn:Ph}
\end{equation}
This distribution  is plotted in Fig. \ref{fig:monospaced}a.
The significant tail extending towards zero  spacings arises because of the large solid angle subtended by highly-inclined 
viewing angles (i.e. with $\theta\approx \pi/2$).  

Then, for a probability distribution  $P_0(h_0)$ of true spacings, we obtain the in-sky distribution
\begin{equation}
P(h)  = \int_0^\infty P(h|h_0) P_0(h_0) \,dh_0 = \frac{2}{\pi}\int_{h}^\infty \frac{P_0(h_0)\,dh_0}{\sqrt{h_0^2-h^2}}\,.
    \label{eqn:Nh}
\end{equation}

In figure \ref{fig:distributions}b, we show the predicted spacings for a log-normal distribution of the true spacings $P_0(h0)$ 
(blue dashed line). 

\begin{equation}
    P_0(h_0) = \frac{1}{\sqrt{2\pi}\,\sigma h_0}
      \,\exp\left(-\frac{\ln^2(h_0/a)}{2\sigma^2}\right)\,.
	\label{eqn:lognormal}
\end{equation}

with $a=20''$ and $\sigma=0.2$, illustrate values chosen so that $P(h)$ is similar to the observed spacing counts.
It is clear that while projection effects reduce the mode of the intrinsic distribution somewhat, the largest effect is to double 
the 16'' FWHM of the true distribution.

\subsection{A Causal  origin of filamentation}

One key question is the origin of the filamentation observed in groupings of Galactic center filaments.  One possibility is that a
single  filament was split by an obstacle into multiple filaments, which then diverge and become more diffuse, as shown in Figures 
\ref{fig:horseshoe}, \ref{fig:sausage} and \ref{fig:feather}. In this picture, an obstacle sets the length scale of the separation 
between the filaments as well as the size of the bundle of filaments. Filamentation implies  a bulk plasma flow along a  filament if  
it is to break up into multiple filaments at the interaction site. One model that we have explored in the past 
to explain the origin of the filaments is that nonthermal radio filaments arise through the interaction of the cosmic-ray driven 
outflow from the disk of the Galactic center and the termination shock of an embedded mass-losing star \citep{zadeh19}. This is 
analogous to the interaction of the solar wind interacting with obstacles, such as comets, Earth and Mars. In these cases, the 
solar wind drapes and sweeps over embedded sources in the solar system.  Many groupings show diverging filaments from a point which 
is similar in appearance to fragmented cometary tails.  


One of the implications of a scenario in which cosmic-ray driven wind interacts  with an obstacle 
is a flow of plasma along the filaments with similar speed to that of the surrounding large-scale cosmic-ray driven outflow.  This 
characteristic combined with decreasing sharpness and increasing thickening along the filaments suggest that the 
relativistic electrons on one end are re-accelerated, and then cool as 
they propagate along the filament. 
In this picture, the variation of the spectral index can be used to make an estimate of the flow speed of plasma. 
To this end, the 
gradients in the spectral index, $d\alpha/ds$, were measured as a function of distance, $s$, along 191 of the longest filaments. 
This selection include filaments that are greater than $165''$ in length and are generally bright. Gradients are estimated using 
weighted linear fits between $s$ and $\alpha$. The weights are uniform for intensities $I > 10^{-4}$ Jy~beam$^{-1}$, and are taken 
to be proportional to $I^2$ at fainter levels. A negative gradient indicates a steepening spectral index as a function of $|b|$, 
i.e. with increasing distance from the Galactic plane. There is seldom an obvious linear gradient along a filament, but by measuring 
gradients in many filaments we can assess if there is any general trend in the population of filaments. The histogram of measured 
gradients is shown in Figure \ref{fig:gradients}.




Using the mean variation of the spectral index along the filaments, -1.1 $deg^{-1}$, the mean spectra index $\alpha\sim-0.83$, and 
the mean equipartition magnetic field, $\sim100$ to 400 $\mu$G depending on the assumed ratio of cosmic-ray protons to electrons, 
we estimate the mean cooling time of the population of cosmic ray particles to be $\sim 0.5-2.7\times10^5$ yr. 
For filaments with lengths ranging between  $\sim5-30$pc, the flow speed is estimated to be about $\sim$100   \kms, 
assuming that shorter filaments have stronger magnetic field.  
This is suggestive of plasma flow 
streaming along filaments rather than diffusion.



\subsection{A Radiative instability  origin of filamentation}

There are a number of groupings of multiple filaments that don't converge to a point and do not appear to show an interaction with an obstacle, such as the 
Harp and the Radio Arc, as shown in Figures \ref{fig:harp}, \ref{fig:radioarc}, respectively. One possibility is that filamentation results from 
synchrotron cooling instability \cite[e.g.,][]{simon1967}. In this picture, the initial cosmic-ray pressure has to at least be comparable to the magnetic 
pressure.  The instability proceeds as follows: synchrotron losses lower the relativistic particle pressure in a magnetized region, which is then 
compressed by the surroundings, increasing the field strength. This drives further losses and ongoing compression. To be effective, this picture requires 
that particles lose energy within a few thousand years implying that the filaments initially had a significant population of TeV electrons emitting X-ray 
synchrotron radiation. Alternatively,  the ISM needed to be permeated by diffuse X-ray synchrotron emission providing a 
reduction in cosmic-ray pressure to locally trigger the instability.
It turns out that this instability causes  filamentation on a scale of $\sim0.5$ pc, as discussed 
below. So, it is possible that both cooling instability and an interaction of nuclear wind with obstacles generating X-ray 
synchrotron emission operate in the Galactic center.


The cause of filamentation could be due to synchrotron cooling instability taking place 
in the magnetized tail generated from  the interaction of cosmic-ray driven nuclear wind  with an obstacle \citep{zadeh19}.
The magnetic field in the filaments is relatively strong compared to the magnetic field in the diffuse region. The lengths of 
the filaments and the scales on which some of the filaments are bent are indicative of the scales of external 
 plasma motion. It is therefore 
reasonable to assume that the magnetic-field energy in the filaments is comparable to the rms energy of the plasma motion, which 
gives an estimate for the magnetic field strength in filaments $B\sim 0.1$~mG, consistent with minimum energy inferred from radio 
synchrotron emission \citep{zadeh22}. 
The regions of enhanced magnetic field where 
filaments originate may be created due to plasma compression or magnetic line stretching in colliding, shearing flows, or in a 
plasma stream interacting with an obstacle. The same regions may be responsible for the electron acceleration, as they may contains 
shocks or sites of magnetic reconnection \citep{drake13,guo20}.

As discussed above, the presence of multi-filament substructure in many radio filaments may indicate that their origin is governed by the so-called cooling instability 
\cite[e.g.,][]{simon1967}. 
Indeed, let us assume that in the regions where magnetic field is amplified, a significant fraction of plasma pressure is provided by ultra-relativistic electrons, while the plasma mass is dominated by the protons. Then plasma compression in the field-perpendicular direction will lead to amplification of the magnetic field, which in turn will lead to enhanced synchrotron radiation, and as a result, to electron cooling and stronger compression. 

As demonstrated in \citep{simon1967}, the growth rate of the instability increases at smaller scales. The maximal growth rate is approximately the inverse synchrotron cooling time, $1/\tau$. This growth rate is achieved at the wavelength $\lambda \approx \pi v_s \tau$, where $v_s$ is the speed of sound in the medium, and it remains at roughly the same value at even smaller wavelengths, that is, it is almost independent of the scale there.     

We now estimate whether the cooling instability may be consistent with the observational width of the filaments, which are on the order of $\sim10^{18}$~cm. We assume that the ultra-relativistic electrons have a power-law distribution over energies, so that their number density scales with the gamma factor as $n(\gamma)d\gamma={n_0}{\gamma^{-p}}d\gamma$, with $p>1$. Assume that the largest energy of the distribution corresponds to $\gamma_*$. Then the total energy density of such cosmic ray electrons is given by
\begin{eqnarray}
U_{cr}=m_ec^2\int\limits_{\gamma_0}^{\gamma_*}n(\gamma)\gamma d\gamma=m_ec^2 n_0\frac{\gamma_*^{2-p}}{2-p}.
\end{eqnarray}
where we assumed that the spectrum is sufficiently hard, $p <2$, which is consistent with acceleration in shocks or sites of magnetic reconnection. It is reasonable to assume that this energy is comparable to the thermal or rms energy of the plasma motion, e.g., $U_{cr}\sim 10^3 \,\mbox{eV}/\mbox{cm}^3$. 

The average power radiated by an ultra-relativistic electron is
\begin{eqnarray}
\langle P\rangle =\frac{4}{3}\sigma_T c \gamma^2 \frac{B^2}{8\pi},
\end{eqnarray}
where 
$\sigma_T\approx 6.65\times 10^{-25}\mathrm{cm}^{2}$ is the Thompson cross section. Therefore, the total radiated power is:
\begin{eqnarray}
L_s=\int\limits_{\gamma_0}^{\gamma_*}\langle P\rangle n(\gamma)d\gamma=\frac{4}{3}\sigma_T c \frac{B^2}{8\pi}\int\limits_{\gamma_0}^{\gamma_*}n(\gamma)\gamma^2 d\gamma= \frac{4}{3}\sigma_T c \frac{B^2}{8\pi}n_0\,\frac{\gamma_*^{3-p}}{3-p}.
\end{eqnarray}
and the characteristic cooling rate of the electrons is estimated as:
\begin{eqnarray}
\frac{1}{\tau}\approx \frac{L_s}{U_{cr}}=\frac{\frac{4}{3}\sigma_T c }{m_e c^2}\,\frac{B^2}{8\pi}\,\left(\frac{2-p}{3-p}\right)\,\gamma_*.
\end{eqnarray}

For $B=0.1$~mG, the cooling time is
\begin{eqnarray}
\tau \approx  2.5\times 10^9 \,\left(\frac{3-p}{2-p}\right)\,\frac{1}{\gamma_*}\, \mbox{years}.
\end{eqnarray}
So, if the electrons can be accelerated to large energies $\gamma_*m_ec^2\sim \mbox{TeV}$, the cooling time may be quite short. For instance, for $p=1.5$ and $\gamma_*=10^7$, we estimate $\tau\approx 750$~years. 

A short cooling time of the plasma can trigger the cooling instability, whose growth rate is comparable to the cooling rate. Indeed, as plasma gets compressed, the magnetic strength increases while the transverse scale of the compressed region, $l$, decreases. The typical time of nonlinear evolution of the compressed plasma structure is $t_{nl}\sim l/v$, where the speed of plasma flow 
in the developing structure 
is comparable to the speed of sound,  $v\sim v_s$. One can estimate $v_s\sim 500\, \mbox{km/s}$ for typical plasma density in the central $150$~pc and energy density of $10^3\,\mbox{eV}/\mbox{cm}^3$. Equating the nonlinear time to the cooling time and assuming $\gamma_*\sim 10^7$, we estimate the transverse size of the plasma region where the cooling instability sets in as $l\leq \pi \tau v_s\sim 3\times 10^{18}$~cm, which is consistent with observations. 

The cooling instability further compresses magnetic field lines and may split a magnetic filament into sub-filaments separated 
from each other in the transverse direction. 
Indeed, for transverse wave numbers $k>1/(\tau v_s)$, the cooling instability growth rate is virtually independent of the wave number, and therefore, even thinner sub-filaments may, in principle, be formed. One, however, may argue that the compressed region still possesses plasma turbulence. Nonlinear evolution rates in turbulence are typically higher at smaller scales, therefore, the thinner substructures would have nonlinear interaction rates exceeding $1/\tau$, while the instability rate is still $1/\tau$. Therefore, the development of the instability would be impeded by turbulent disruption at smaller scales. It is, therefore, reasonable to assume that the typical space between sub-filaments should be comparable to the thickness of a filament region itself, $\Delta l\sim l\sim v_s\tau$, that is, only a few sub-filaments should be expected in each filamentary structure.  

Another  possible filament forming mechanism 
 that does not require rapid cooling involves an active turbulent medium. 
In this picture,  turbulent compression of  the
magnetic field in the cosmic-ray driven nuclear wind, 
can  explain the synchrotron emissivity.  Such compressed field regions could  arise 
as a result of active turbulent medium  \citep{boldyrev06}. 
In the turbulent picture, the spatial distribution of the magnetic field energy is highly intermittent, and the regions of strong 
field have filamentary structure. Finally, a recent study of filament groupings that exhibit sorting by filament length suggested that the 
synchrotron-emitting electrons are produced by a time-dependent injecting source or crossing spatially intermittent magnetic 
bundles \citep{thomas20}.  While this model may be applicable to some systems, many examples do not exhibit this ordering in their 
filament lengths and there are no obvious compact sources running across them. 

\subsection{Summary}

We have concentrated on morphological and spectral studies of some of the most spectacular groupings of filaments in 
the Galactic center. 
The  filaments run parallel to each other, shift  sideways together  or  converge to a point 
suggesting  that the filaments in a grouping are physically associated with each other rather than 
being chance line-of-sight associations. Furthermore, 
the morphology of individual grouping 
suggests that the  magnetic field of the filaments change direction coherently implying
that they are parts of a large-scale magnetic structure.
We determined  a new characteristic of the population of the filaments in a grouping in that they are spaced 
from each other with the mean value of $\sim16''$.  We also determined the mean width span of all filament groupings is 
$\sim27''$. We modeled and reproduced  the observed  distribution of filament spacing assuming that the 
filament groupings are randomly oriented  with 
respect to the line of sight. 
This assumes  that  the Galactic center filaments are 
one-dimensional  and  are not  limb-brightened 
sheet-like  structures  viewed  edge-on. 


We note a number of striking examples of two-pronged forked filaments with a compact source detected 
at the junction.  The spectral index variation along the length of the filaments 
and morphological changes such as the width, curvature and brightness 
support the scenario of a flow of thermal plasma  being carried along the length of the filaments. 
Another striking result is the spectral index variation of harp-like filaments,  
where there is a tendency 
for filaments with the longest lengths to have steeper spectra. This correlation between length and spectral index 
variation 
implies that cosmic ray particles in short filaments are more recently accelerated than longer filaments.

As for the cause of filamentation observed throughout the Galactic center, we explored a new possibility that 
filamentation occurs because of synchrotron cooling instability. 
Rapid synchrotron cooling can  trigger the instability and forming sub-filaments parallel to each other with a spacing that 
is similar scale to observed values  of $\sim0.5-1$ pc.   
Another possibility is that filamentation is triggered  as 
a result of a large-scale, cosmic-ray driven nuclear wind  with an obstacle. In this picture, the obstacle,  which could be 
stellar wind bubble, an HII region or a cloud, compresses the magnetic field and forms a filament in the direction opposite to the 
motion of the obstacle with respect to the direction of the nuclear wind.  
Thus, the  structure of the obstacle is imprinted  on the number of parallel filaments. 
Future high-resolution observations will test the obstacle picture. 

\section{Data Availability}

All the data including  VLA and MeerKAT  that we used here are available online and are not proprietary.
We have reduced and calibrated these data and are available if  requested.




\section{Acknowledgments}
Work by R.G.A. was supported by NASA under award number 80GSFC21M0002. FYZ is partially supported by the grant AST-0807400 from the the National Science Foundation. The MeerKAT 
telescope is operated by the South African Radio Astronomy Observatory, which is a facility of the National Research Foundation, an agency of the Department of Science and 
Innovation. The authors acknowledge the Center for High Performance Computing (CHPC), South Africa, for providing computational resources to this research project. 
The National Radio Astronomy Observatory is a facility of the 
National Science Foundation operated under cooperative agreement by Associated Universities, Inc. IH acknowledges support from the UK Science and Technology Facilities Council 
[ST/N000919/1], and from the South African Radio Astronomy Observatory which is a facility of the National Research Foundation (NRF), an agency of the Department of Science and
Innovation.  The work of SB was partly supported by NSF Grant PHY-2010098, by NASA Grant 80NSSC18K0646, and by the Wisconsin Plasma Physics Laboratory (US Department of Energy Grant DE-SC0018266).


\facilities{VLA, MeerKAT}

\bibliographystyle{aasjournal}

\appendix
\section{Appendix}

 To examine projection effects on the filament spacings, 
consider a pair of filaments with perpendicular separation $h_0$, represented by the lines 
$(x, h_0, 0)$ and $(x, 0, 0)$ 
in a Cartesian coordinate system $(x,y,z)$.  Suppose that the direction to the observer is $(\phi,\theta)$ in the 
associated spherical polar coordinates.  To find the apparent separation of the lines as projected in the plane of the observer's 
sky, we construct another coordinate system $(x',y',z')$ having the same origin, with the $z'$-axis directed to the observer so 
that the sky-plane coordinates are $(x',y')$.  We may choose the $y'$-axis to be parallel to the projection of the $y$-axis on the 
plane of the observer's sky; then the coordinate systems are related by the transformation \begin{equation}
    \left(\begin{matrix}
    x' \\ y' \\ z'
    \end{matrix}\right) \;=\;
     \left(\begin{matrix}
    \cos\theta\cos\phi & \cos\theta\sin\phi & \sin\theta \\
                  -\sin\phi & \cos\phi               & 0               \\
    \sin\theta\cos\phi & \sin\theta\sin\phi & \cos\theta
    \end{matrix}\right)
    \left(\begin{matrix}
    x \\ y \\ z
    \end{matrix}\right)\,.    \label{eqn:transformation}
  \end{equation}
The coordinates tracing the filaments in the observer's sky are
\begin{eqnarray}
    x' &=& \cos\theta(x\cos\phi +a\sin\phi)\\
    y' &=& -x\sin\phi + a \cos\phi 
    \label{eqn:sky-loci}
\end{eqnarray}
i.e.
\begin{equation}
    y' = -x'\tan\phi\sec\theta+a\sec\phi
    \label{eqn:}
\end{equation}
with $a=0$ and $h_0$.  Then the apparent plane-of-sky distance between the two filaments is 
\begin{equation}
	h = \frac{h_0}{\sqrt{1+\sin^2\phi\,\tan^2\theta}}\,.
\end{equation}

Assuming that the filament systems are randomly oriented, the joint probability distribution of the observer's viewing angles is
\begin{equation}
    P(\phi,\theta) = \frac{\sin\theta}{4\pi}\,,
    \label{eqn:Pangular}
\end{equation}
which we convert to a joint probability density in $(\phi,h)$:
\begin{equation}
    P(\phi,h) \;=\; P(\phi,\theta)\left|\frac{\partial h}{\partial \theta}\right|^{-1}  = \frac{2}{\pi}\frac{|\sin\phi|}{\sqrt{h_0^2-h^2\cos^2\phi}}\,,
    \label{eqn:Ph_phi}
\end{equation}
and then integrate over $\phi$ to obtain the probability density $h$ given $h_0$, given in equation (1).

\begin{deluxetable}{rlrrrrrr}
\tablecaption{The mean angular separation of filaments in groups}
\tablehead{
\colhead{Group} &
\colhead{Group } &
\colhead{$l$} &
\colhead{$b$} &
\colhead{Span} &
\colhead{Number of } &
\colhead{Mean Separation} &
\colhead{Figure}\\
\colhead{Number} &
\colhead{Name} &
\colhead{(deg)} &
\colhead{(deg)} &
\colhead{(asec)} &
\colhead{filaments} &
\colhead{(asec)} &
\colhead{Number}
}
\startdata
 0 & Horseshoe &  358.743 &   -0.215 &      18.0 &        2 &     18.0& \ref{fig:horseshoe} \\
 1 & Pelican &  358.828 &    0.471 &      45.1 &        4 &     15.0& \ref{fig:pelican} \\
 2 & Arrow &  359.128 &    0.634 &      45.1 &        2 &     45.1& \ref{fig:arrow} \\
 3 & Snake-1 &  359.132 &   -0.296 &      33.2 &        2 &     33.2& \ref{fig:snake} \\
 4 & Snake-2 &  359.159 &   -0.111 &      28.8 &        3 &     14.4& \ref{fig:snake} \\
 5 & Snake-3 &  359.196 &   -0.474 &     131.8 &        5 &     33.0& \ref{fig:snake} \\
 6 & Candle &  359.221 &   -0.129 &      25.2 &        4 &      8.4& \ref{fig:candle} \\
 7 & Hummingbird-1 &  359.300 &   -0.175 &      15.6 &        2 &     15.6& \ref{fig:hummingbird} \\
 8 & Sausage &  359.317 &   -0.430 &      21.0 &        2 &     21.0& \ref{fig:sausage} \\
 9 & Hummingbird-2 &  359.366 &   -0.140 &      82.1 &        5 &     20.5& \ref{fig:hummingbird} \\
10 & River &  359.397 &   -0.187 &      21.6 &        2 &     21.6& \ref{fig:cataract} \\
11 & Forceps &  359.399 &   -0.213 &      18.0 &        2 &     18.0& \ref{fig:cataract} \\
12 & Feather &  359.411 &   -0.709 &      35.5 &        3 &     17.7& \ref{fig:feather} \\
13 & Sgr C-1 &  359.425 &    0.043 &      41.0 &        3 &     20.5& \ref{fig:sgrc} \\
14 & French Knife &  359.429 &    0.132 &      21.4 &        3 &     10.7& \ref{fig:fknife} \\
15 & Sgr C-2 &  359.446 &   -0.005 &      25.8 &        2 &     25.8& \ref{fig:sgrc} \\
16 & Bent Harp &  359.484 &    0.122 &      53.1 &        5 &     13.3& \ref{fig:bentharp} \\
17 & Concorde &  359.495 &    0.188 &      26.2 &        3 &     13.1& \ref{fig:fknife} \\
18 & Cataract &  359.504 &   -0.321 &     289.8 &        8 &     41.4& \ref{fig:cataract} \\
19 & Ripple &  359.520 &    0.237 &      36.0 &        3 &     18.0& \ref{fig:ripple} \\
20 & Edge-on Spiral &  359.563 &    0.249 &      28.1 &        2 &     28.1& \ref{fig:ripple} \\
21 & Broken Harp-1 &  359.625 &    0.298 &      31.8 &        4 &     10.6& \ref{fig:brokenharp} \\
22 & Broken Harp-2 &  359.640 &    0.308 &      77.4 &        6 &     15.5& \ref{fig:brokenharp} \\
23 & Waterfall &  359.687 &   -0.435 &      90.0 &        6 &     18.0& \ref{fig:waterfall} \\
24 & Hilltop &  359.702 &   -0.517 &      29.7 &        5 &      7.4& \ref{fig:waterfall} \\
25 & Flamingo-1 &  359.717 &    0.228 &      33.2 &        3 &     16.6& \ref{fig:flamingo} \\
26 & Knot &  359.738 &   -0.792 &      21.0 &        2 &     21.0& \ref{fig:knot} \\
27 & Flamingo-2 &  359.808 &    0.130 &      32.4 &        3 &     16.2& \ref{fig:flamingo} \\
28 & Eyebrow &  359.830 &    0.085 &      29.2 &        3 &     14.6& \ref{fig:flamingo} \\
29 & Harp &  359.851 &    0.374 &      91.5 &        8 &     13.1& \ref{fig:harp} \\
30 & Cleaver &  359.890 &   -0.289 &      23.1 &        2 &     23.1& \ref{fig:cleaver} \\
31 & Comet Tail &  359.992 &   -0.572 &      54.9 &        3 &     27.5& \ref{fig:comet} \\
32 & Radio Arc S-1 &    0.138 &   -0.299 &     249.3 &       12 &     22.7& \ref{fig:radioarcs} \\
33 & Radio Arc &    0.160 &   -0.115 &     224.5 &       17 &     14.0& \ref{fig:radioarc} \\
34 & Ring &    0.168 &    0.142 &      96.9 &        7 &     16.2& \ref{fig:ring} \\
35 & Radio Arc S-2 &    0.172 &   -0.423 &      18.4 &        2 &     18.4& \ref{fig:radioarcs} \\
36 & Space Shuttle &    0.228 &    0.812 &      29.0 &        3 &     14.5& \ref{fig:shuttle} \\
37 & Porcupine-1 &    0.288 &   -0.259 &      20.3 &        4 &      6.8& \ref{fig:porcupine} \\
38 & Porcupine-2 &    0.297 &   -0.266 &      62.8 &        7 &     10.5& \ref{fig:porcupine} \\
39 & Porcupine-3 &    0.306 &   -0.273 &      28.6 &        3 &     14.3& \ref{fig:porcupine} \\
40 & Contrail &    0.405 &   -0.277 &      31.0 &        3 &     15.5& \ref{fig:contrail} \\
41 & Bent Fork &    0.687 &    0.147 &       7.5 &        2 &      7.5& \ref{fig:bentfork} \\
42 & Meteor Trail &    0.859 &   -0.597 &      25.5 &        2 &     25.5& \ref{fig:meteor} \\
\hline
& Robust mean & & & 27$\pm$8 & & 16$\pm$7
\enddata
\end{deluxetable}

\label{tab:table1}

\begin{deluxetable}{lrrlr}
\tabletypesize{\scriptsize}
\tablecaption{Compact sources and locations of interest}
\tablehead{
\colhead{Name} &
\colhead{$l$ (deg)} &
\colhead{$b$ (deg)} &
\colhead{Type} & 
\colhead{Figure}
}
\startdata
G358.693-0.227 &  358.693 &   -0.227 & Source & \ref{fig:horseshoe} \\
G358.733-0.229 &  358.733 &   -0.229 & Source & \ref{fig:horseshoe} \\
G358.776+0.454 &  358.776 &    0.454 & Source & \ref{fig:pelican} \\
G358.794+0.479 &  358.794 &    0.479 & Source & \ref{fig:pelican} \\
G358.798+0.468 &  358.798 &    0.468 & Source & \ref{fig:pelican} \\
G359.072+0.735 &  359.072 &    0.735 & Source & \ref{fig:arrow} \\
G359.120-0.265 &  359.120 &   -0.265 & Source & \ref{fig:snake} \\
G359.132-0.200 &  359.132 &   -0.200 & Source & \ref{fig:snake} \\
G359.211-0.085 &  359.211 &   -0.085 & Source & \ref{fig:candle} \\
G359.214-0.100 &  359.214 &   -0.100 & Source & \ref{fig:candle} \\
G359.224-0.136 &  359.224 &   -0.136 & Source & \ref{fig:candle} \\
G359.321-0.158 &  359.321 &   -0.158 & Location & \ref{fig:hummingbird} \\
G359.321-0.431 &  359.321 &   -0.431 & Source & \ref{fig:sausage} \\
G359.324-0.430 &  359.324 &   -0.430 & Source & \ref{fig:sausage} \\
G359.344-0.416 &  359.344 &   -0.416 & Source & \ref{fig:sausage} \\
G359.401-0.238 &  359.401 &   -0.238 & Location & \ref{fig:cataract} \\
G359.405-0.259 &  359.405 &   -0.259 & Location & \ref{fig:cataract} \\
G359.416-0.706 &  359.416 &   -0.706 & Source & \ref{fig:feather} \\
G359.419-0.583 &  359.419 &   -0.583 & Source & \ref{fig:feather} \\
G359.420-0.660 &  359.420 &   -0.660 & Location & \ref{fig:feather} \\
G359.423-0.664 &  359.423 &   -0.664 & Source & \ref{fig:feather} \\
G359.439+0.006 &  359.439 &    0.006 & Location & \ref{fig:sgrc} \\
G359.439+0.103 &  359.439 &    0.103 & Location & \ref{fig:fknife} \\
G359.453-0.033 &  359.453 &   -0.033 & Location & \ref{fig:sgrc} \\
G359.455-0.055 &  359.455 &   -0.055 & Source & \ref{fig:sgrc} \\
G359.467-0.171 &  359.467 &   -0.171 & Source & \ref{fig:cataract} \\
G359.475+0.127 &  359.475 &    0.127 & Source & \ref{fig:bentharp} \\
G359.485+0.121 &  359.485 &    0.121 & Source & \ref{fig:bentharp} \\
G359.491+0.121 &  359.491 &    0.121 & Source & \ref{fig:bentharp} \\
G359.522+0.097 &  359.522 &    0.097 & Source & \ref{fig:bentharp} \\
G359.557+0.140 &  359.557 &    0.140 & Source & \ref{fig:ripple} \\
G359.729-0.811 &  359.729 &   -0.811 & Location & \ref{fig:knot} \\
G359.734-0.801 &  359.734 &   -0.801 & Location & \ref{fig:knot} \\
G359.746-0.784 &  359.746 &   -0.784 & Source & \ref{fig:knot} \\
G359.748-0.781 &  359.748 &   -0.781 & Source & \ref{fig:knot} \\
G359.758+0.198 &  359.758 &    0.198 & Location & \ref{fig:flamingo} \\
G359.808+0.117 &  359.808 &    0.117 & Source & \ref{fig:flamingo} \\
G359.838+0.375 &  359.838 &    0.375 & Source & \ref{fig:harp} \\
G359.841+0.366 &  359.841 &    0.366 & Source & \ref{fig:harp} \\
G359.844+0.358 &  359.844 &    0.358 & Source & \ref{fig:harp} \\
G359.848+0.298 &  359.848 &    0.298 & Source & \ref{fig:harp} \\
G359.851+0.293 &  359.851 &    0.293 & Source & \ref{fig:harp} \\
G359.874-0.264 &  359.874 &   -0.264 & Source & \ref{fig:cleaver} \\
G359.886+0.102 &  359.886 &    0.102 & Source & \ref{fig:flamingo} \\
G359.921-0.283 &  359.921 &   -0.283 & Location & \ref{fig:cleaver} \\
G359.950-0.259 &  359.950 &   -0.259 & Source & \ref{fig:cleaver} \\
G359.992-0.555 &  359.992 &   -0.555 & Location & \ref{fig:comet} \\
G0.023-0.626 &    0.023 &   -0.626 & Location & \ref{fig:comet} \\
G0.030-0.660 &    0.030 &   -0.660 & Source & \ref{fig:comet} \\
G0.034-0.661 &    0.034 &   -0.661 & Source & \ref{fig:comet} \\
G0.045-0.668 &    0.045 &   -0.668 & Source & \ref{fig:comet} \\
G0.225+0.843 &    0.225 &    0.843 & Source & \ref{fig:shuttle} \\
G0.226+0.792 &    0.226 &    0.792 & Source & \ref{fig:shuttle} \\
G0.264-0.197 &    0.264 &   -0.197 & Source & \ref{fig:porcupine} \\
G0.420-0.297 &    0.420 &   -0.297 & Location & \ref{fig:contrail} \\
G0.421-0.290 &    0.421 &   -0.290 & Location & \ref{fig:contrail} \\
G0.443-0.324 &    0.443 &   -0.324 & Location & \ref{fig:contrail} \\
G0.697+0.148 &    0.697 &    0.148 & Source & \ref{fig:bentfork} \\
\enddata
\end{deluxetable}

\label{tab:table2}

\begin{sidewaysfigure}
\plottwo{{cross_section_filtered_G359.33-0.42}.eps}{{cross_section_G359.33-0.42}.eps}
\plottwo{{detected_G359.33-0.42}.eps}{{alpha_G359.33-0.42}.eps}
\caption{(The Sausage) Same as Figure \ref{fig:horseshoe} except G359.317-0.430 filaments are displayed.
\label{fig:sausage}}
\end{sidewaysfigure}

\begin{sidewaysfigure}
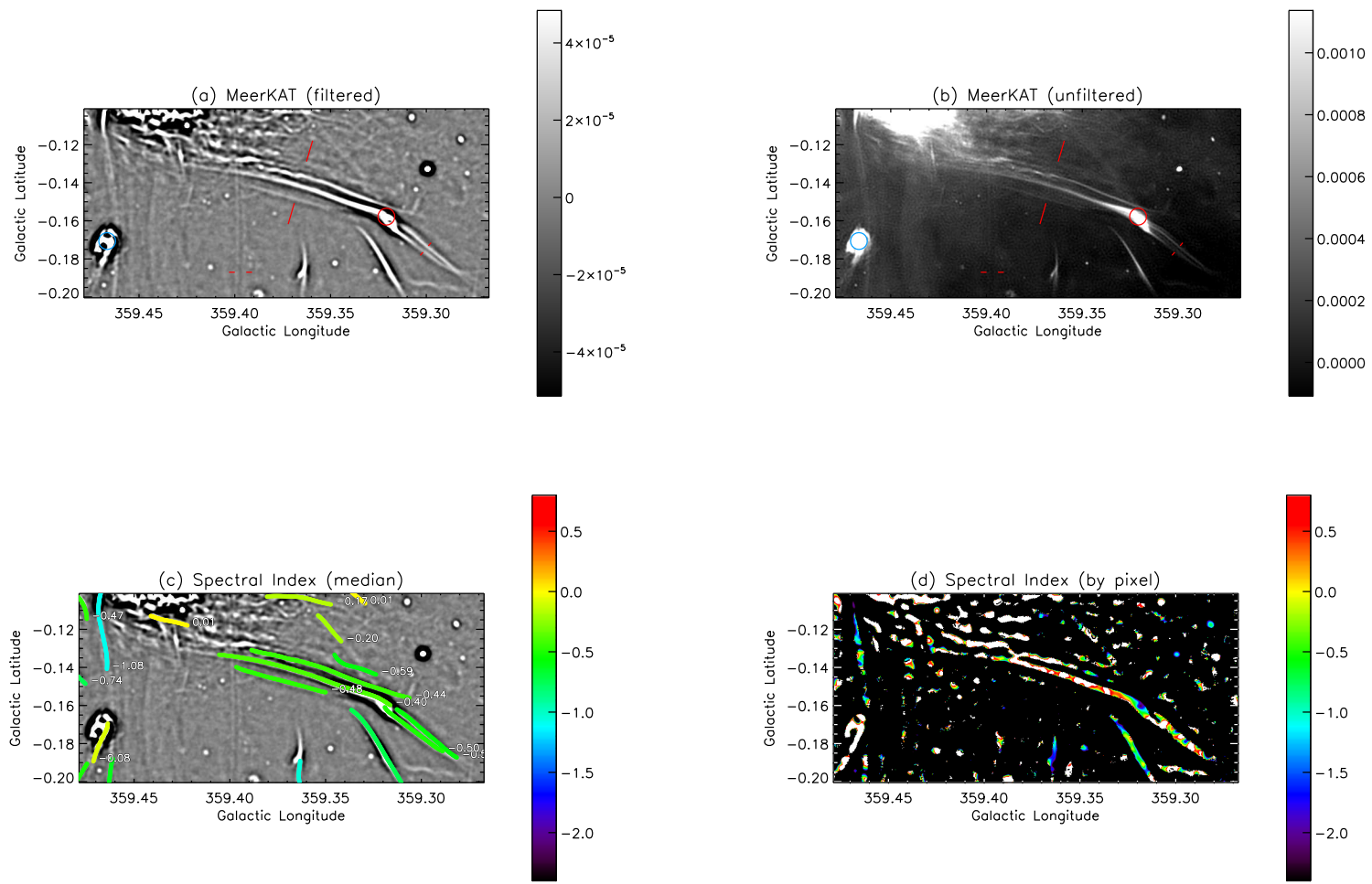

\plottwo{{cross_section_filtered_G359.37-0.15}.eps}{{cross_section_G359.37-0.15}.eps}
\plottwo{{detected_G359.37-0.15}.eps}{{alpha_G359.37-0.15}.eps}
\caption{(The Hummingbird) Same as Figure \ref{fig:horseshoe} except G359.300-0.175, G359.366-0.140, and G359.397-0.187  filaments are displayed.
\label{fig:hummingbird}}
\end{sidewaysfigure}

\begin{sidewaysfigure}
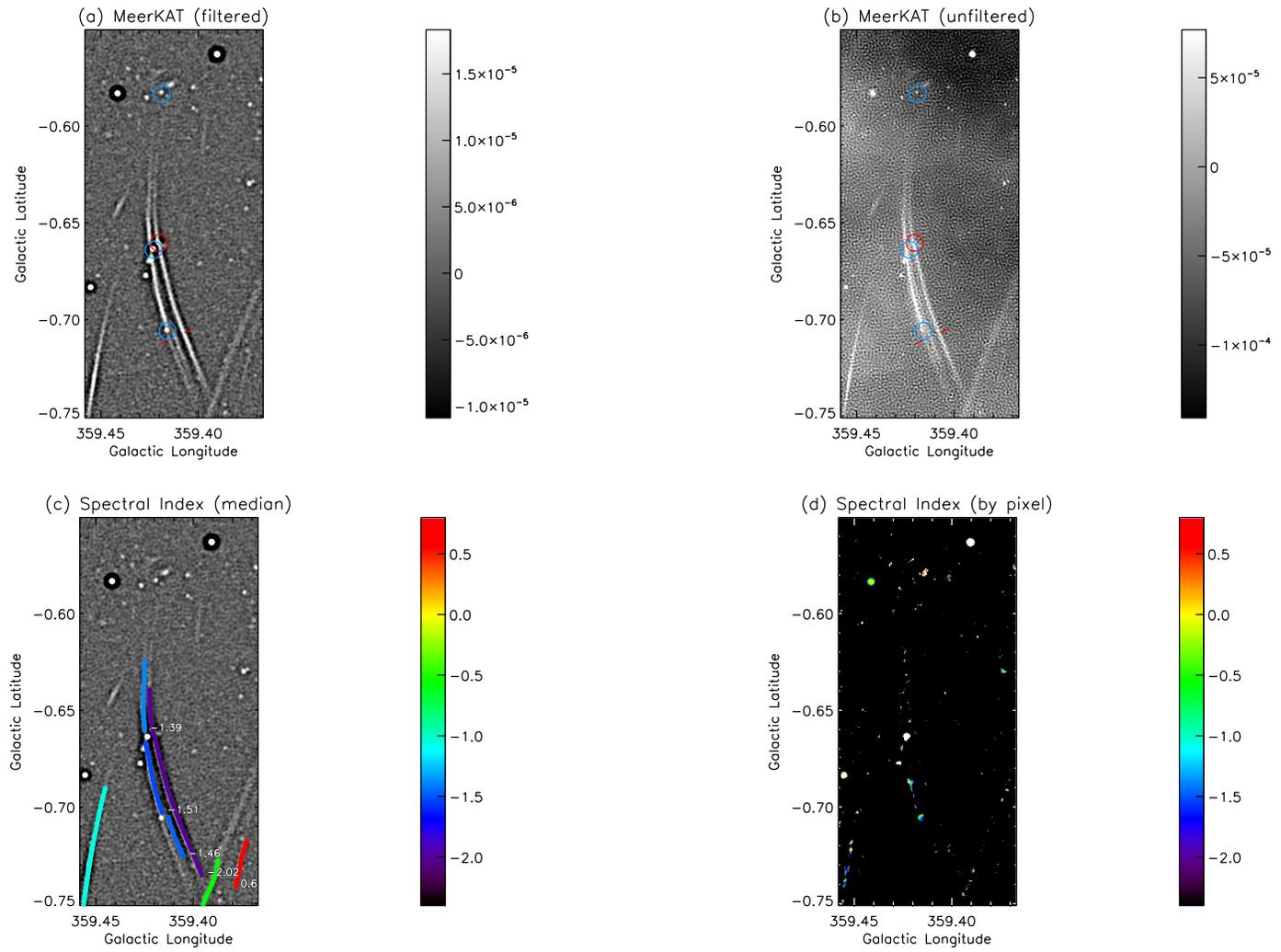

\plottwo{{cross_section_filtered_G359.41-0.65}.eps}{{cross_section_G359.41-0.65}.eps}
\plottwo{{detected_G359.41-0.65}.eps}{{alpha_G359.41-0.65}.eps}
\caption{(The Feather) Same as Figure \ref{fig:horseshoe} except G359.411-0.709 filaments are displayed.
\label{fig:feather}}
\end{sidewaysfigure}

\begin{sidewaysfigure}
\plottwo{{cross_section_filtered_G359.43+0.00}.eps}{{cross_section_G359.43+0.00}.eps}
\plottwo{{detected_G359.43+0.00}.eps}{{alpha_G359.43+0.00}.eps}
\epsscale{0.4}
\plotone{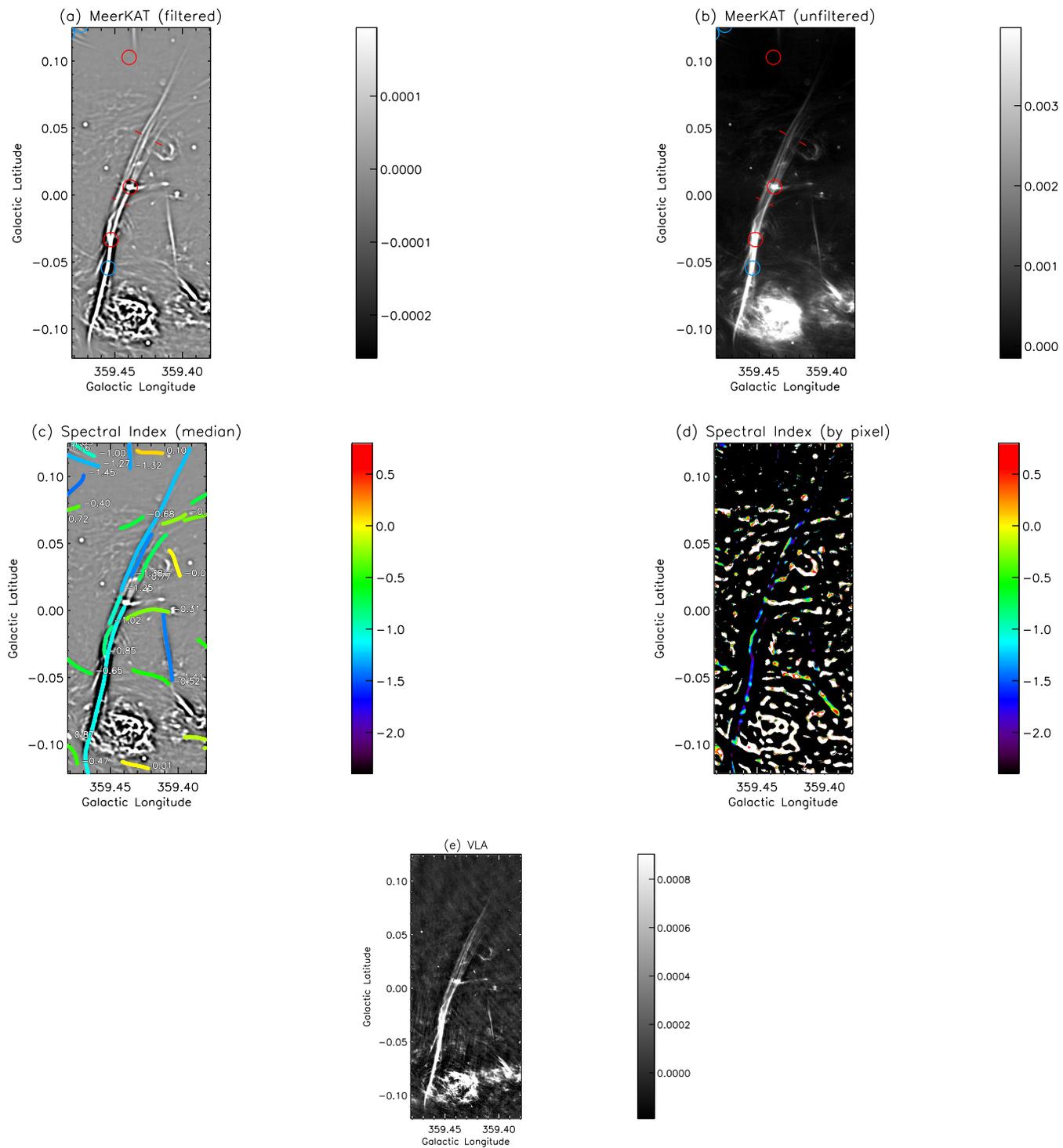}
\caption{(The SgrC) Top four panels are the same as Figure \ref{fig:horseshoe} except G359.425+0.043 and G359.446-0.005 filaments displayed. The bottom panel shows a 20cm VLA image of the same region with $\sim1''$ spatial resolution. 
\label{fig:sgrc}}
\end{sidewaysfigure}


\begin{sidewaysfigure}
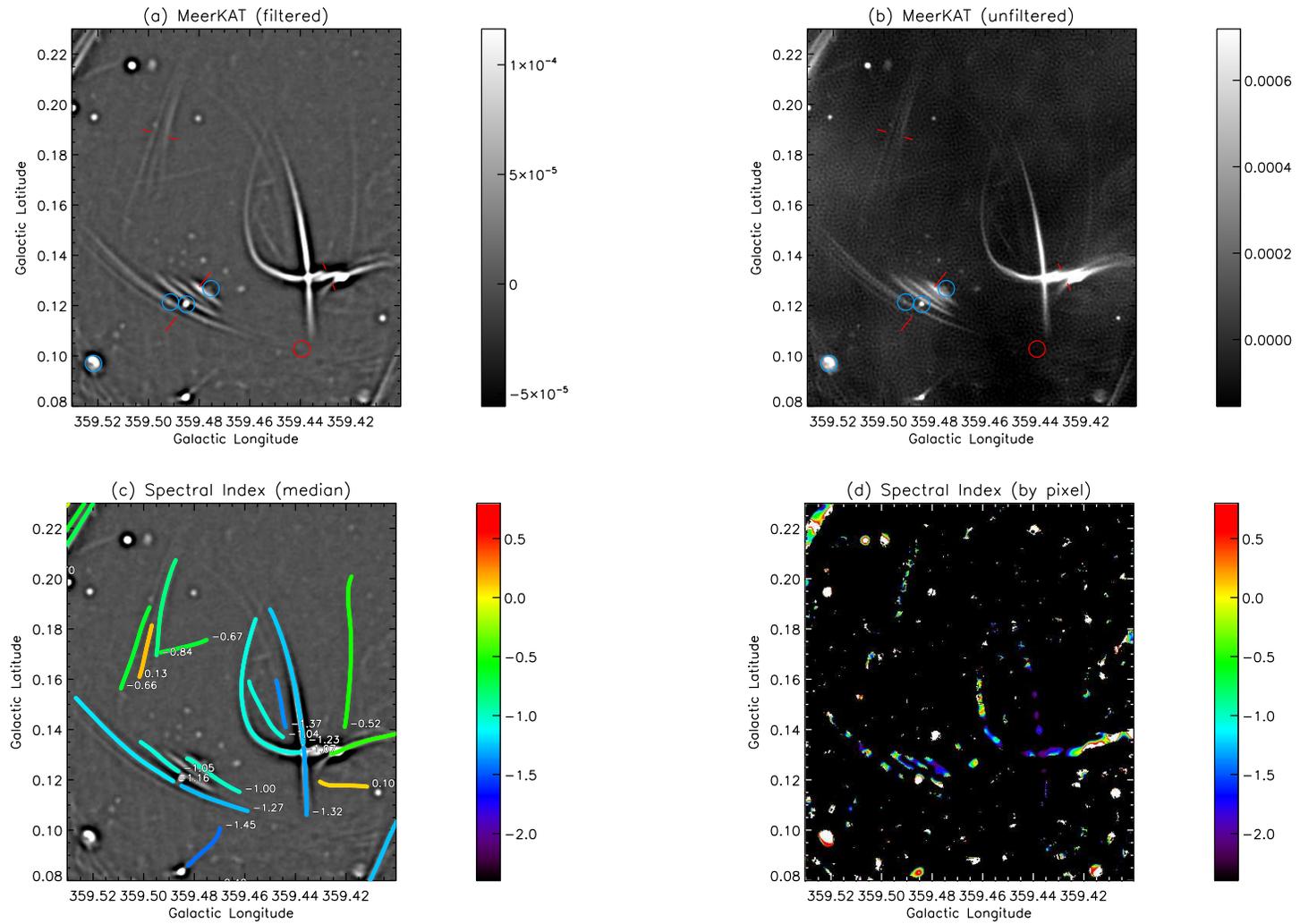

\plottwo{{cross_section_filtered_G359.46+0.16}.eps}{{cross_section_G359.46+0.16}.eps}
\plottwo{{detected_G359.46+0.16}.eps}{{alpha_G359.46+0.16}.eps}
\caption{(The French Knife and the Concorde) Same as Figure \ref{fig:horseshoe} except G359.429+0.132 filaments are displayed
\label{fig:fknife}}
\end{sidewaysfigure}

\begin{sidewaysfigure}
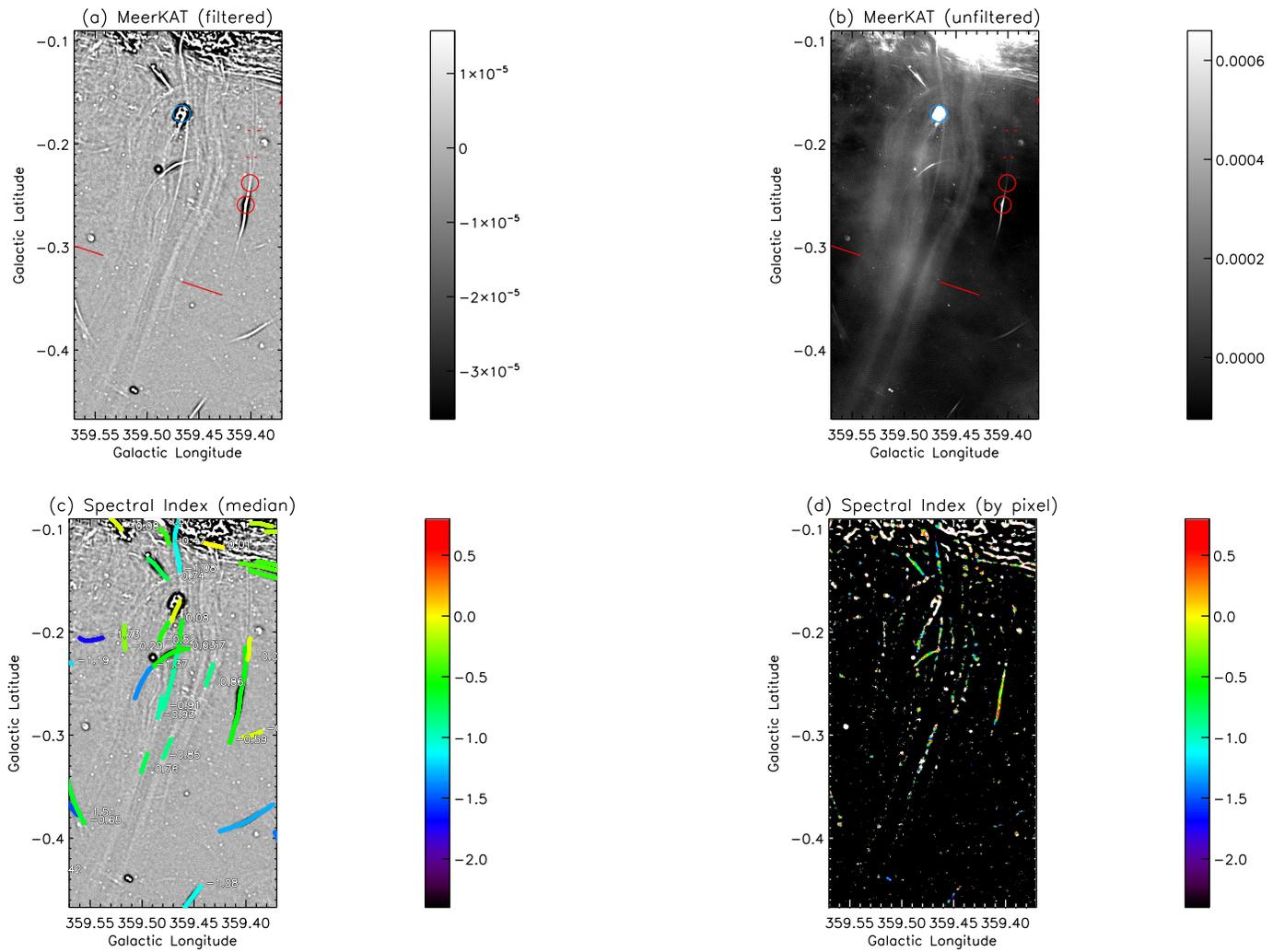

\plottwo{{cross_section_filtered_G359.47-0.28}.eps}{{cross_section_G359.47-0.28}.eps}
\plottwo{{detected_G359.47-0.28}.eps}{{alpha_G359.47-0.28}.eps}
\caption{(The Cataract, the Forceps and the River) Same as Figure \ref{fig:horseshoe} except G359.504-0.321 filaments are displayed.
\label{fig:cataract}}
\end{sidewaysfigure}

\begin{sidewaysfigure}
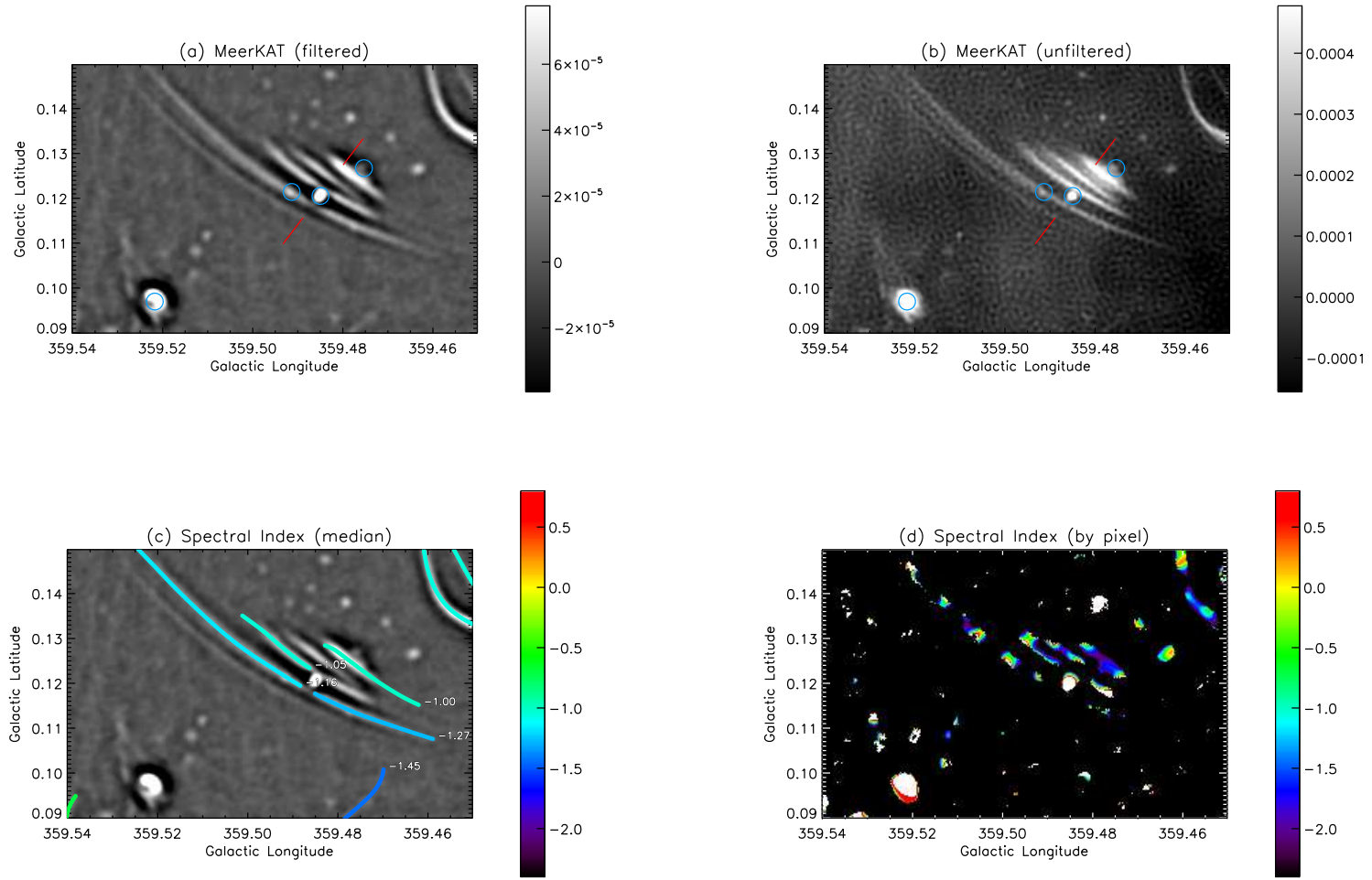

\plottwo{{cross_section_filtered_G359.49+0.12}.eps}{{cross_section_G359.49+0.12}.eps}
\plottwo{{detected_G359.49+0.12}.eps}{{alpha_G359.49+0.12}.eps}
\caption{(The Bent Harp) Same as Figure \ref{fig:horseshoe} except G-0.50+0.12 filaments are displayed
\label{fig:bentharp}}
\end{sidewaysfigure}

\begin{sidewaysfigure}
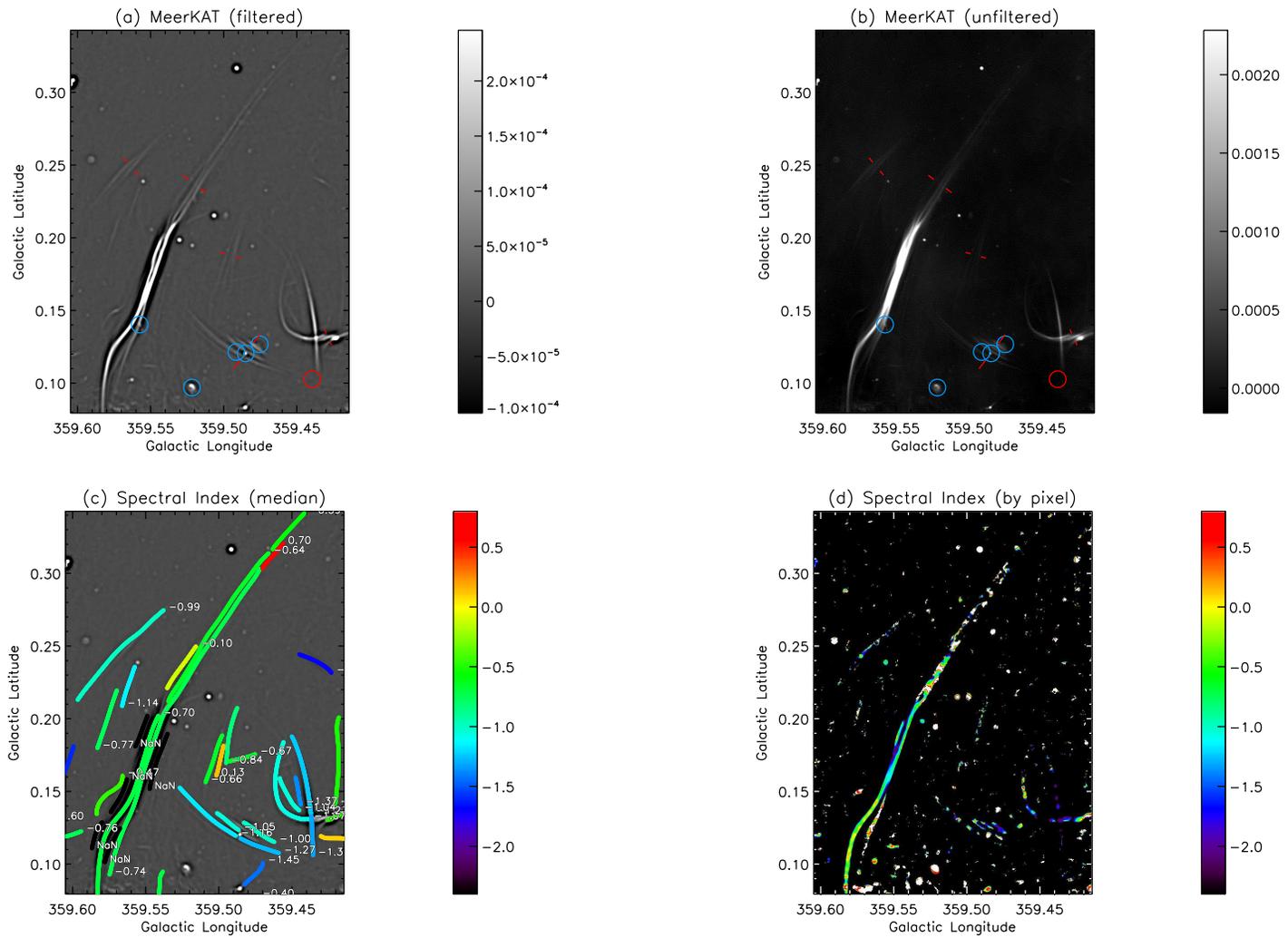

\plottwo{{cross_section_filtered_G359.51+0.21}.eps}{{cross_section_G359.51+0.21}.eps}
\plottwo{{detected_G359.51+0.21}.eps}{{alpha_G359.51+0.21}.eps}
\caption{(The Ripple and the Edge-on Spiral) Same as Figure \ref{fig:horseshoe} except G-0.49+0.21 filaments are displayed.
\label{fig:ripple}}
\end{sidewaysfigure}

\begin{sidewaysfigure}
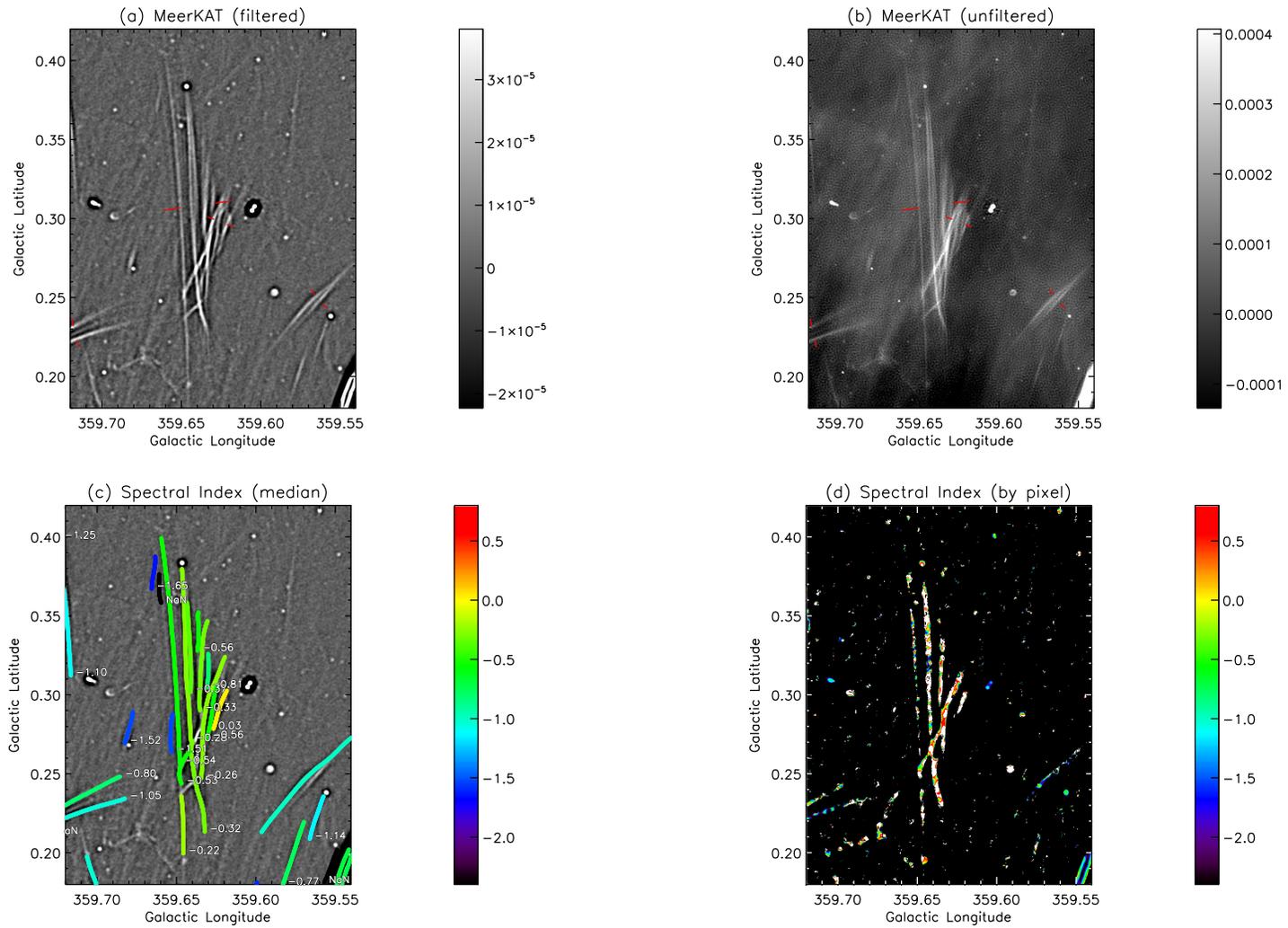

\plottwo{{cross_section_filtered_G359.63+0.30}.eps}{{cross_section_G359.63+0.30}.eps}
\plottwo{{detected_G359.63+0.30}.eps}{{alpha_G359.63+0.30}.eps}
\caption{(The Broken Harp) Same as Figure \ref{fig:horseshoe} except G359.563+0.249, G359.625+0.298 and G359.640+0.308 filaments are displayed.
\label{fig:brokenharp}}
\end{sidewaysfigure}


\begin{sidewaysfigure}
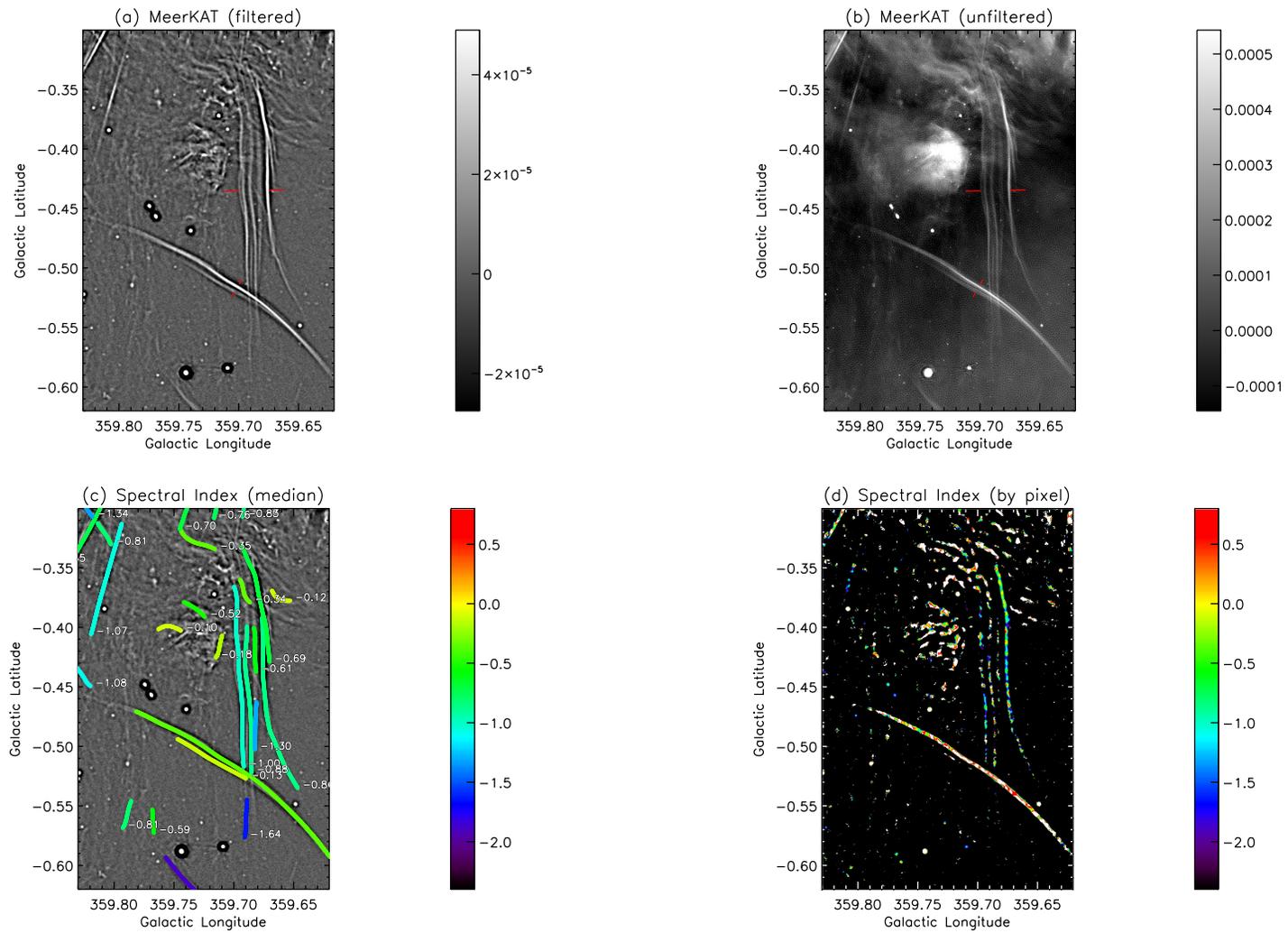

\plottwo{{cross_section_filtered_G359.72-0.46}.eps}{{cross_section_G359.72-0.46}.eps}
\plottwo{{detected_G359.72-0.46}.eps}{{alpha_G359.72-0.46}.eps}
\caption{(The Waterfall) Same as Figure \ref{fig:horseshoe} except G359.687-0.435 and G359.702-0.517 filaments are displayed.
\label{fig:waterfall}}
\end{sidewaysfigure}


\begin{sidewaysfigure}
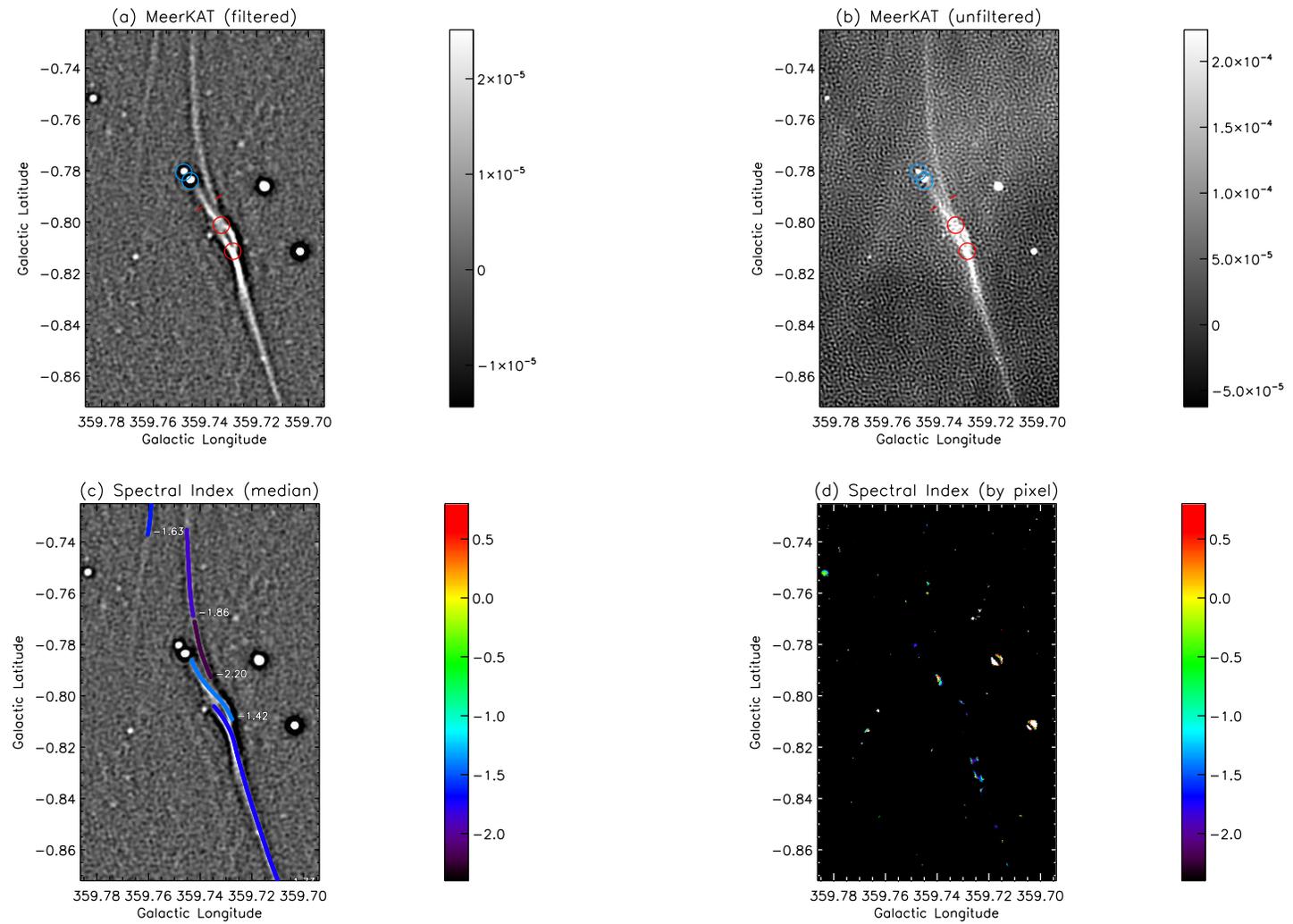

\plottwo{{cross_section_filtered_G359.74-0.80}.eps}{{cross_section_G359.74-0.80}.eps}
\plottwo{{detected_G359.74-0.80}.eps}{{alpha_G359.74-0.80}.eps}
\caption{(The Knot) Same as Figure \ref{fig:horseshoe} except G359.738-0.792 filaments are displayed.
\label{fig:knot}}
\end{sidewaysfigure}

\begin{sidewaysfigure}
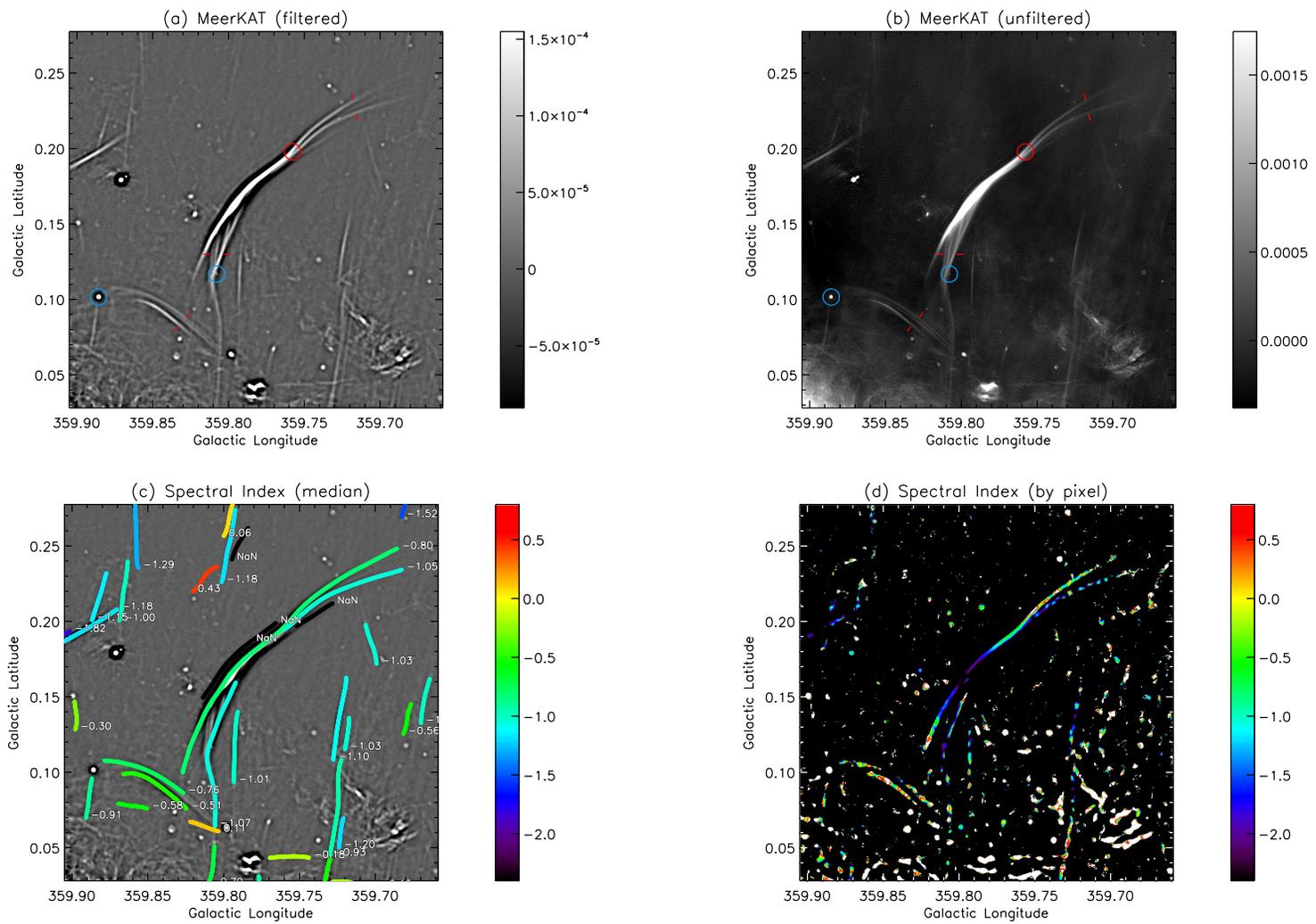

\plottwo{{cross_section_filtered_G359.78+0.15}.eps}{{cross_section_G359.78+0.15}.eps}
\plottwo{{detected_G359.78+0.15}.eps}{{alpha_G359.78+0.15}.eps}
\caption{(The Flamingo) Same as Figure \ref{fig:horseshoe} except G359.808+0.130 and G359.717+0.228 filaments are displayed.
\label{fig:flamingo}}
\end{sidewaysfigure}

\begin{sidewaysfigure}
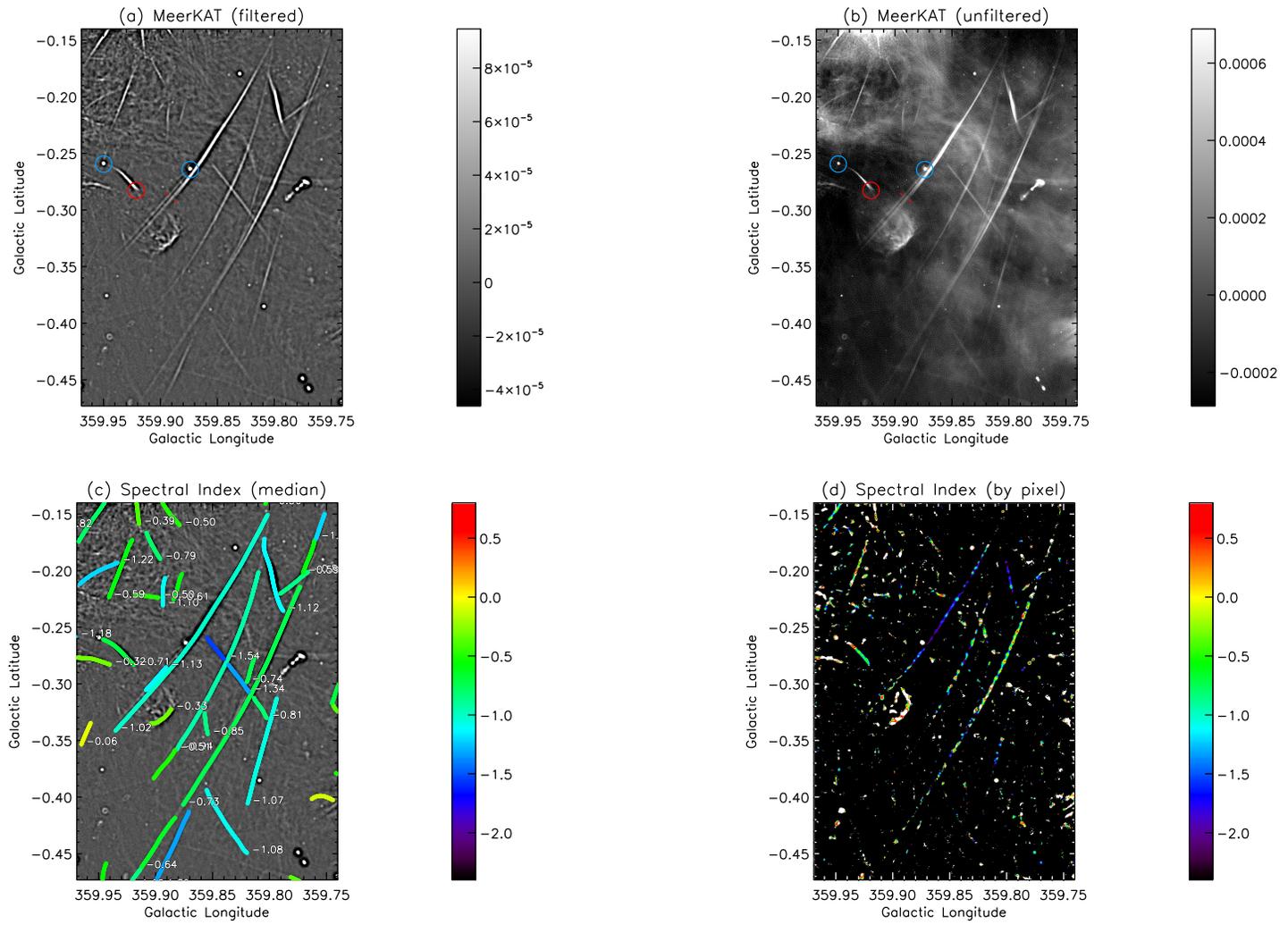

\plottwo{{cross_section_filtered_G359.85-0.31}.eps}{{cross_section_G359.85-0.31}.eps}
\plottwo{{detected_G359.85-0.31}.eps}{{alpha_G359.85-0.31}.eps}
\caption{(The Cleaver Knife) Same as Figure \ref{fig:horseshoe} except G359.890-0.289 filaments are displayed.
\label{fig:cleaver}}
\end{sidewaysfigure}

\begin{sidewaysfigure}
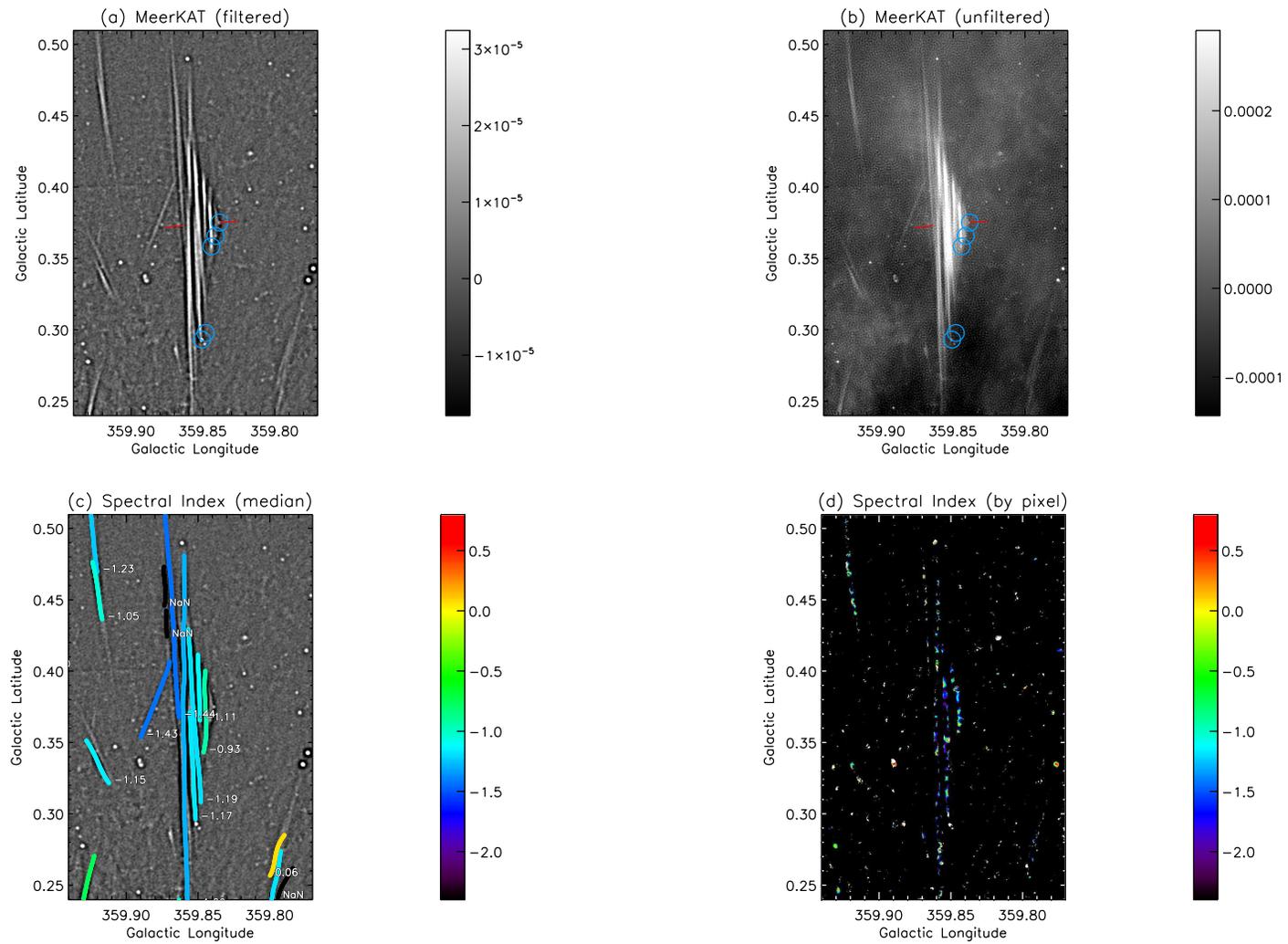

\plottwo{{cross_section_filtered_G359.85+0.38}.eps}{{cross_section_G359.85+0.38}.eps}
\plottwo{{detected_G359.85+0.38}.eps}{{alpha_G359.85+0.38}.eps}
\caption{(The Harp) Same as Figure \ref{fig:horseshoe} except G359.851+0.374 filaments are displayed.
\label{fig:harp}}
\end{sidewaysfigure}

\begin{sidewaysfigure}
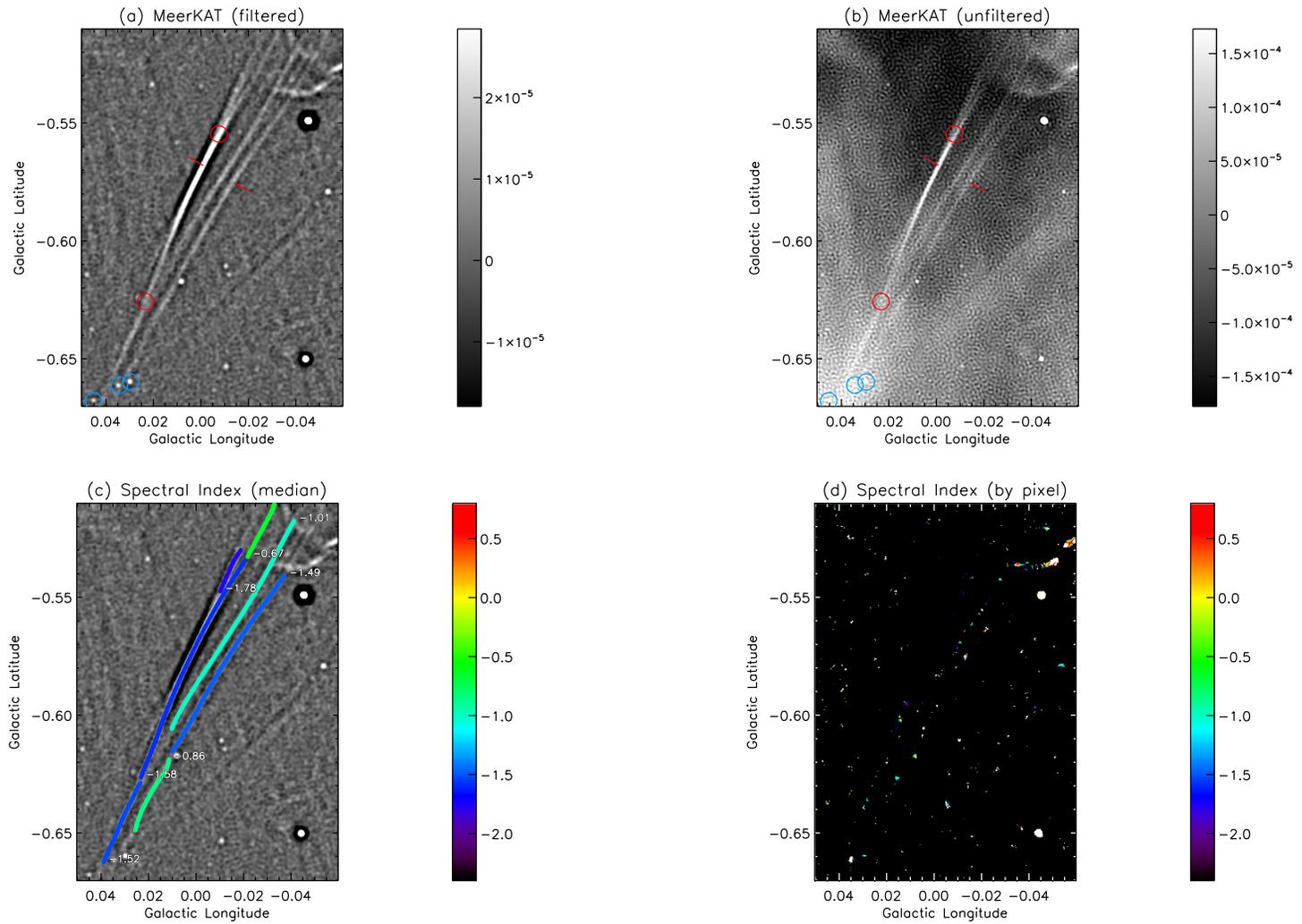

\plottwo{{cross_section_filtered_G359.99-0.59}.eps}{{cross_section_G359.99-0.59}.eps}
\plottwo{{detected_G359.99-0.59}.eps}{{alpha_G359.99-0.59}.eps}
\caption{(The Comet) Same as Figure \ref{fig:horseshoe} except G359.992-0.572 filaments are displayed.
\label{fig:comet}}
\end{sidewaysfigure}

\begin{sidewaysfigure}
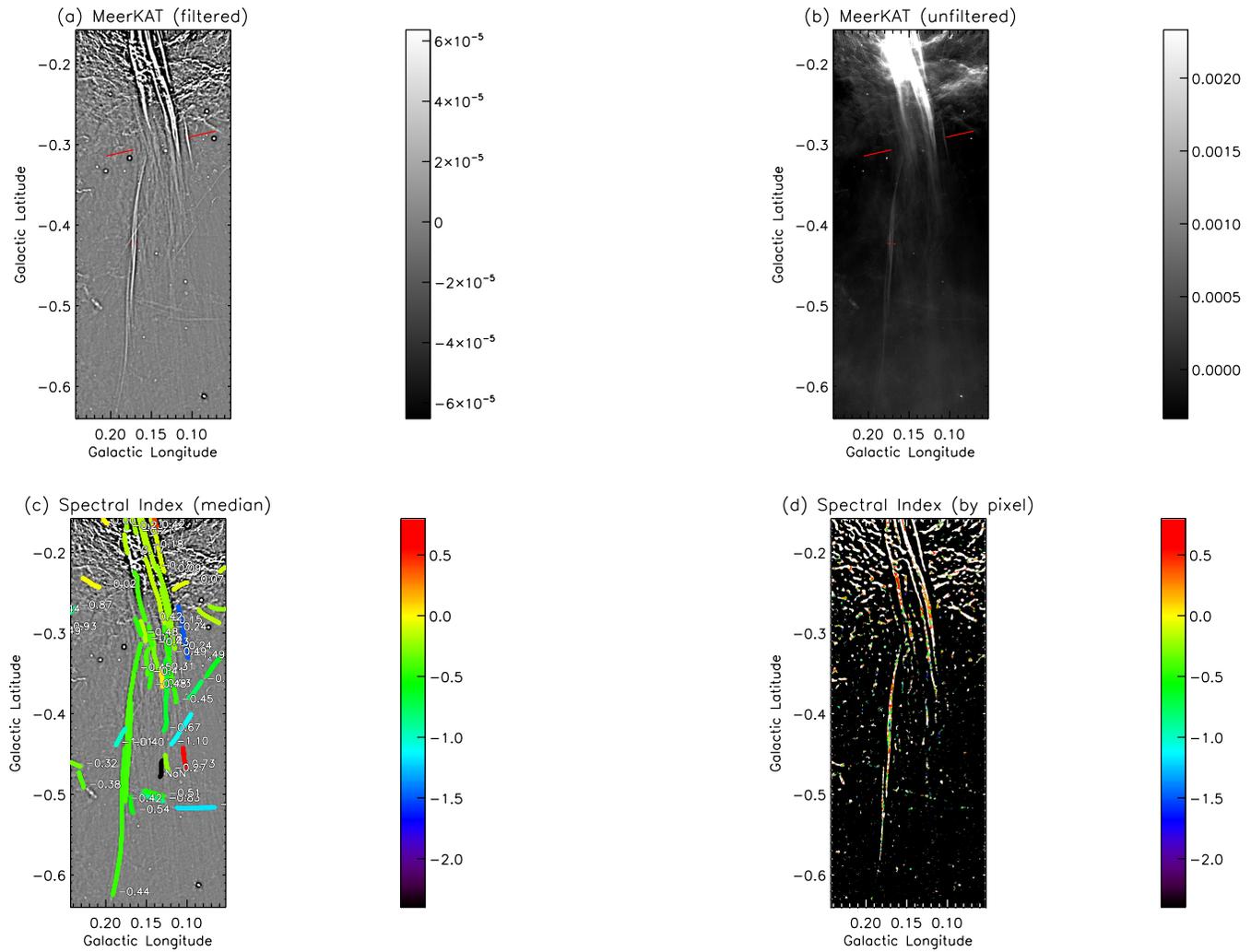

\plottwo{{cross_section_filtered_G0.15-0.40}.eps}{{cross_section_G0.15-0.40}.eps}
\plottwo{{detected_G0.15-0.40}.eps}{{alpha_G0.15-0.40}.eps}
\caption{(The Radio Arc S) Same as Figure \ref{fig:horseshoe} except G0.138-0.299 and G0.172-0.423 filaments are displayed.
\label{fig:radioarcs}}
\end{sidewaysfigure}

\begin{sidewaysfigure}
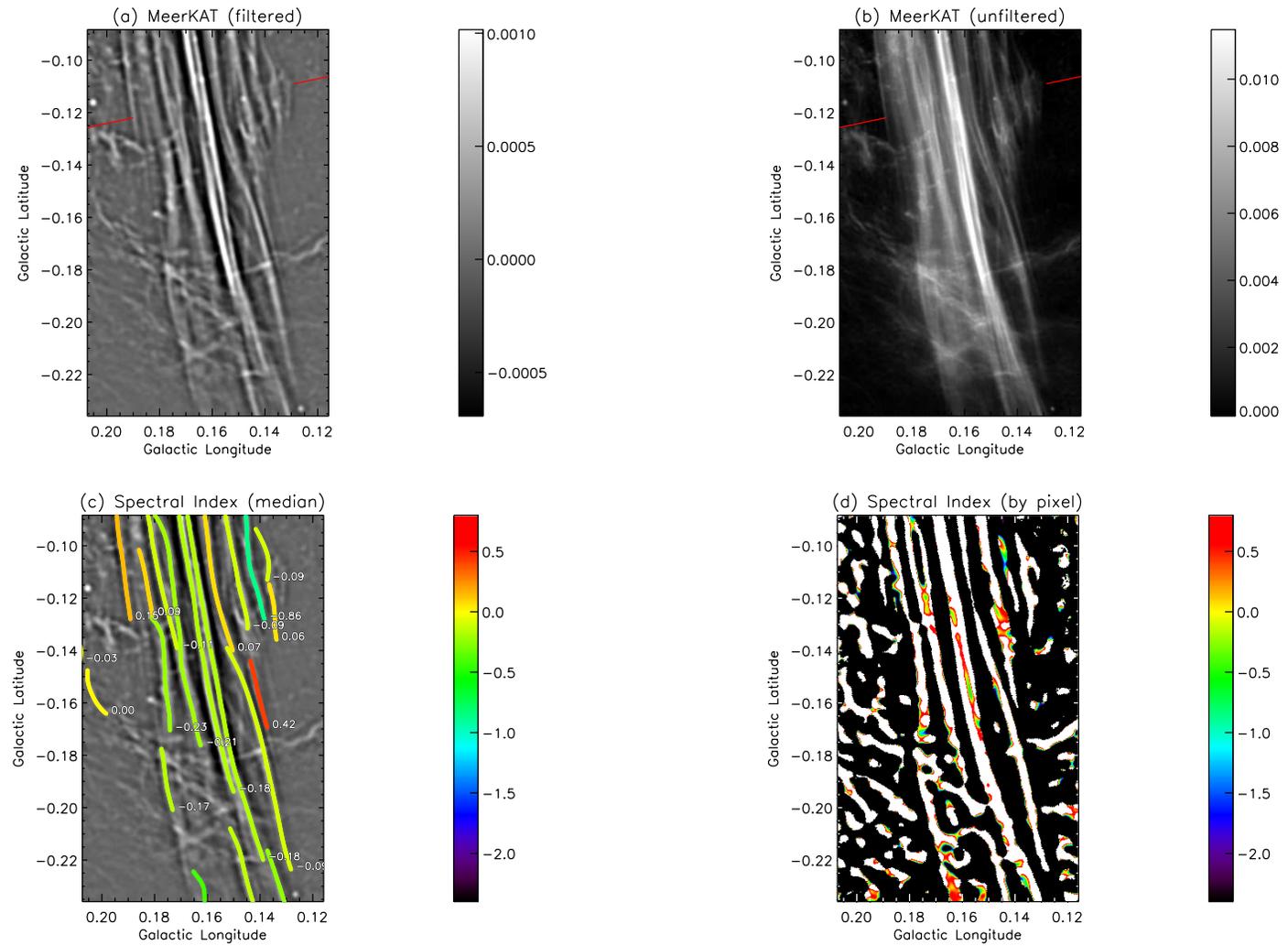

\plottwo{{cross_section_filtered_G0.16-0.16}.eps}{{cross_section_G0.16-0.16}.eps}
\plottwo{{detected_G0.16-0.16}.eps}{{alpha_G0.16-0.16}.eps}
\caption{(The Radio Arc) Same as Figure \ref{fig:horseshoe} except G0.16-0.16 filaments are displayed.
\label{fig:radioarc}}
\end{sidewaysfigure}

\begin{sidewaysfigure}
\plottwo{{cross_section_filtered_G0.16+0.16}.eps}{{cross_section_G0.16+0.16}.eps}
\plottwo{{detected_G0.16+0.16}.eps}{{alpha_G0.16+0.16}.eps}
\caption{(The Ring) Same as Figure \ref{fig:horseshoe} except G0.16+0.16 filaments are displayed.
\label{fig:ring}}
\end{sidewaysfigure}

\begin{sidewaysfigure}
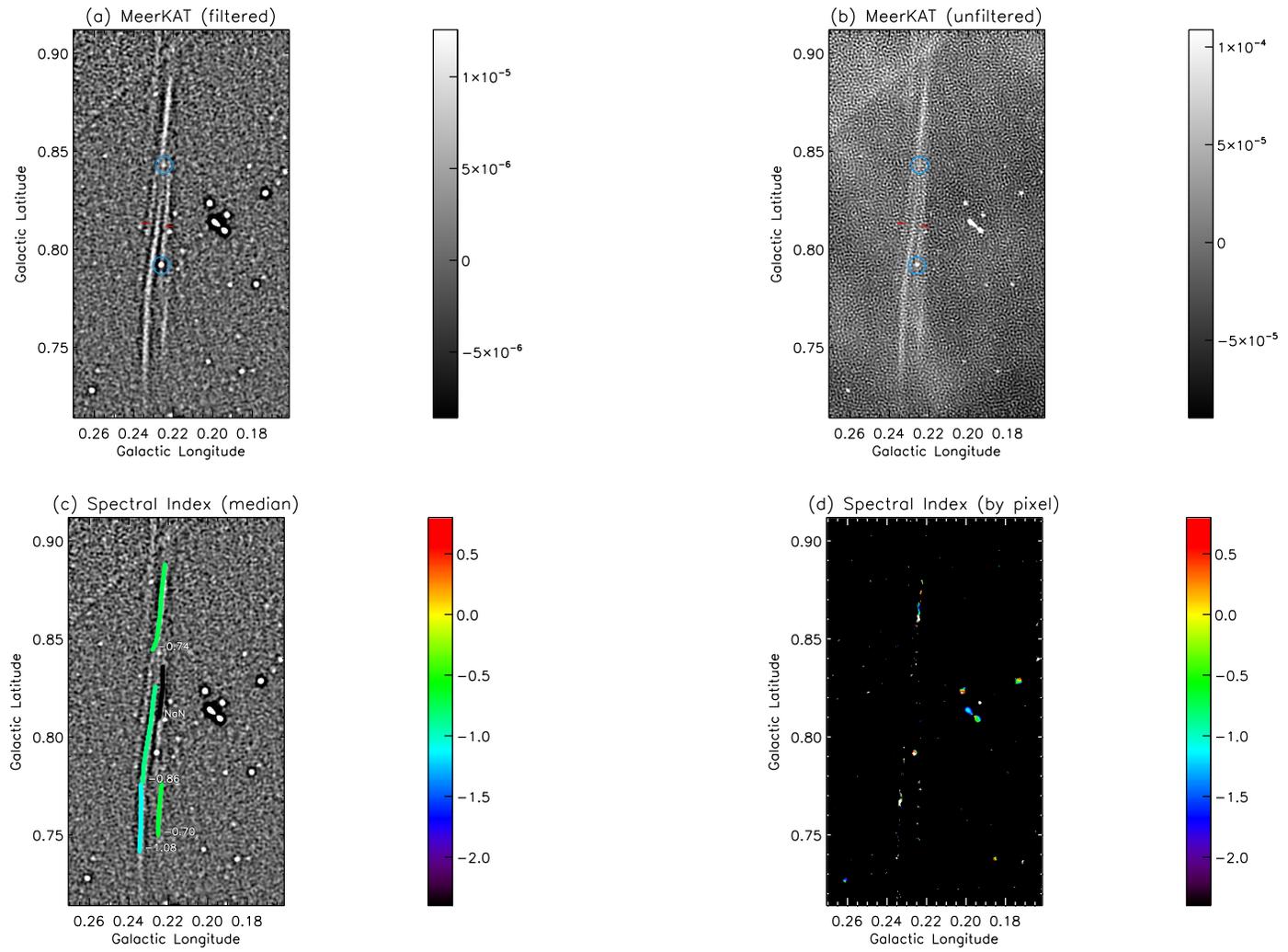

\plottwo{{cross_section_filtered_G0.22+0.81}.eps}{{cross_section_G0.22+0.81}.eps}
\plottwo{{detected_G0.22+0.81}.eps}{{alpha_G0.22+0.81}.eps}
\caption{(The Space Shuttle) Same as Figure \ref{fig:horseshoe} except G0.22+0.81 filaments are displayed.
\label{fig:shuttle}}
\end{sidewaysfigure}

\begin{sidewaysfigure}
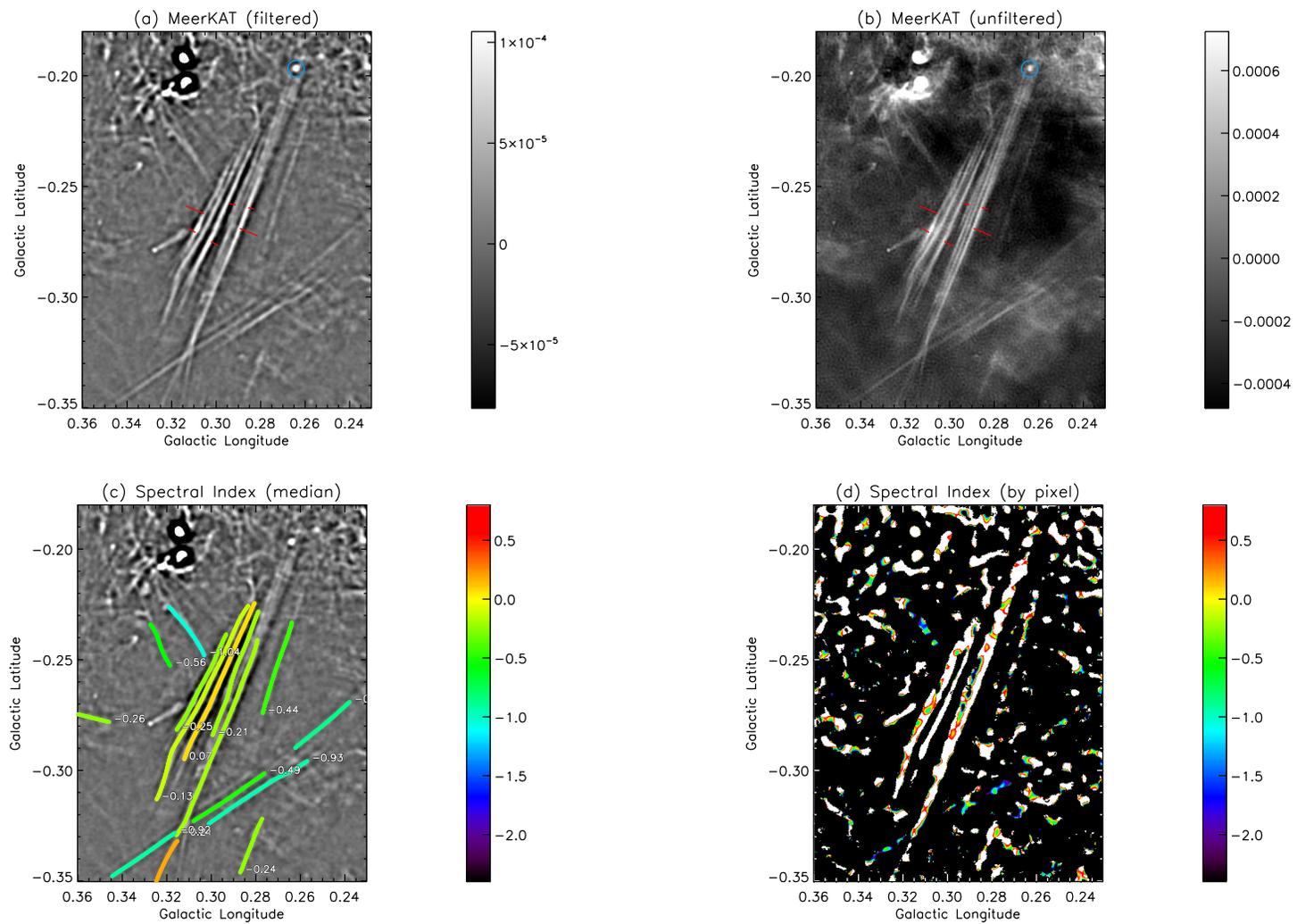

\plottwo{{cross_section_filtered_G0.30-0.26}.eps}{{cross_section_G0.30-0.26}.eps}
\plottwo{{detected_G0.30-0.26}.eps}{{alpha_G0.30-0.26}.eps}
\caption{(The Porcupine) Same as Figure \ref{fig:horseshoe} except G0.30-0.26 filaments are displayed.
\label{fig:porcupine}}
\end{sidewaysfigure}

\begin{sidewaysfigure}
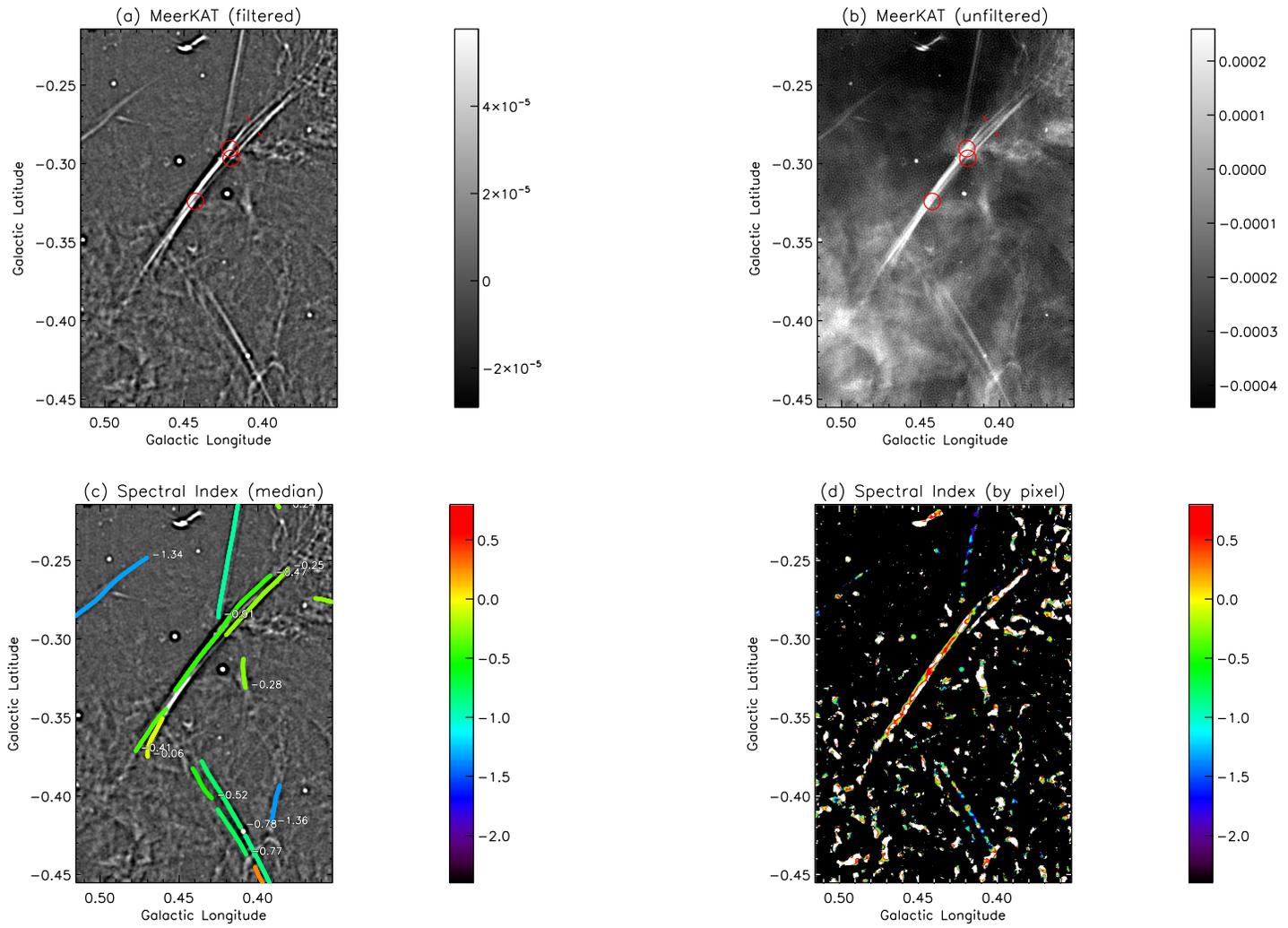

\plottwo{{cross_section_filtered_G0.43-0.33}.eps}{{cross_section_G0.43-0.33}.eps}
\plottwo{{detected_G0.43-0.33}.eps}{{alpha_G0.43-0.33}.eps}
\caption{(The Contrail) Same as Figure \ref{fig:horseshoe} except G0.43-0.33 filaments are displayed.
\label{fig:contrail}}
\end{sidewaysfigure}

\begin{sidewaysfigure}
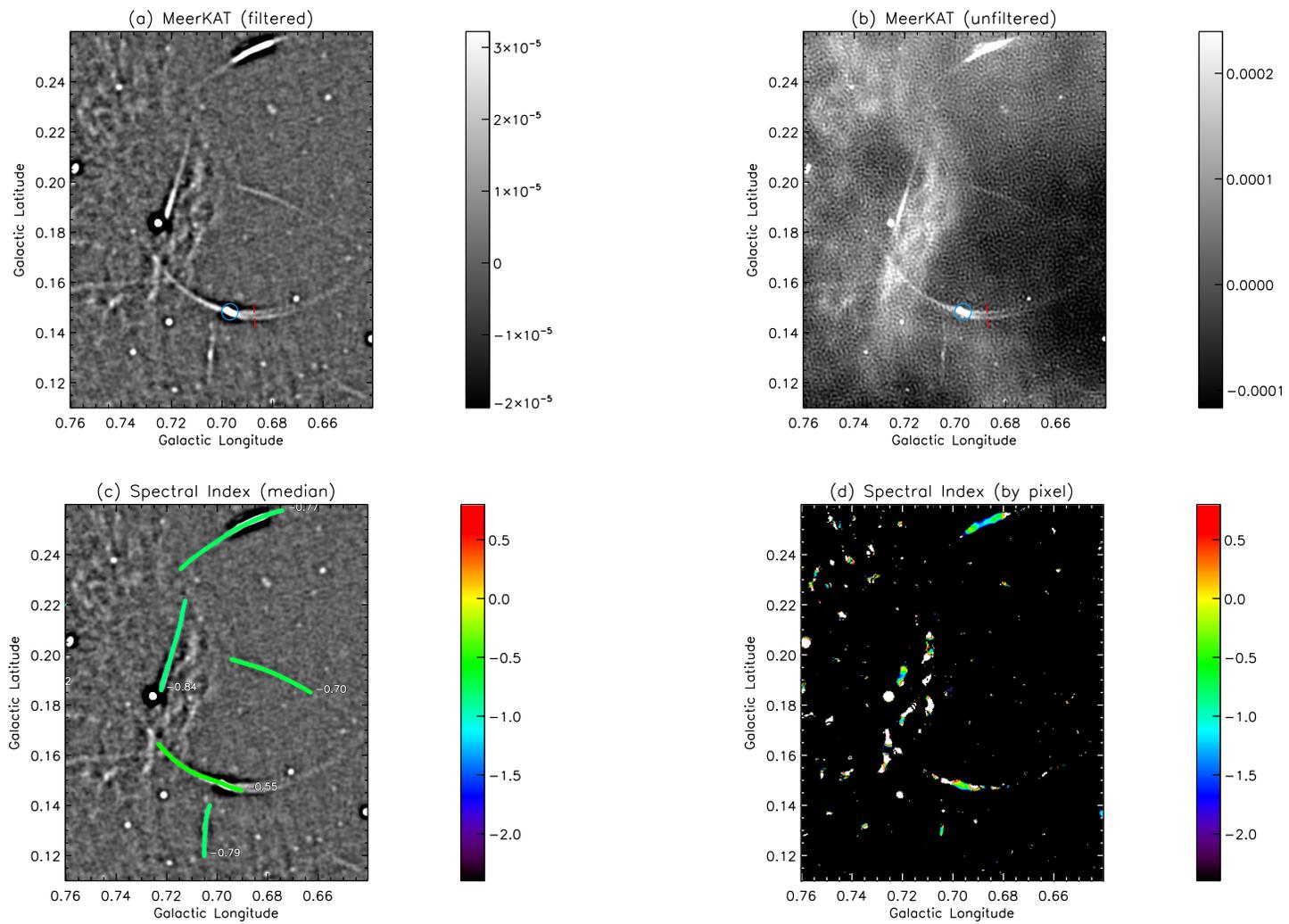

\plottwo{{cross_section_filtered_G0.70+0.19}.eps}{{cross_section_G0.70+0.19}.eps}
\plottwo{{detected_G0.70+0.19}.eps}{{alpha_G0.70+0.19}.eps}
\caption{(The Bent Fork) Same as Figure \ref{fig:horseshoe} except G0.70+0.19 filaments are displayed.
\label{fig:bentfork}}
\end{sidewaysfigure}

\begin{sidewaysfigure}
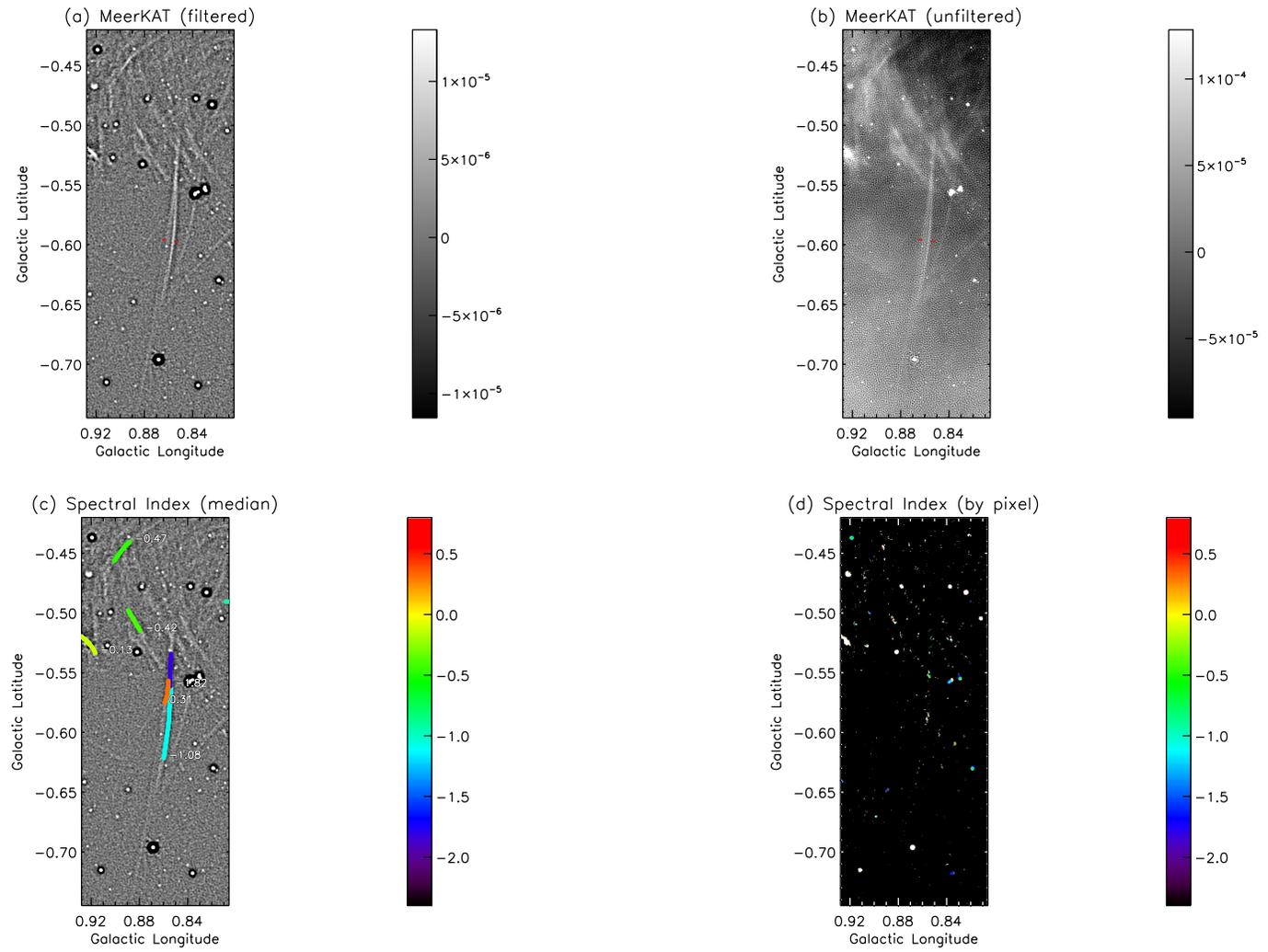

\plottwo{{cross_section_filtered_G0.87-0.58}.eps}{{cross_section_G0.87-0.58}.eps}
\plottwo{{detected_G0.87-0.58}.eps}{{alpha_G0.87-0.58}.eps}
\caption{(The Meteor) Same as Figure \ref{fig:horseshoe} except G0.87-0.58 filaments are displayed.
\label{fig:meteor}}
\end{sidewaysfigure}


\begin{sidewaysfigure}
\plottwo{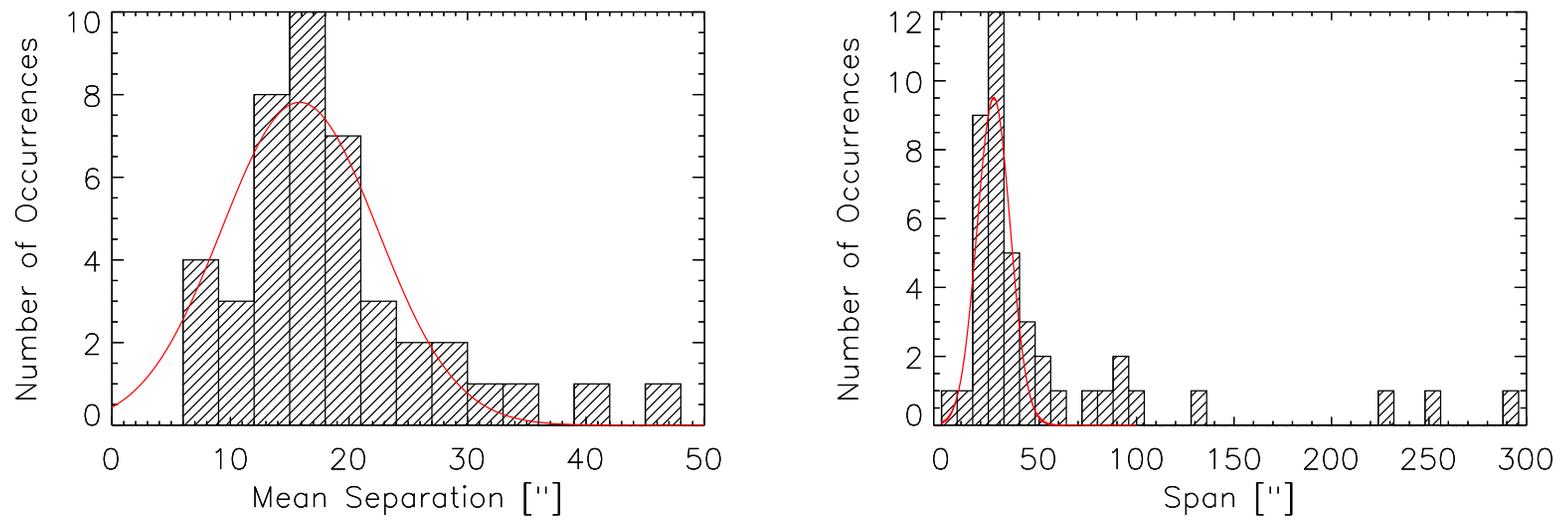}{{hist_span}.eps}
\caption{Histograms of the mean separation of filaments (left) and the total span or width (right) for the 42 cross 
sections listed in Table 1 and indicated in the previous figures. 
Gaussian fits to characterize the distributions are shown as red lines. 
Parameters of the Gaussian fits are listed at the bottom of Table 1. 
\label{fig:separation}}
\end{sidewaysfigure}

\begin{sidewaysfigure}
    \centering\vspace*{-0cm}
\includegraphics[width=8cm]{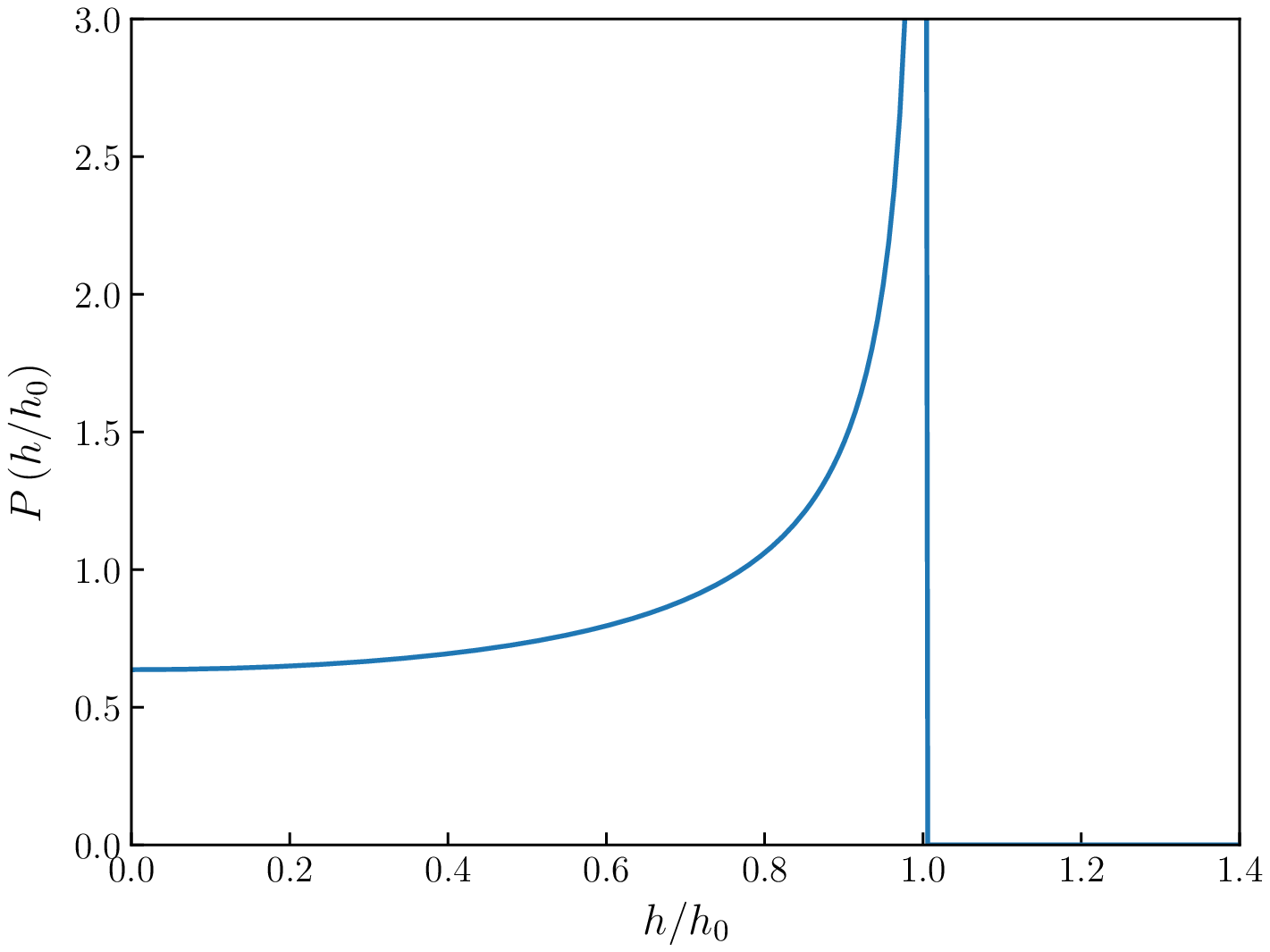}
   \includegraphics[width=8cm]{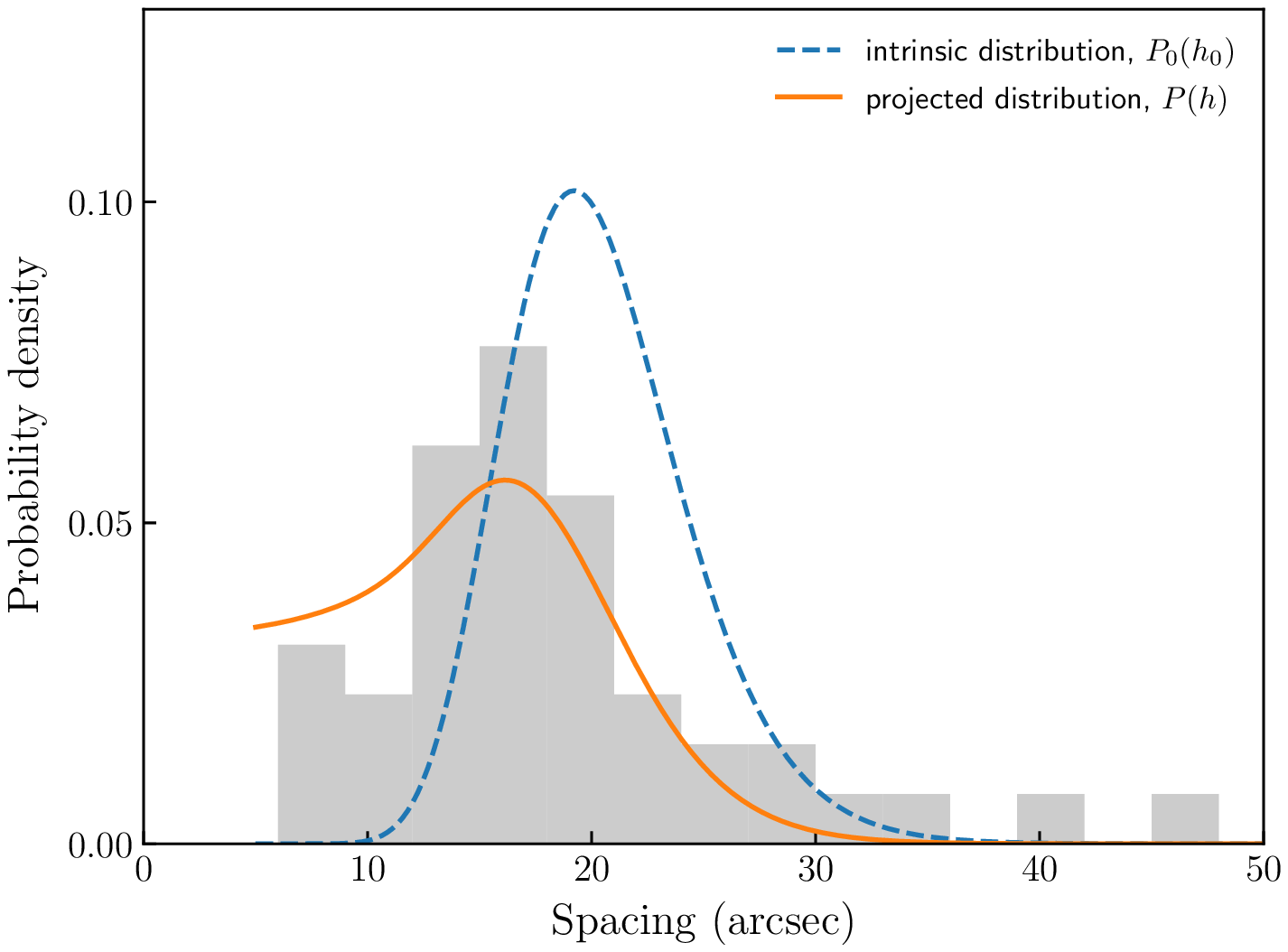}
    \caption{Distribution of apparent filament spacings (ie. as observed on the sky) for a filament population with fixed true spacing $h_0$. Comparison of a lognormal intrinsic filament spacing distribution (blue dashed curve), the resulting distribution of spacings on the sky (orange curve) and the observed distribution.} 
    \label{fig:monospaced}
    \label{fig:distributions}
\end{sidewaysfigure}

\begin{sidewaysfigure}
    \centering\vspace*{-0cm}
\includegraphics[width=8cm]{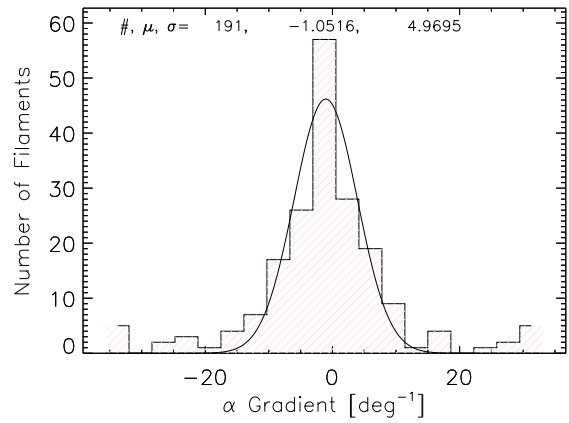}
    \caption{ 
A histogram of the linear gradients in spectral index, $\alpha$, as a function of position along the filament for 
191 long filaments in the Galactic center. The distribution is characterized by a Gaussian function with a mean ($\mu$) of $-1.1$ 
deg$^{-1}$ and a dispersion ($\sigma$) of $5$ deg$^{-1}$. The negative value of the mean gradient indicates an overall tendency for 
filaments to steepen with distance from the Galactic plane. The weights are uniform for intensities $I > 10^{-4}$ Jy~beam$^{-1}$, 
and are taken to be proportional to $I^2$ at fainter levels. Gradients are estimated using weighted linear fits between $s$ and 
$\alpha$. 
The negative value of the mean gradient indicates an overall tendency for 
filaments to steepen with distance from the Galactic plane, although exceptions to this trend are abundant.
} 
    \label{fig:gradients}
    \label{fig:gradientb}
\end{sidewaysfigure}


\end{document}